\pdfoutput=1
\documentclass[iop, apj, twocolappendix, numberedappendix]{emulateapj}

\newif\iflocal
\localfalse

\usepackage{xspace}
\usepackage{amsmath}
\usepackage{framed} 
\usepackage{txfonts}
\usepackage{epstopdf}
\usepackage{color}
\usepackage{rotating}
\usepackage{natbib}
\usepackage{ulem}
\usepackage{xspace}
\usepackage[colorlinks=true,urlcolor=black,linkcolor=blue,citecolor=blue]{hyperref}
\usepackage{sistyle}
\SIthousandsep{,}

\iflocal

\newcommand{\kpch}{\>{h^{-1}{\rm kpc}}}
\newcommand{\mpch}{\>h^{-1}{\rm {Mpc}}}

\newcommand{\msunh}{\>h^{-1} M_\odot}

\def\gcm3{\mathrm{g} / \mathrm{cm}^3}

\def\LCDM{$\Lambda$CDM\ }

\def\mvir{M_{\rm vir}}
\def\rvir{R_{\rm vir}}

\def\mtom{M_{\rm 200m}}
\def\rtom{R_{\rm 200m}}

\def\vtom{v_{\rm 200m}}
\def\ntom{N_{\rm 200m}}
\def\nutom{\nu_{\rm 200m}}

\def\mtoc{M_{\rm 200c}}
\def\rtoc{R_{\rm 200c}}

\def\mdelta{M_{\Delta}}
\def\rdelta{R_{\Delta}}

\def\vdelta{v_{\Delta}}

\def\rhoc{\rho_{\rm c}}
\def\rhom{\rho_{\rm m}}

\def\rhodelta{\rho_{\Delta}}

\def\sparta{\textsc{Sparta}\xspace}
\def\shellfish{\textsc{Shellfish}\xspace}

\def\gtsima{$\; \buildrel > \over \sim \;$}
\def\ltsima{$\; \buildrel < \over \sim \;$}
\def\prosima{$\; \buildrel \propto \over \sim \;$}
\def\gsim{\lower.7ex\hbox{\gtsima}}
\def\lsim{\lower.7ex\hbox{\ltsima}}
\def\simgt{\lower.7ex\hbox{\gtsima}}
\def\simlt{\lower.7ex\hbox{\ltsima}}
\def\simpr{\lower.7ex\hbox{\prosima}}

\newcommand{\avg}[1]{\langle #1 \rangle}

\def\tifl{t_{\rm ifl}}
\def\tdyn{t_{\rm dyn}}
\def\thubble{t_{\rm H}}

\def\gammadyn{\Gamma_{\rm dyn}}

\def\vr{v_{\rm r}}
\def\vt{v_{\rm t}}

\def\rsp{R_{\rm sp}}
\def\rrsp{r_{\rm sp}}
\def\tsp{t_{\rm sp}}
\def\rspmean{R_{\rm sp}^{\rm mn}}
\def\rspmed{R_{\rm sp}^{50\%}}
\def\rspsf{R_{\rm sp}^{75\%}}
\def\rspes{R_{\rm sp}^{87\%}}
\def\rspnn{R_{\rm sp}^{99\%}}
\def\rsptom{\rsp/\rtom}

\def\msp{M_{\rm sp}}
\def\mmsp{m_{\rm sp}}

\def\msptom{\msp/\mtom}

\def\gammarsp{$\Gamma$--$\rsp$ relation\xspace}

\def\smrmax{{\rm SMR}_{\rm max}}
\def\sigmadyn{\sigma_{\rm dyn}}

\usepackage{etoolbox}
\makeatletter
\patchcmd{\NAT@citex}
  {\@citea\NAT@hyper@{\NAT@nmfmt{\NAT@nm}\NAT@date}}
  {\@citea\NAT@nmfmt{\NAT@nm}\NAT@hyper@{\NAT@date}}
  {}
  {}
\patchcmd{\NAT@citex}
  {\@citea\NAT@hyper@{%
     \NAT@nmfmt{\NAT@nm}%
     \hyper@natlinkbreak{\NAT@aysep\NAT@spacechar}{\@citeb\@extra@b@citeb}%
     \NAT@date}}
  {\@citea\NAT@nmfmt{\NAT@nm}%
   \NAT@aysep\NAT@spacechar%
   \NAT@hyper@{\NAT@date}}
  {}
  {}
\patchcmd{\NAT@citex}
  {\@citea\NAT@hyper@{%
     \NAT@nmfmt{\NAT@nm}%
     \hyper@natlinkbreak{\NAT@spacechar\NAT@@open\if*#1*\else#1\NAT@spacechar\fi}%
       {\@citeb\@extra@b@citeb}%
     \NAT@date}}
  {\@citea\NAT@nmfmt{\NAT@nm}%
   \NAT@spacechar\NAT@@open\if*#1*\else#1\NAT@spacechar\fi%
   \NAT@hyper@{\NAT@date}}
  {}
  {}
\makeatother

\else

\fi

\shorttitle{The Splashback Radius}
\shortauthors{Diemer}

\journalinfo{The Astrophysical Journal Supplement, {\rm 231:5 (20pp), 2017 July}}
\submitted{Received 2017 March 28; revised 2017 June 2; accepted 2017 June 12; published 2014 July 14}

\begin{document}


\iflocal
\def\figdir{figs}
\else
\def\figdir{.}
\fi


\defcitealias{diemer_13_scalingrel}{DKM13}
\defcitealias{diemer_14_profiles}{DK14}
\defcitealias{diemer_15_cm}{DK15}
\defcitealias{diemer_17_rsp}{Paper II}


\title{The splashback radius of halos from particle dynamics. I. The SPARTA algorithm}
\author{Benedikt Diemer}

\affil{
Institute for Theory and Computation, Harvard-Smithsonian Center for Astrophysics, 60 Garden St., Cambridge, MA 02138, USA; \href{mailto:benedikt.diemer@cfa.harvard.edu}{benedikt.diemer@cfa.harvard.edu}
}


\begin{abstract}
Motivated by the recent proposal of the splashback radius as a physical boundary of dark matter halos, we present a parallel computer code for Subhalo and PARticle Trajectory Analysis (\sparta). The code analyzes the orbits of all simulation particles in all host halos, billions of orbits in the case of typical cosmological $N$-body simulations. Within this general framework, we develop an algorithm that accurately extracts the location of the first apocenter of particles after infall into a halo, or splashback. We define the splashback radius of a halo as the smoothed average of the apocenter radii of individual particles. This definition allows us to reliably measure the splashback radii of 95\% of host halos above a resolution limit of $1000$ particles. We show that, on average, the splashback radius and mass are converged to better than 5\% accuracy with respect to mass resolution, snapshot spacing, and all free parameters of the method.
\end{abstract}



\section{Introduction}
\label{sec:intro}

According to the widely accepted $\Lambda$CDM paradigm, dark matter (DM) collapses into a cosmic web of walls, filaments, and halos \citep{zeldovich_70, bond_96_filaments}. The collapse becomes highly non-linear, making numerical simulations the predominant tool for studying the formation and structure of DM halos \citep{klypin_83, davis_85_clustering, efstathiou_85}. The DM component in such simulations is almost always represented by virtual particles, whether the simulations include hydrodynamics or not \citep[e.g.,][]{hockney_81, kravtsov_97_art, springel_01_gadget, springel_10_arepo, hopkins_15_gizmo}. As the densities and gravitational accelerations are highest at the centers of halos, a significant part of the computing time is typically spent on the orbits of particles around halo centers. 

Despite this computational effort, the orbits of individual halo particles in cosmological volumes are practically never investigated in detail. There are several reasons for neglecting the orbital trajectories, namely that they have no directly observable impact on galaxies and that they make for a large, cumbersome dataset that is difficult to interpret. Some effort was invested in understanding how particle orbits are connected to the observed universal density profile of halos, such as through the radial orbit instability \citep{merritt_85, huss_99, macmillan_06, gajda_15}, but most studies directly investigate integrated quantities such as density and angular momentum profiles. Subhalo orbits have received somewhat more attention, largely because they can be observed as the distribution of satellite galaxies around a larger host \citep{kravtsov_04_satellites, knebe_04, reed_05_subs, benson_05, kuhlen_07, sales_07, ludlow_09, lovell_11, jiang_15}. However, modeling subhalo orbits correctly is difficult, due to physical effects such as dynamical friction \citep{vandenbosch_99, boylankolchin_08} and numerical effects such as over-merging \citep{moore_96, vandenbosch_16}.

Given that particle orbits in cosmological simulations have not been investigated in detail, they could open a new window into the processes by which halos form and evolve. For example, the splashback radius, $\rsp$, has recently been proposed as a physically motivated definition of the halo boundary \citep{diemer_14_profiles, adhikari_14, more_15}. Conventionally, halo radii such as $\rvir$ are defined to contain a particular overdensity contrast inspired by the spherical top-hat collapse model \citep{gunn_72_sphericalcollapse}, but such definitions do not generally correspond to any feature in the dynamics of particles or in the density profiles (see \citealt{diemer_17_rsp}, hereafter \citetalias{diemer_17_rsp}, for a more detailed discussion). The splashback radius, however, is directly connected to particle dynamics: it is the radius where particles reach the apocenter of their first orbit after infall. By analogy to the spherical collapse model, this radius is a well-motivated halo boundary as it separates infalling matter from matter that is orbiting in the halo potential \citep{fillmore_84, bertschinger_85, adhikari_14, shi_16}. Moreover, the splashback radius has recently been detected observationally in the stacked member galaxy density profiles of massive galaxy clusters \citep[][see also \citealt{zu_17} and \citealt{busch_17}, as well as the related observational studies of \citealt{tully_15}, \citealt{patej_16}, \citealt{adhikari_16_df}, and \citealt{umetsu_17}]{more_16, baxter_17}. 

Despite these successes, some fundamental theoretical and observational challenges remain. First, both the calibration of \citet{more_15} and the observational detections have been based on stacked density profiles from large samples of halos. Instead, we would like to determine $\rsp$ for individual simulated halos, for example in order to measure the halo-to-halo scatter in $\rsp$. \citet{mansfield_17} successfully measured nonspherical splashback shells in individual halos, but their algorithm demands a relatively high resolution of \num{50000} particles per halo and encounters difficulties at the lowest mass accretion rates. Thus, we have yet to measure $\rsp$ for the majority of halos in a cosmological simulation volume. Second, we have yet to understand the relation between the apocenter of particles' first orbits and the density profile. Particles enter halos with a distribution of energies and angular momenta, presumably causing a distribution of apocenters that we wish to relate to the steepening in the density profile. While \citet{mansfield_17} confirmed that their splashback shells are consistent with the motion of particles near the shell, we do not know what percentile of splashbacks (if any) their measurement corresponds to.

In this work, we present the first computational framework for the systematic analysis of particle orbits in $N$-body simulations, including cosmological volumes containing a large number of halos. This code is called Subhalo and PARticle Trajectory Analysis (\sparta) and will be described in detail in a future publication. Here, we focus on those parts of the algorithm that that are relevant for measuring the splashback radius. Although \sparta tracks subhalos as well as particles, we refrain from using subhalo orbits in this investigation because they suffer from dynamical friction and numerical resolution issues. We carefully test the convergence of the algorithm with mass resolution and snapshot spacing, and we show that it is capable of reliably measuring $\rsp$ in the vast majority of halos with more than $1000$ particles. In \citetalias{diemer_17_rsp}, we will present the statistics of the measured splashback radii as a function of mass accretion rate, halo mass, redshift, and cosmology.

We begin by defining the various symbols used throughout the paper in Section~\ref{sec:defs} and describing our numerical simulations in Section~\ref{sec:sims}. In Section~\ref{sec:sparta}, we describe the algorithm to measure splashback radii. We present convergence tests of the algorithm in Section~\ref{sec:res}.  We discuss the physical interpretation of our results in Section~\ref{sec:discussion}, and we summarize our conclusions in Section~\ref{sec:conclusion}.

 
\begin{deluxetable*}{lccccccccccl}
\tablecaption{$N$-body Simulations
\label{table:sims}}
\tablewidth{0pt}
\tablehead{
\colhead{Name} &
\colhead{$L$} &
\colhead{$N^3$} &
\colhead{$m_{\rm p}$} &
\colhead{$\epsilon$} &
\colhead{$\epsilon / (L / N)$} &
\colhead{$z_{\rm initial}$} &
\colhead{$z_{\rm final}$} &
\colhead{$N_{\rm snaps}$} &
\colhead{$z_{\rm f-snap}$} &
\colhead{Cosmology} &
\colhead{Reference}
}
\startdata
L2000      & $2000$  & $1024^3$ & $5.6 \times 10^{11}$ & $65$   & $1/30$  & $49$ & $0$    & $100$ & $20$ & $WMAP$ (Bolshoi) & \citetalias{diemer_15_cm} \\
L1000      & $1000$  & $1024^3$ & $7.0 \times 10^{10}$ & $33$   & $1/30$  & $49$ & $0$    & $100$ & $20$ & $WMAP$ (Bolshoi) & \citetalias{diemer_13_scalingrel} \\
L0500      & $500$   & $1024^3$ & $8.7 \times 10^{9}$  & $14$   & $1/35$  & $49$ & $0$    & $100$ & $20$ & $WMAP$ (Bolshoi) & \citetalias{diemer_14_profiles} \\
L0250      & $250$   & $1024^3$ & $1.1 \times 10^{9}$  & $5.8$  & $1/42$  & $49$ & $0$    & $100$ & $20$ & $WMAP$ (Bolshoi) & \citetalias{diemer_14_profiles} \\
L0125      & $125$   & $1024^3$ & $1.4 \times 10^{8}$  & $2.4$  & $1/51$  & $49$ & $0$    & $100$ & $20$ & $WMAP$ (Bolshoi) & \citetalias{diemer_14_profiles} \\
L0063      & $62.5$  & $1024^3$ & $1.7 \times 10^{7}$  & $1.0$  & $1/60$  & $49$ & $0$    & $100$ & $20$ & $WMAP$ (Bolshoi) & \citetalias{diemer_14_profiles} \\
L0031      & $31.25$ & $1024^3$ & $2.1 \times 10^{6}$  & $0.25$ & $1/122$ & $49$ & $2$    & $64$  & $20$ & $WMAP$ (Bolshoi) & \citetalias{diemer_15_cm} \\
TestSim200 & $62.5$  & $256^3$  & $1.1 \times 10^{9}$  & $5.8$  & $1/42$  & $49$ & $-0.1$ & $193$ & $9$  & $WMAP$ (Bolshoi) & This work \\
TestSim100 & $62.5$  & $256^3$  & $1.1 \times 10^{9}$  & $5.8$  & $1/42$  & $49$ & $-0.1$ & $96$  & $9$  & $WMAP$ (Bolshoi) & This work \\
TestSim50  & $62.5$  & $256^3$  & $1.1 \times 10^{9}$  & $5.8$  & $1/42$  & $49$ & $-0.1$ & $48$  & $9$  & $WMAP$ (Bolshoi) & This work
\enddata
\tablecomments{The $N$-body simulations used in this paper, where $L$ denotes the box size in comoving $\mpch$, $N^3$ the number of particles, $m_{\rm p}$ the particle mass in $\msunh$, $\epsilon$ the force softening length in physical $\kpch$, $z_{\rm initial}$ and $z_{\rm final}$ the redshift range of the simulation, $N_{\rm snaps}$ the number of snapshots written to disk, and $z_{\rm f-snap}$ the redshift of the first snapshot. The references correspond to \citet[][\citetalias{diemer_13_scalingrel}]{diemer_13_scalingrel}, \citet[][\citetalias{diemer_14_profiles}]{diemer_14_profiles}, and \citet[][\citetalias{diemer_15_cm}]{diemer_15_cm}.}
\end{deluxetable*}

\section{Definitions}
\label{sec:defs}

Throughout the paper, we assume a flat $\Lambda$CDM cosmology with a mean matter (cold dark matter and baryon) density $\rho_{\rm m}$ and a critical density $\rho_{\rm c}$. 

\subsection{Halo Radii and Masses}
\label{sec:method:defs:rm}

We denote three-dimensional radii measured from the halo center as $r$, and reserve capital $R$ for specific radii used to define the halo boundary. The spherical overdensity mass of a halo is defined as the mass within the radius enclosing a density $\rhodelta$ where $\Delta$ is an overdensity with respect to $\rho_{\rm m}$ or $\rho_{\rm c}$ at a particular redshift, such that
\begin{equation}
\label{eq:mdelta_m}
M_{\Delta \rm m} = M(<R_{\Delta \rm m})= \frac{4 \pi}{3} \Delta\rho_{\rm m}(z)R^3_{\Delta {\rm m}} \,,
\end{equation}
for example $\rtom$ and $\mtom$, or 
\begin{equation}
\label{eq:mdelta_c}
M_{\Delta \rm c} = M(<R_{\Delta \rm c})= \frac{4 \pi}{3} \Delta\rho_{\rm c}(z)R^3_{\Delta {\rm c}} \,,
\end{equation}
for example $\rtoc$ and $\mtoc$. The labels $\mvir$ and $\rvir$ indicate a varying overdensity $\Delta_{\rm vir}(z)$ which we compute using the approximation of \citet{bryan_98_virial}.

In keeping with this scheme, we use $\rrsp$ for the splashback radius (i.e., the apocenter of the first orbit) of individual tracers such as particles and $\rsp$ for the overall splashback radius of a halo. The corresponding masses $\mmsp$ and $\msp$ are defined as the masses enclosed by those radii. We will derive $\rsp$ from a distribution of the $\rrsp$ of the particles, and we investigate a number of definitions with respect to this distribution. We use superscripts to denote their relation to the particle distribution, such as $\rspmean$ for the mean, $\rspmed$ for the median, and $\rspsf$ for the $75$th percentile of the distribution.

The splashback radius, and the outer profiles of halos in general, are most universal in units of any radius defined with respect to the mean density of the universe, particularly $\rtom$ \citep{diemer_14_profiles, lau_15}. Thus, our method for determining $\rsp$ is largely based on $\rtom$ and $\mtom$, and we use those quantities throughout the paper unless otherwise noted. We sometimes express halo mass as peak height, $\nu$, which is defined as
\begin{equation}
\nu \equiv \nutom \equiv \frac{\delta_{\rm c}}{\sigma(\mtom, z)} = \frac{\delta_{\rm c}}{\sigma(\mtom, z = 0) \times D_+(z)}
\label{eq:nu}
\end{equation}
where $\delta_{\rm c} = 1.686$ is the critical overdensity for top-hat collapse \citep[][ignoring a weak dependence on cosmology and redshift]{gunn_72_sphericalcollapse}, $D_+(z)$ is the linear growth factor normalized to unity at $z = 0$ \citep[e.g.][]{eisenstein_99}, and $\sigma(M)$ denotes the rms density fluctuation in a sphere whose radius is the Lagrangian radius corresponding to mass $M$. We use the fitting function of \citet{eisenstein_98} to compute the linear power spectrum on which the variance is based.

\subsection{Dynamical Time and Mass Accretion Rate}
\label{sec:method:defs:dyn}

The dynamical time of halos, $\tdyn$, will be useful as a fundamental time unit for various purposes. However, numerous different definitions are used throughout the literature and are often not carefully distinguished. Generally, $\tdyn$ is defined as the ratio of a characteristic size and a characteristic velocity, where the size can either be the halo radius $\rdelta$ (time to pericenter), its diameter $2 \rdelta$ (crossing time), or its circumference $2 \pi \rdelta$ (orbital time), and the velocity is
\begin{equation}
\label{eq:vdelta}
\vdelta \equiv \sqrt{\frac{G\mdelta}{\rdelta}} \,.
\end{equation}
Here, we are interested in the time it takes a particle to reach the apocenter of its first orbit after infall, that is, the time until splashback. Thus, we define the dynamical time as
\begin{equation}
\label{eq:tdyn}
\tdyn(z) \equiv t_{\rm cross}(z) = \frac{2 \rdelta}{\vdelta} = 2^{3/2} \thubble(z) \left( \frac{\rhodelta(z)}{\rhoc(z)} \right)^{-1/2}
\end{equation}
where $\thubble$ is the Hubble time, 
\begin{equation}
\label{eq:thubble}
\thubble(z) \equiv \frac{1}{H(z)} = \sqrt{\frac{3}{8 \pi G \rhoc(z)}} \,.
\end{equation}
We note that at high redshift $\rhom \approx \rhoc$, so that for $\Delta = {\rm 200m}$ the dynamical time is a fixed fraction of the Hubble time, $\tdyn \approx (2^{3/2}/\sqrt{200}) \thubble = \thubble / 5$.

The dynamical time is the basis for our definition of the mass accretion rate. While theoretical models often use an instantaneous (or constant) accretion rate $s \equiv d \log(M)/d\log(a)$, such a definition is of little practical use when dealing with simulation data because the noisy nature of the mass accretion histories means that the scatter in $s$ increases without upper bound as the time interval decreases. Thus, \citet{diemer_14_profiles} defined the mass accretion rate over a finite range of time:
\begin{equation}
\label{eq:gammadk14}
\Gamma_{\rm DK14} \equiv \frac{\Delta \log(M)}{\Delta \log(a)} = \frac{\log(M_1) - \log(M_0)}{\log(a_1) - \log(a_0)} \,,
\end{equation}
where the $a_0$-$a_1$ pairs were chosen manually to correspond to roughly a crossing time \citep[see also][]{lau_15, more_15, mansfield_17}. As we expect $\rsp$ to be sensitive to the accretion history during a particle's orbit, we adjust their definition by setting the time interval to the dynamical time:
\begin{equation}
\label{eq:gammadyn}
\gammadyn(t) \equiv \frac{\log[M(t)] - \log[M(t-\tdyn)]}{\log[a(t)] - \log[a(t - \tdyn)]}
\end{equation}
where $M = \mtom$.


\section{$N$-body Simulations}
\label{sec:sims}

We apply our algorithm to a suite of dissipationless $N$-body simulations of different box sizes and resolutions (Table~\ref{table:sims}). Our fiducial cosmology is the same as that of the Bolshoi simulation \citep{klypin_11_bolshoi} and is consistent with $WMAP7$ \citep{komatsu_11}, namely a flat \LCDM cosmology with $\Omega_{\rm m} = 0.27$, $\Omega_{\rm b} = 0.0469$, $\sigma_8 = 0.82$, and $n_{\rm s} = 0.95$. The initial power spectrum was generated using the \textsc{Camb} code \citep{lewis_00_camb}, and translated into initial conditions using the second-order Lagrangian perturbation theory code \textsc{2LPTic} \citep{crocce_06_2lptic}. The simulations were started at redshift $z = 49$, sufficiently high to avoid transient effects \citep{crocce_06_2lptic}. The simulations were run with the publicly available code \textsc{Gadget2} \citep{springel_05_gadget2}. 

We use the phase--space halo finder \textsc{Rockstar} \citep{behroozi_13_rockstar} to extract halos and subhalos from each simulation and the \textsc{Consistent-Trees} code \citep{behroozi_13_trees} to establish subhalo relations and assemble merger trees. The halo catalogs and merger trees used in this paper differ from those used in the papers listed in Table~\ref{table:sims}, as they were run with the most recent versions of \textsc{Rockstar} and \textsc{Consistent-Trees}, and, most importantly, used $\rtom$ as the halo radius. This definition matters because a halo is defined to be a host halo if it is not within $\rtom$ of another, larger halo. The radius used for these host and subhalo definitions was computed using only bound particles because subhalo masses can otherwise include large, spurious contributions from their hosts, leading to errors in the merger trees. For the remainder of the paper, however, we generally use $\rtom$ as computed from all particles, bound and unbound, and explicitly state when using bound-only masses and radii. For the vast majority of host halos, the difference between the two masses is small.

\begin{figure}
\centering
\includegraphics[trim = 4mm 7mm 5mm 1mm, clip, scale=0.62]{\figdir/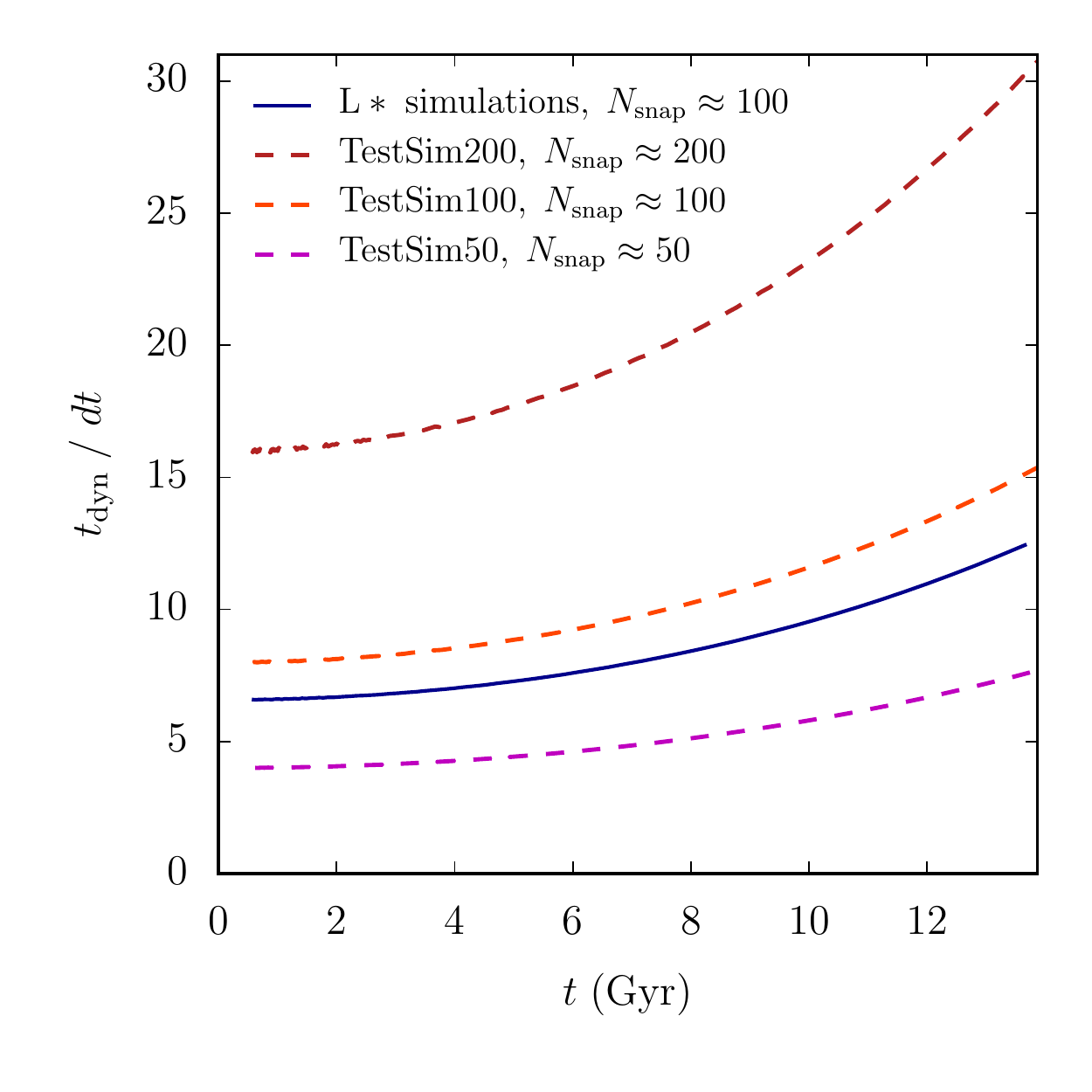}
\caption{Snapshot spacing in the simulations used in this paper, expressed as the number of snapshots per dynamical time. The solid blue line shows the spacing of the main simulations (L0031--L2000, see Table~\ref{table:sims}), which all share the same snapshot spacing, leading to a resolution of about $6.5$ snapshots per dynamical time at the earliest times and about $12$ at $z=0$. The dashed lines show the spacing for TestSim200 and its subsampled variants.}
\label{fig:conv_timeres}
\end{figure}

Typically, one worries about two types of resolution issues when dealing with $N$-body simulations: mass resolution (the number of particles in a halo) and force resolution (the length scale below which forces are non-Newtonian). With the algorithm presented in this paper, we need to consider a third resolution scale, namely the time spacing of the snapshots. In general, only about 100--200 snapshots of a simulation are saved, resulting in relatively poor time sampling. Figure~\ref{fig:conv_timeres} gives an overview of the snapshot spacing for the simulations used in this paper. In units of the dynamical time, our main simulation suite exhibits a snapshot spacing of between $0.15 \tdyn$ at early times and a little less than $0.1 \tdyn$ at $z=0$. In order to investigate the convergence properties of our algorithm with time spacing, we ran a smaller test simulation (see Table~\ref{table:sims}) with roughly twice the number of snapshots. We then subsampled the snapshots of this TestSim200 by factors of $2$ and $4$, creating test cases with realistic (TestSim100) and extremely poor snapshot resolution (TestSim50). We note that the \textsc{consistent-trees} merger tree algorithm is not expected to work reliably below about $100$ snapshots, depending on how the snapshots are spaced \citep{behroozi_13_trees}. We study the convergence of our algorithm with snapshot spacing in Section~\ref{sec:res:timeres}.


\section{Algorithm}
\label{sec:sparta}

In this section, we describe our algorithm to measure the splashback radii of simulated halos. Many of the figures used to illustrate our method and its convergence are based on our test simulations to save computational expense, and because the large number of outputs allows us to study the convergence of the algorithm with snapshot spacing. We have verified that the results (e.g., the \gammarsp) from the test simulation do not, in any important or systematic way, differ from the full-size simulations. 

\subsection{General Overview}
\label{sec:sparta:overview}

We have implemented our splashback algorithm within the somewhat more general code framework \sparta (an acronym for ``Subhalo and PARticle Trajectory Analysis'') which will be presented in a separate paper (Diemer 2017, in preparation). Here, we restrict ourselves to its most basic components and the algorithms that are relevant for determining the splashback radius and mass of halos. 

\sparta is an MPI-parallelized C code designed to follow the trajectories of dynamical tracers in particle-based simulations. It tracks halos (as defined by some halo finder and merger tree code) in a time-forward manner, starting at the first snapshot. The domain can be decomposed in two ways: either into slabs of adjustable size in each dimension or using a space-filling curve. Each process is concerned only with the halos and subhalos within its volume, with overlapping regions due to the spatial extent of halos. The volumes are rectilinear in the case of slabs and arbitrarily shaped in the case of a space-filling curve. The most important concepts in \sparta are as follows:
\begin{enumerate}
\item A halo provides the largest unit of memory, and each halo is uniquely assigned to one process. Subhalos are not treated as halos, but as dynamical tracers within a host halo (Section~\ref{sec:sparta:halos}).
\item Dynamical tracers include particles and subhalos and are assigned to a host halo. A tracer can exist in multiple halos at the same time (Sections~\ref{sec:sparta:tcrptl} and~\ref{sec:sparta:tcrsho}). At each snapshot, the positions of the tracers in each halo are connected to their previous trajectories. Only four time bins of each trajectory are kept in memory.
\item The trajectories snippets are analyzed for particular ``events'' (also referred to as ``results'') such as infall into a halo or apocentric passage (Sections~\ref{sec:sparta:resifl} and~\ref{sec:sparta:ressbk}). The algorithm presented here works on one-dimensional trajectories (i.e., radius and radial velocity).
\item When a halo ceases to exist (e.g., because it merges or because the end of the simulation is reached), the events connected to its tracers are analyzed, for example to compute the splashback radius (Section~\ref{sec:sparta:rsp}).
\end{enumerate}
The following sections describe these concepts in detail. \sparta writes all results to an hdf5 file\footnote{HDF5 stands for ``hierarchical data format'' and is a popular file format that can be read using most commonly used programming languages. See \hyperlink{http://www.hdfgroup.org}{hdfgroup.org} for more information.} and contains a python module to aid with the reading and analysis of such output files.

\subsection{Halos and Subhalos}
\label{sec:sparta:halos}

\sparta tracks all halos and subhalos in a catalog, where each halo is uniquely assigned to a process. At each snapshot, \sparta connects halos to their descendants by matching their unique IDs, saving certain halo properties and updating others. Whenever a record with no progenitor is found in the halo catalog, a new halo is created and sent to the process that is responsible for its location in space. Thus, \sparta does not have to be started at the first snapshot, but its intended mode of operation is to follow entire halo histories. Furthermore, \sparta explicitly tracks the relation between host and subhalos as defined by the merger trees. Subhalos are forced to live on the same process as their host and keep a pointer\footnote{In the context of \sparta, a ``pointer'' refers to an array index rather than a memory pointer, as memory allocation is dynamic and the actual location of objects can change at any time.} to their host, while hosts keep a list of pointers to all their subhalos. Inconsistencies in these relations cause the code to abort, ensuring that no subhalo relations are omitted by accident. The host or sub status of a halo is recorded for its full history.

At each snapshot, each process computes the boundaries of the rectilinear, potentially periodic volume that contains all its halos, including a particular search radius around the halo centers. This radius depends on the tracers in each halo and various settings (see Section~\ref{sec:sparta:tcrptl} for details). All particles contained within the rectilinear volume are loaded from snapshot files, and a tree is constructed from their positions \citep[we use the tree implementation of the \textsc{Rockstar} halo finder,][]{behroozi_13_rockstar}. For each halo, the particles within its search radius are found using a tree search, and the mass profile is constructed from those particles. 

While the halo catalog may already specify $\rtom$, \sparta computes it directly from the mass profile. If the search radius is too small to find $\rtom$, it is increased iteratively until a solution has been found. We emphasize that the values of $\rtom$ are computed from the full particle distribution, including bound and unbound particles, in order to avoid the ambiguities inherent in any unbinding procedure. For host halos, the difference between the bound and full mass profiles is generally small. For subhalos, however, the difference can be large because the density around a subhalo may never reach sufficiently low values, meaning that material from its host is included in the spherical overdensity $\rtom$ which leads to an unphysically high radius and mass. Thus, whenever we need to consider subhalo masses and radii, we use the bound-only values from the halo catalog. Specifically, when computing the mass ratio of a subhalo to its host halo, we divide the subhalo's bound-only peak mass (the highest mass it ever attained) by the host halo's bound-only mass.

\subsection{Particle Tracers}
\label{sec:sparta:tcrptl}

\begin{figure*}
\centering
\includegraphics[trim = 10mm 0mm 2mm 1mm, clip, scale=0.54]{\figdir/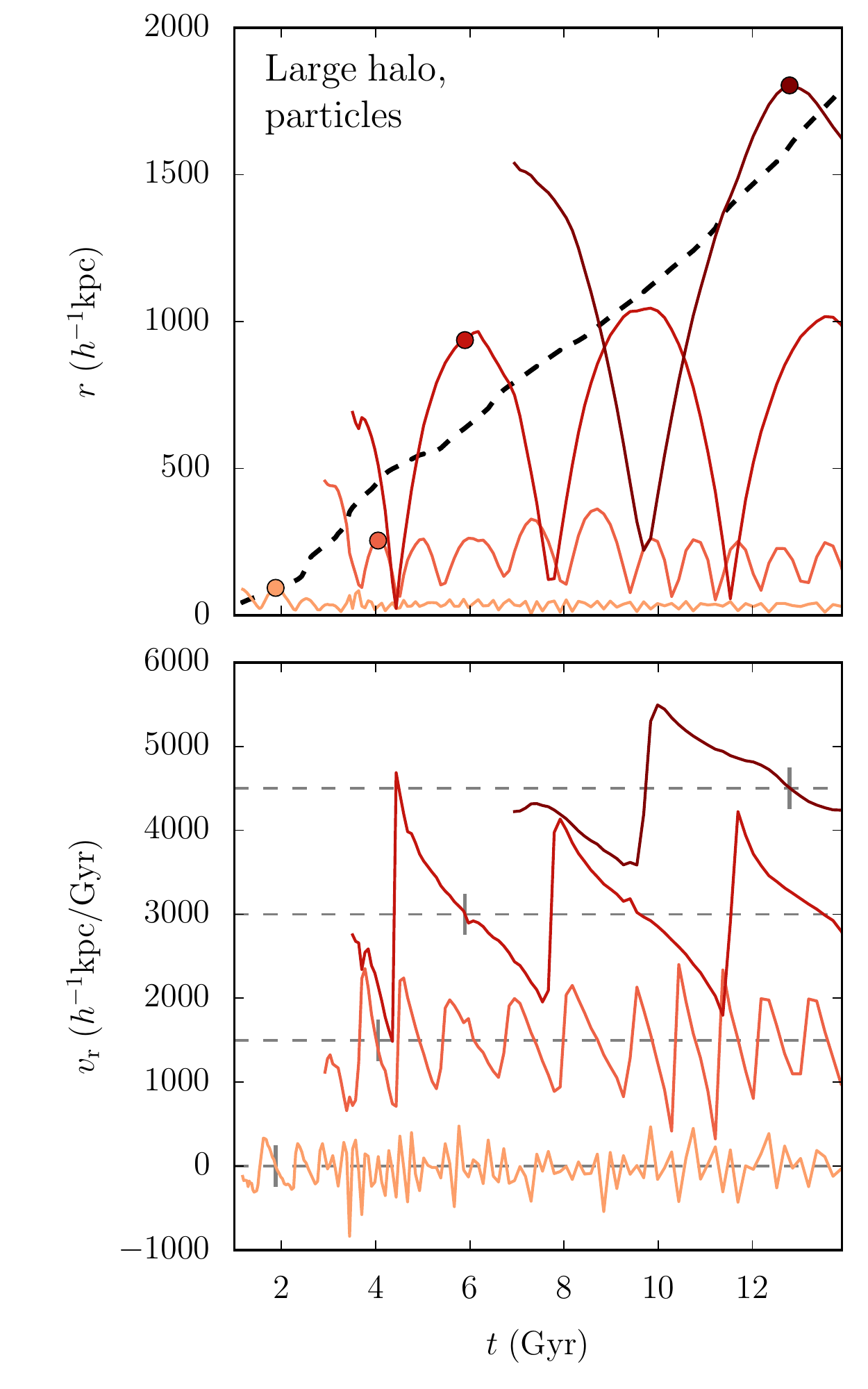}
\includegraphics[trim = 31mm 0mm 0mm 1mm, clip, scale=0.54]{\figdir/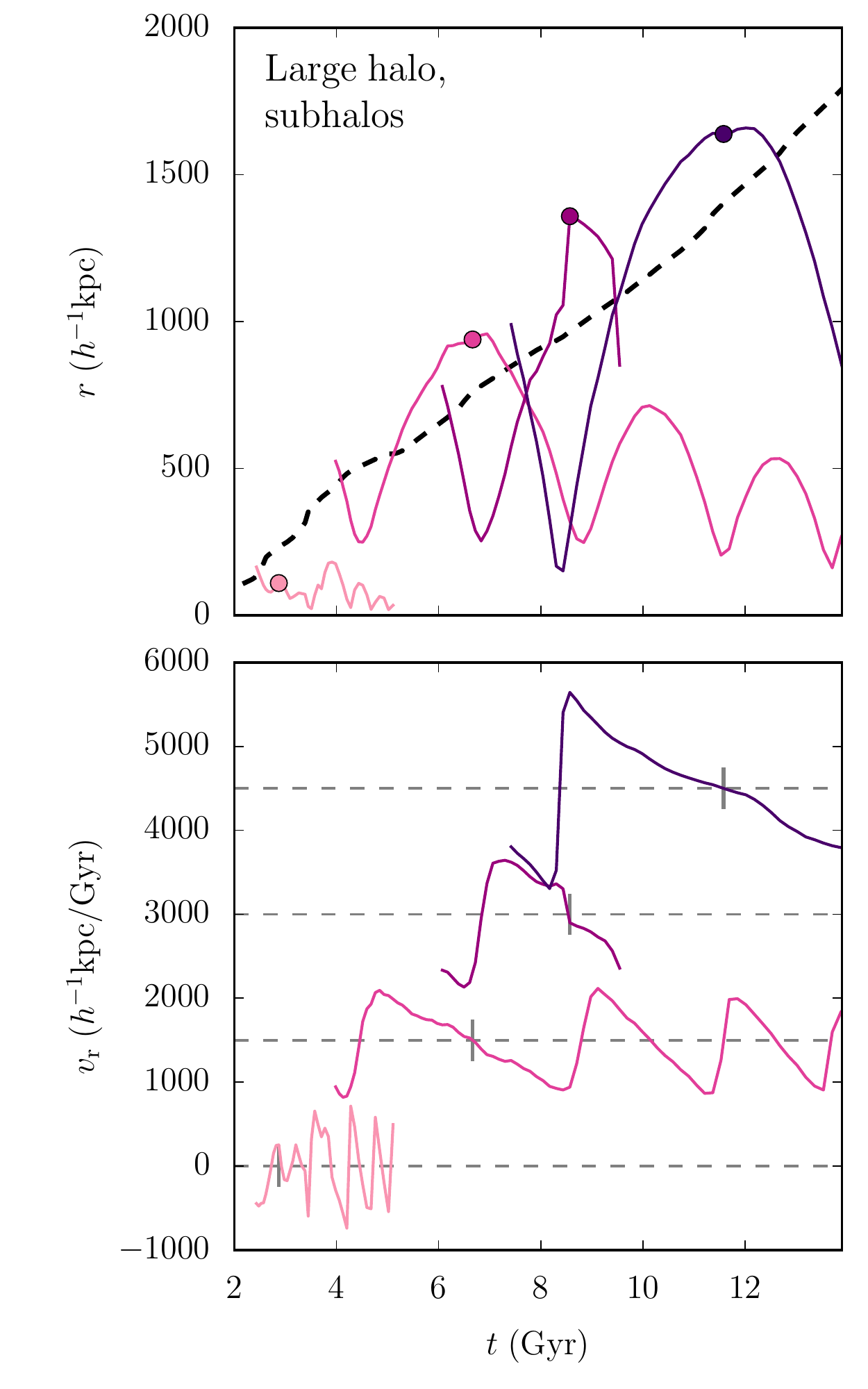}
\includegraphics[trim = 20mm 0mm 2mm 1mm, clip, scale=0.54]{\figdir/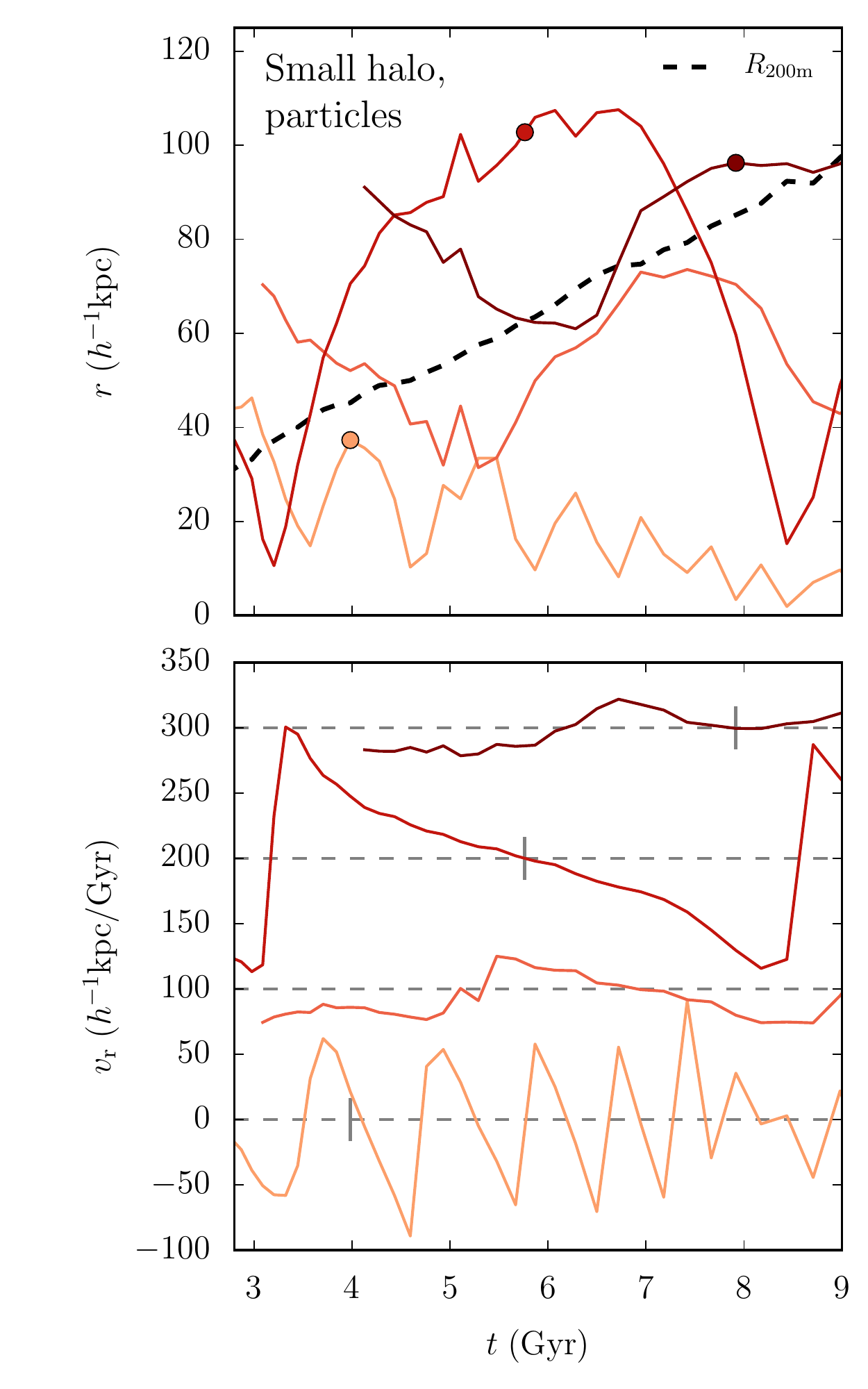}
\caption{Characteristic particle and subhalo trajectories in radius (top row) and radial velocity (bottom row). The radial velocities  are offset from each other for clarity, and the dashed lines mark zero velocity for each orbit. The columns show well-resolved particle orbits in a large halo (left, based on TestSim200), subhalo orbits in the same halo (center), and particle orbits in a much smaller halo that are more poorly resolved in time (right, based on TestSim100). The points in the top panels mark $\rrsp$ as determined by the algorithm described in Section~\ref{sec:sparta:ressbk}, and the corresponding times are marked with vertical gray lines in the bottom panels. For one of the trajectories shown in the right panels, the splashback radius was not determined because of noise in the trajectory near the orbit's pericenter.}
\label{fig:traj_halo}
\end{figure*}

A dynamical tracer is simply defined as an object (e.g., a dark matter particle or subhalo) that follows some orbit with respect to a halo center (a ``trajectory''; see Figure~\ref{fig:traj_halo} for examples). Particle tracers can be part of multiple halos at the same time because each halo keeps a separate, dynamically allocated array of tracers of each type. Particle tracers are created whenever a particle first comes within $r_{\rm create} = 2\rtom$ of a halo's center, and \sparta follows its trajectory from that point onward. Due to the enormous number of particles in a simulation, storing the full trajectories of all tracers in memory is impossible. Thus, \sparta keeps only a certain number of snapshots (four for the purposes of this paper). The particle properties that are stored are adjusted depending on the chosen output variables (for example, radius and radial velocity rather than the full three-dimensional position, unless the latter is necessary for a particular purpose).

At each snapshot, all particle trajectories are analyzed for certain events, including first infall into the halo (crossing $\rtom$, hereafter ``infall'') and reaching the apocenter of its first orbit (hereafter ``splashback''). When such an event is detected, a tracer result (hereafter simply ``result'' or ``event'') is recorded and stored in the halo's result arrays. In principle, a tracer can create any number of results, though infall and splashback, by definition, only occur at most once for each tracer in a given halo. Events are stored in separate arrays but carry the ID of their generating tracer so that they can be reconnected to each other later. 

A particle tracer's life can end for one of three reasons, namely (1) when the analysis of its trajectory has finished (i.e., when the search for both an infall and a splashback event has succeeded or failed), (2) when it strays too far from the halo to be considered a dynamical tracer of the halo potential ($r > r_{\rm delete} = 3 \rtom$), or (3) when the halo itself ceases to exist according to the catalog. At this point, the tracer object is deleted from memory, and only its results remain. If the deleted tracer had entered within $\rtom$ of the halo, we add its ID to a list of tracers to be ignored in the future. Such a list is kept by each halo for each tracer type and checked before creating new tracers to avoid accidentally treating a returning tracer as infalling for the first time. 

Our choice of the maximum radius for a particle tracer, $r_{\rm delete}$, matters somewhat, as it sets an upper limit for the size of the splashback radius. On the other hand, we cannot set $r_{\rm delete}$ to arbitrarily high values because particle tracers dominate \sparta's memory consumption, and because a large search radius would force each process to consider a large fraction of the simulation box. We discuss our choice of $r_{\rm delete} = 3\rtom$ in Section~\ref{sec:res:rdelete} and show that splashbacks at larger radii typically belong to halos that are being disrupted.

This complication also highlights why tracers are kept only for host halos. In subhalos, we cannot tell whether a particle belongs to the subhalo or its host. If the latter, the particle's motion about the subhalo center will make no sense dynamically, as the particle does not execute a trajectory in the subhalo's potential. Thus, when a halo falls into a larger host and becomes a subhalo, we remove all its tracers and add them to its ``ignore'' list. The halo keeps all its previous tracer results, but no new tracers are added while it is a subhalo. If it becomes a host halo again (i.e., if it is a ``backsplash halo''), we begin adding new tracers once again.

A special case occurs for particles that are already in the halo when it is first found in the catalog, for particles in subhalos that become host halos again, and (more generally) any particle that appears within a halo's $\rtom$ without previously having been tracked. Without going back in time, we cannot determine whether such particles are on their first or not. Thus, we ignore all particles within $\rtom$ of a newly created halo or halos that are ``reborn'' as host halos. We note that this algorithm still allows a small fraction of erroneously identified first orbits: if a particle is outside $\rtom$ when a halo is first created or reborn, but had orbited the halo before, we classify its next orbit as being its first infall. In practice, this issue (as well as the entire algorithm dealing with preexisting particles) affects only a small fraction of particles.

\subsection{Subhalo Tracers}
\label{sec:sparta:tcrsho}

Subhalo tracers are treated in much the same way as particles, except that their positions and velocities are determined by the halo finder. Subhalo tracers are created whenever a new subhalo is added to a host, but are {\it not} deleted if the subhalo leaves the host. Instead, the trajectory of the (former) subhalo is traced until the halo merges away or the simulation ends. Thus, subhalo tracers are allowed to stray far away from their previous host, which poses no performance problem as it does not change the simulation volume for which particles have to be loaded (the halo catalog is loaded in its entirety anyway).

Infall events for subhalos are recorded whenever they cross $\rtom$. As the halo merger trees are based on $\rtom$ as computed only from bound host halo particles, some subhalos may already lie within the $\rtom$ of all particles when they first become subhalos. However, infall events can be constructed from their saved trajectories in most cases. One exception occurs when halos are newly created as subhalos: in this case, the subhalo never technically crossed into its host, and no infall result is recorded. Subhalo trajectories are analyzed for splashback events in exactly the same way as particle trajectories. Figure~\ref{fig:traj_halo} shows some examples of subhalo orbits.

In our computation of the splashback radius of halos, we will entirely ignore subhalo trajectories because they suffer from dynamical friction and resolution effects (Section~\ref{sec:sparta:rsp:selection}). However, subhalos host satellites and are thus an important observational tracer of the density profile and $\rsp$ \citep{more_16}. We leave a more detailed investigation of subhalo dynamics for future work.

\subsection{Infall Results and Subhalo Tagging}
\label{sec:sparta:resifl}

The first type of result we record about a tracer corresponds to its first infall into the host halo, specifically the time when the tracer crosses $\rtom$ (i.e. has a radius $r(t_1) > \rtom(t_1)$ and $r(t_2) < \rtom(t_2)$ at a later snapshot). The exact infall time $\tifl$ is determined by linear interpolation. At first sight, the information contained in such an infall result may seem trivial: the time of infall, which, by definition, determines the radius where this infall happened as $\rtom(\tifl)$. However, these events also contain more advanced information, such as the radial and tangential velocities at infall, $\vr/\vtom$ and $\vt/\vtom$. 

Another property of particle infall will turn out to be important in determining $\rsp$, namely whether a particle fell in as part of a subhalo, and if so, what the subhalo-to-host mass ratio (SMR) of the subhalo was. Whenever a subhalo falls into a new host, its particles are ``tagged'' with this SMR, which is in turn passed to the tracer's infall result. However, the question of which particles should be tagged is somewhat tricky because at infall the subhalo's $\rtom$ already contains a significant fraction of host halo material. One could imagine a number of algorithms to select particles to tag, including the following:
\begin{enumerate}
\item All particles within a certain fraction of $\rtom$ of the subhalo. However, this selection includes many host particles who happen to be within the subhalo at the time of infall.
\item All bound particles. Given the subhalo's bound-only mass and concentration, one can estimate its potential based on a Navarro--Frenk--White (NFW) density profile \citep{navarro_95, navarro_96, navarro_97_nfw}. Unfortunately, subhalos tend to be already somewhat disrupted at infall \citep[e.g.,][]{behroozi_14_infall}. Thus, this tagging scheme captures particles with low kinetic energy that will fall in with the subhalo and tend to splash back at small radii, but fails to capture particles with low binding energy that tend to be flung out of the halo and splash back at very large radii.
\item All particles that have orbited the subhalo for some time as determined from the age of the corresponding particle tracer objects.
\item All particles that have resided {\it inside} the subhalo for some time. We look for an already existing infall result and only tag the particle if this infall result occurred a certain amount of time ago. 
\end{enumerate}
After much experimentation, we choose the last selection criterion because it makes the most physical sense and turns out to be relatively robust. In particular, when a subhalo enters a host for the first time, we perform a tree search for all particles within $2 \rtom$ of the subhalo center (where $\rtom$ is computed from only particles bound to the subhalo in order to avoid host halo contributions). For each of the found particles, we check for an existing infall result. If found, we compare the infall time to the current time. In order to avoid erroneously tagging host halo particles, we set the minimum time since the infall of a subhalo particle to $1/2\ \tdyn$. During this time, the infalling subhalo should roughly traverse $\rtom$ of the subhalo if it travels at a speed of $\vtom$. Thus, particles that ``fall'' into the subhalo at separations smaller than $\approx 2 \rtom$ of the host will not be tagged. There is one final issue we need to address: not all particles that belong to the subhalo have recorded infall times, for example those particles that were already part of a halo when it first appeared in the halo catalog. Thus, if no infall event is found, we check whether the particle ID is on the subhalo's ``ignore'' list. If it is, we assume that the particle has belonged to the subhalo for some time and tag it.

If the particle belongs to the subhalo according to these criteria, we find the corresponding tracer object in the host halo and tag it with the new sub-to-host ratio, but only if it has not previously been tagged with a larger ratio. In some cases, the tracer particle can enter the host before the subhalo infall is recorded, leaving behind an infall event without a sub-to-host ratio tag. We look for previously completed infall results of tagged particles and tag those results if found. However, if an infall event is older than $1/2\ \tdyn$ at the time of tagging, we do not tag it because the particle clearly was not part of the infalling subhalo but entered the host independently.

We find that this algorithm converges in the sense that tagging particles more aggressively (e.g., tagging particles that entered the subhalo less than $1/2\ \tdyn$ ago) does not significantly change the averaged $\rsp$. This convergence indicates that the additional particles tagged are not biased in their $\rrsp$ because they do not truly belong to the subhalo. We further discuss the convergence of our particle tagging algorithm in Section~\ref{sec:res:smr}.

\subsection{Tracer Splashback Radii}
\label{sec:sparta:ressbk}

\begin{figure*}
\centering
\includegraphics[trim = 10mm 20mm 60mm 0mm, clip, scale=0.50]{\figdir/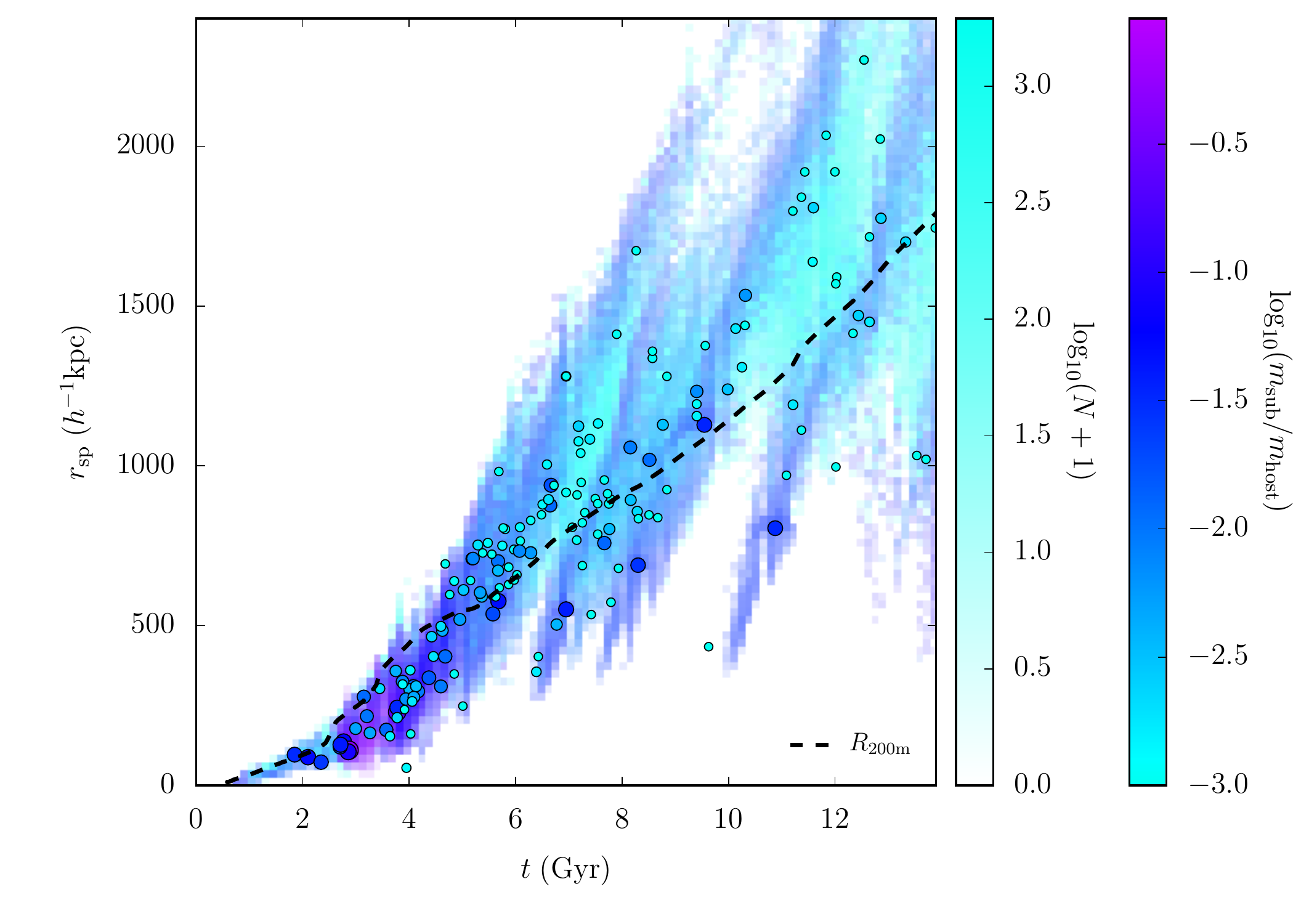}\llap{\makebox[61mm][l]{\raisebox{3.54cm}{\includegraphics[trim = 31mm 21mm 61mm 2mm, clip, scale=0.22]{\figdir/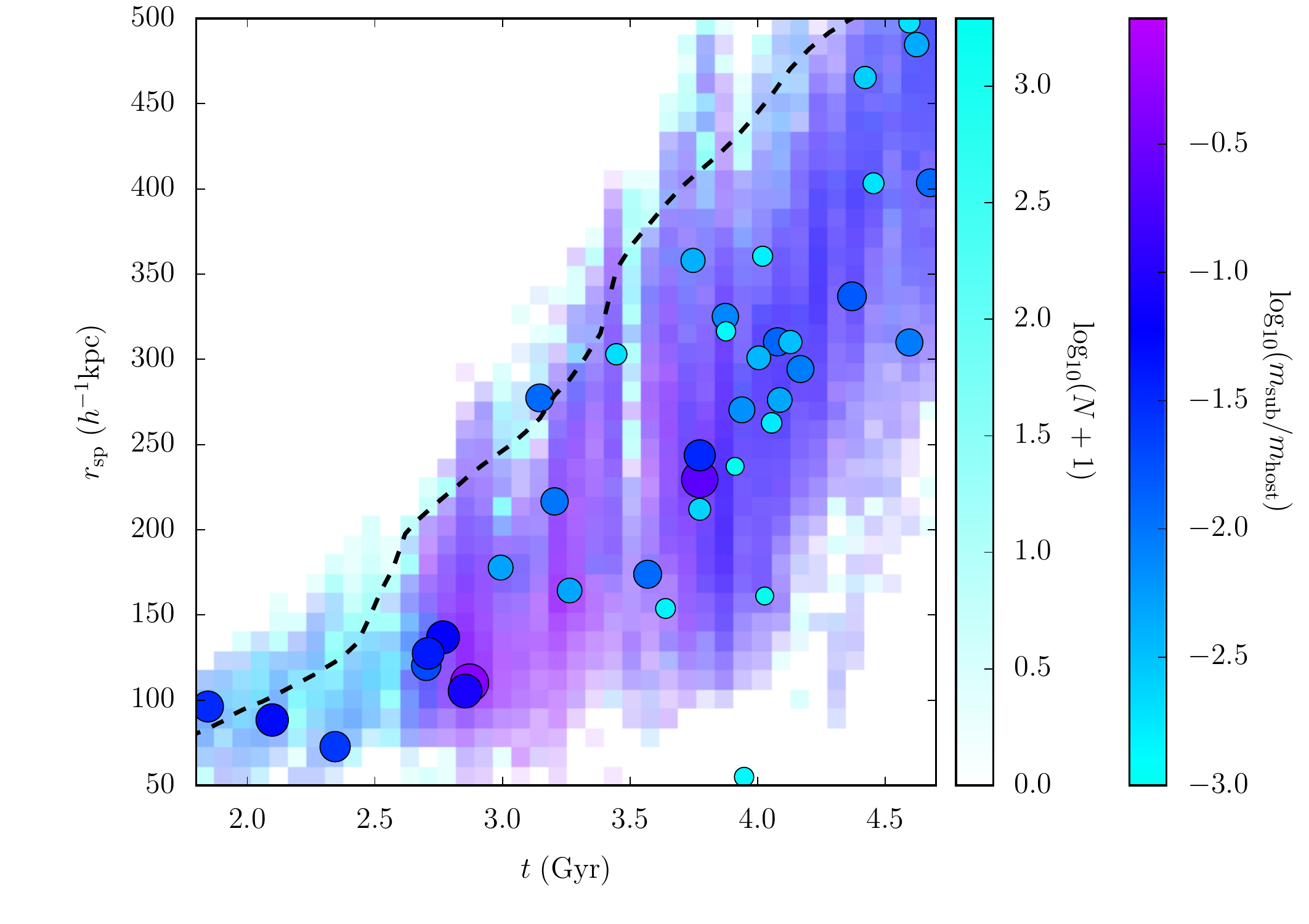}}}}
\includegraphics[trim = 0mm 20mm 3mm 0mm, clip, scale=0.50]{\figdir/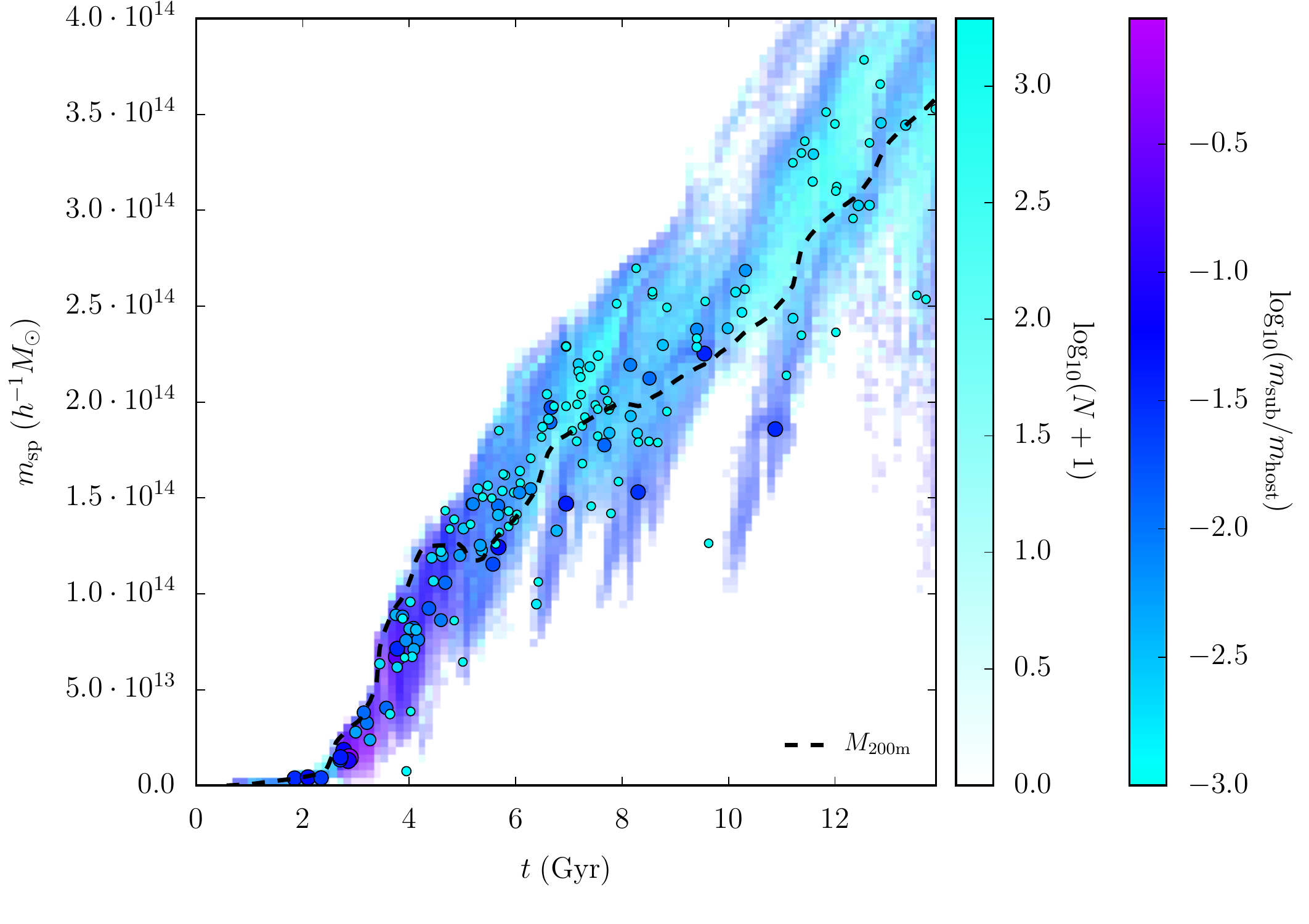}\llap{\makebox[89.4mm][l]{\raisebox{3.54cm}{\includegraphics[trim = 31mm 21mm 61mm 2mm, clip, scale=0.22]{\figdir/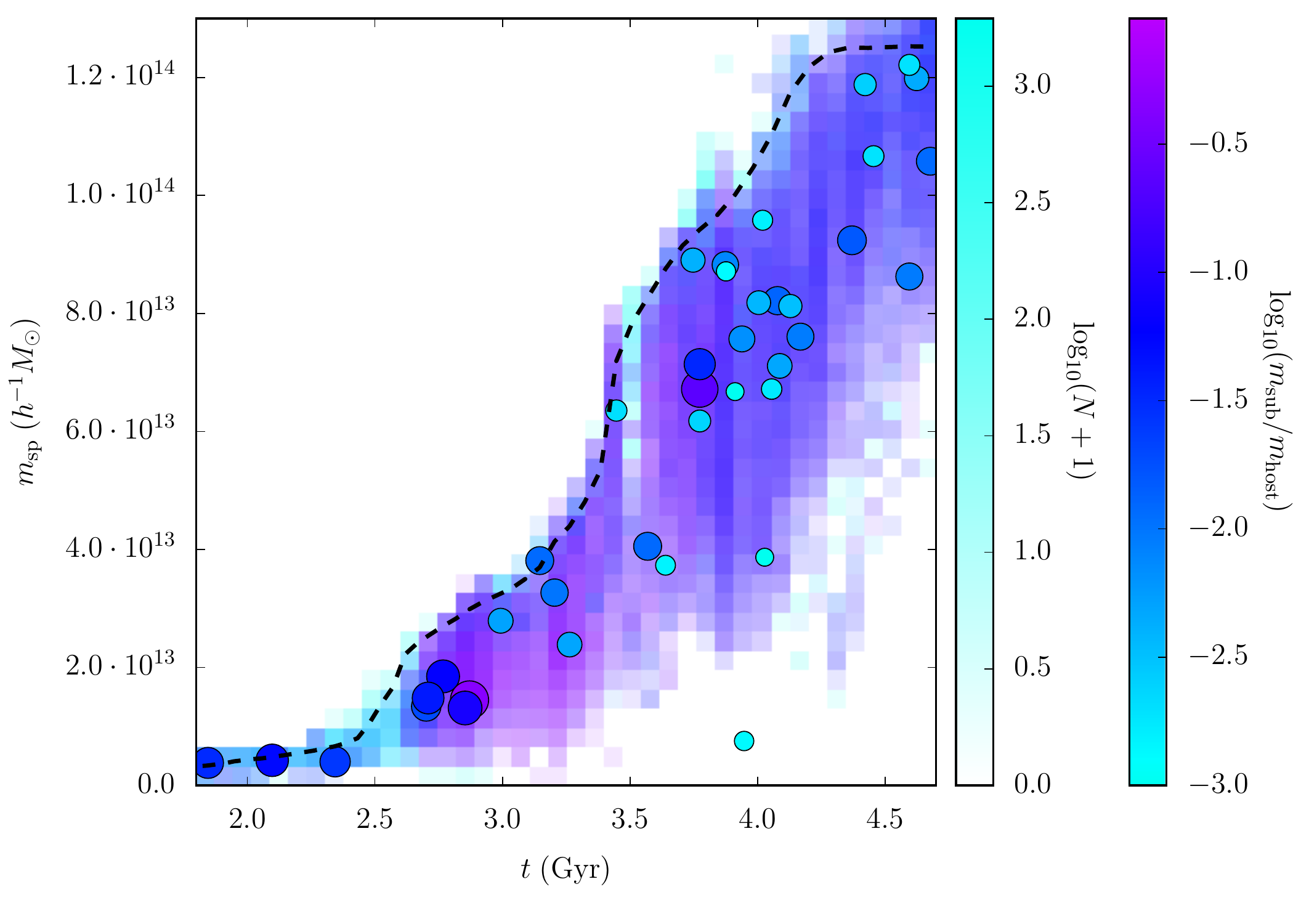}}}}
\includegraphics[trim = 10mm 3mm 60mm 0mm, clip, scale=0.50]{\figdir/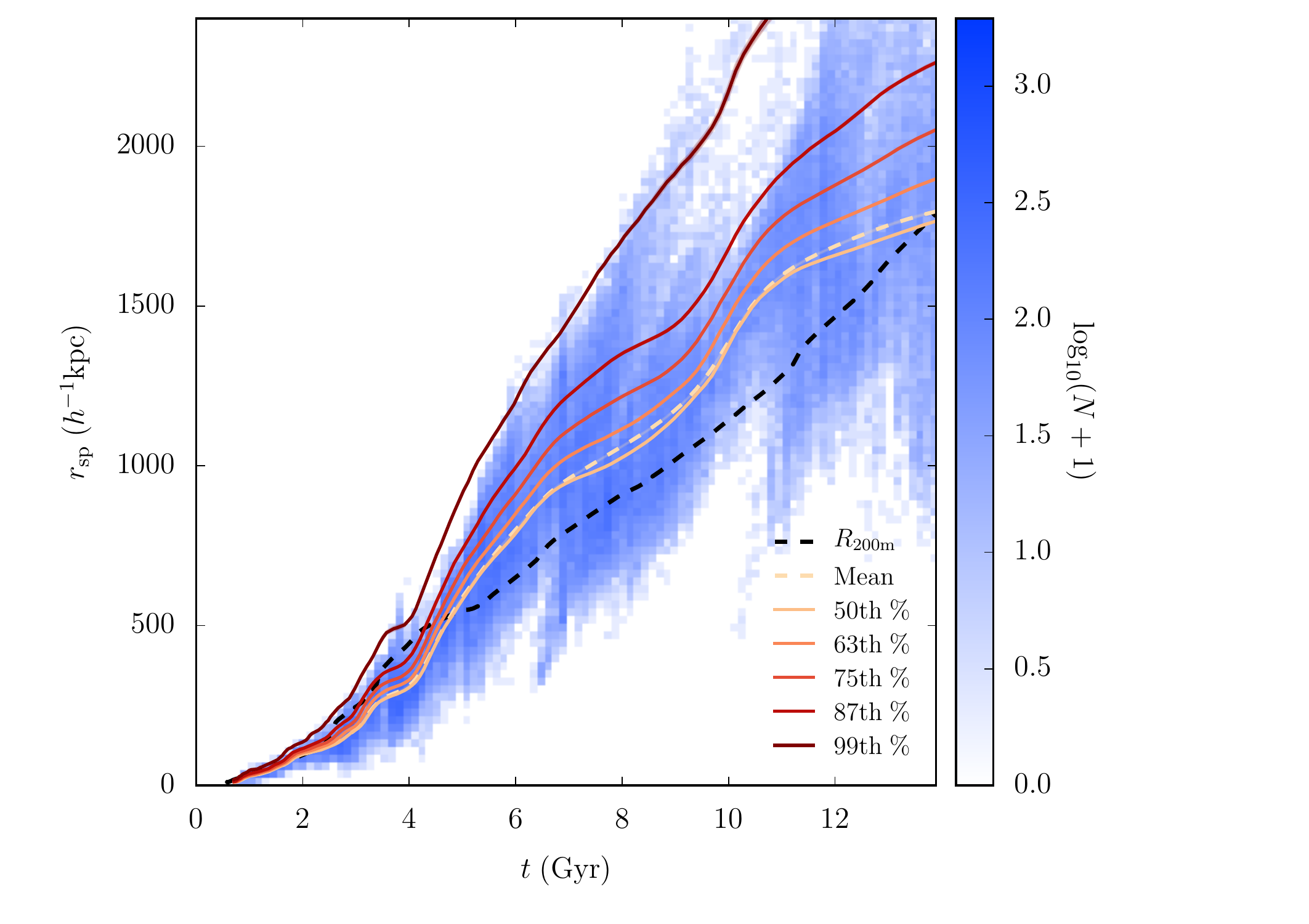}\llap{\makebox[61mm][l]{\raisebox{4.35cm}{\includegraphics[trim = 31mm 21mm 61mm 2mm, clip, scale=0.22]{\figdir/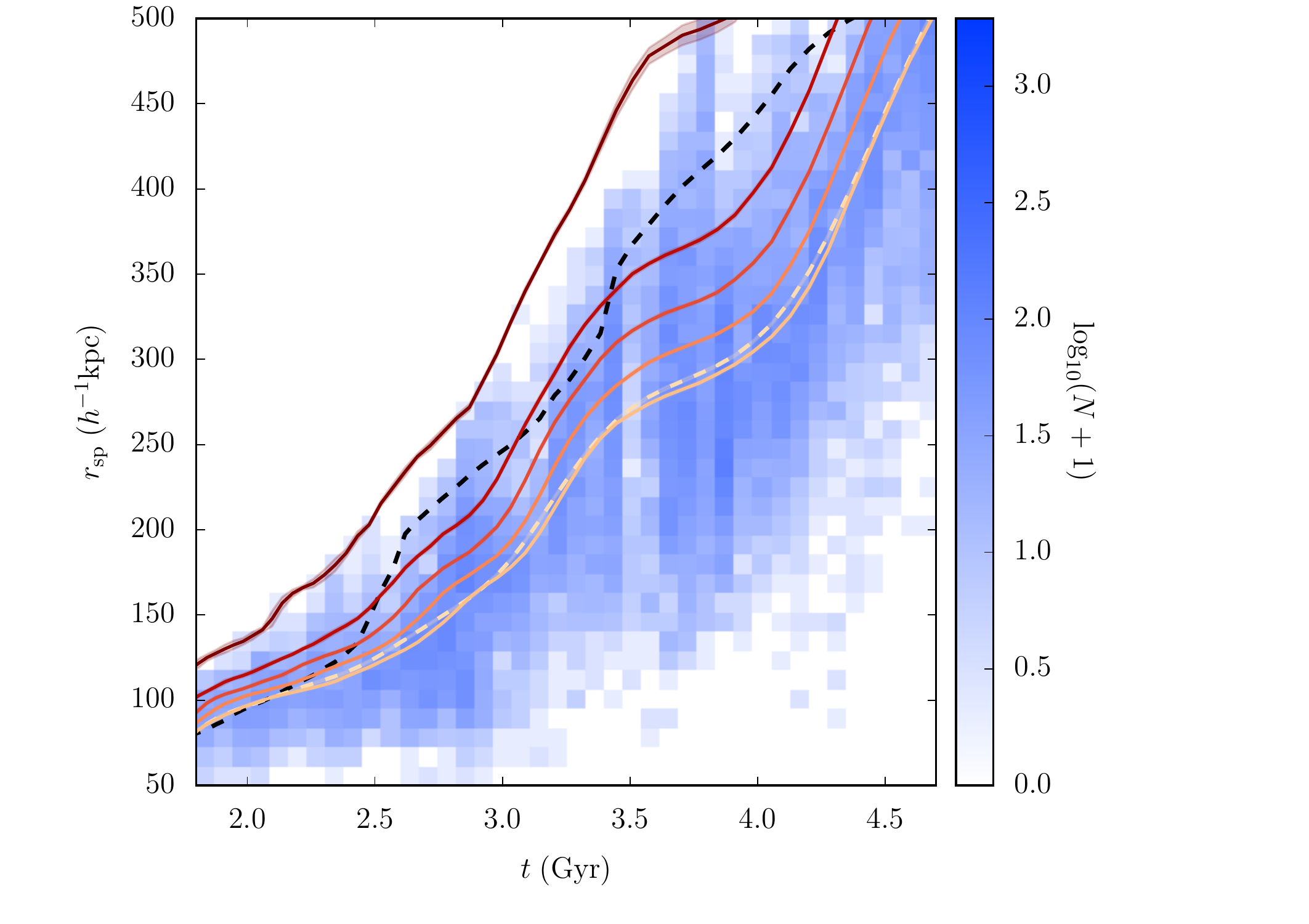}}}}
\includegraphics[trim = 0mm 3mm 3mm 0mm, clip, scale=0.50]{\figdir/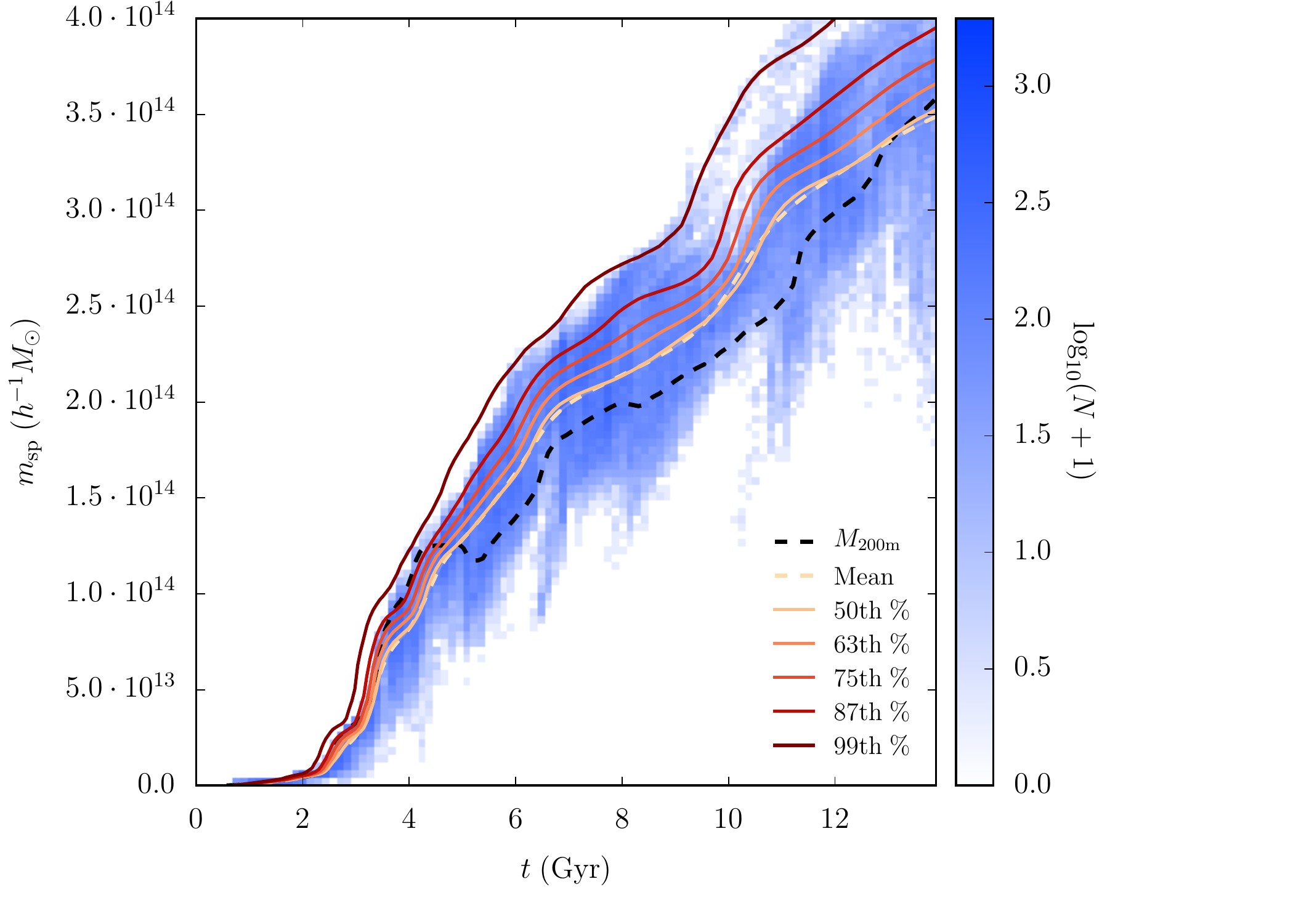}\llap{\makebox[89.4mm][l]{\raisebox{4.35cm}{\includegraphics[trim = 31mm 21mm 61mm 2mm, clip, scale=0.22]{\figdir/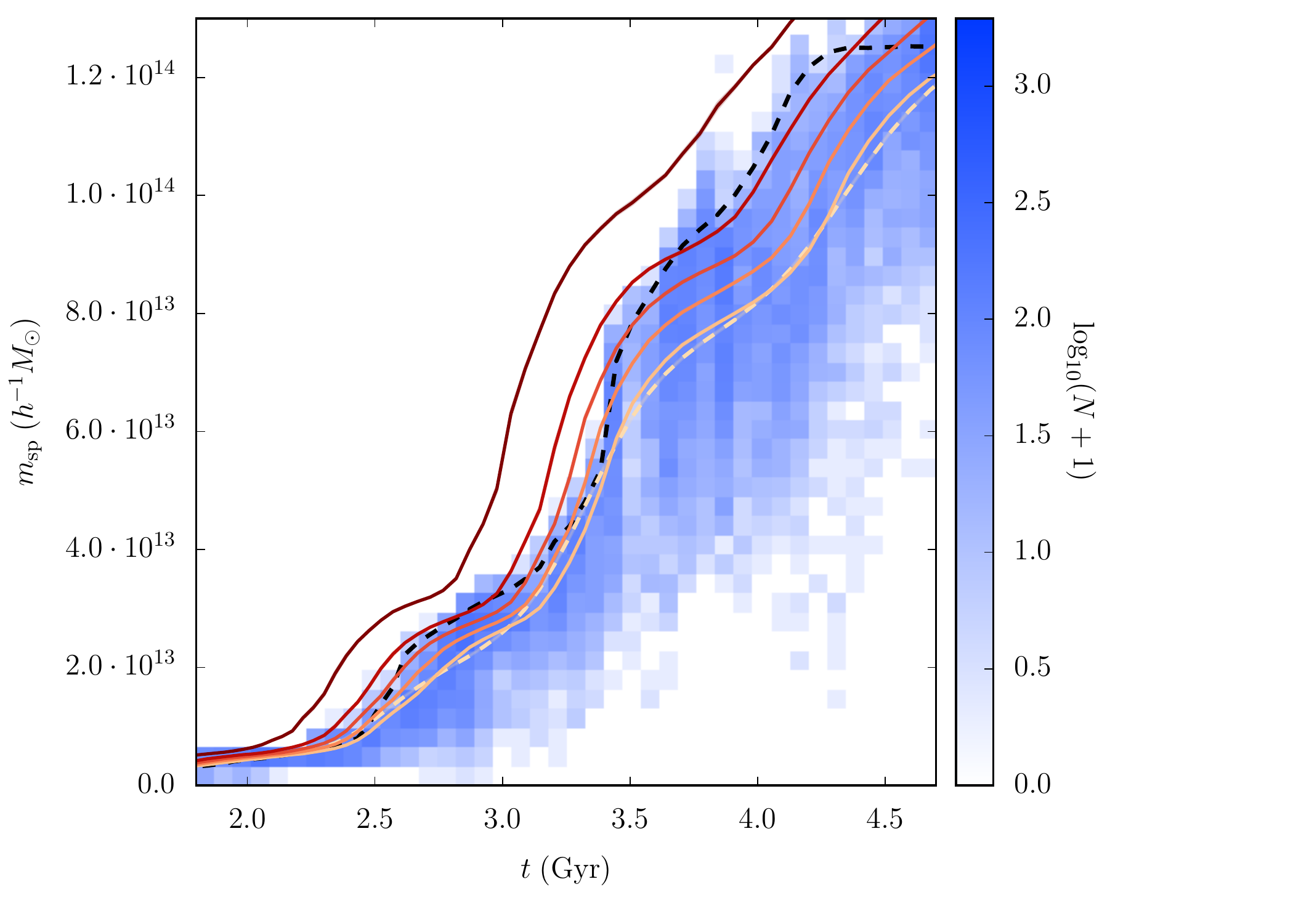}}}}
\caption{Distribution of radius (left) and enclosed mass (right) of particle and subhalo splashback events for a representative example halo. The intensity of the shading indicates the logarithmic count of particle splashback events in each bin, and the black dashed lines show the halo's $\rtom$ and $\mtom$. In the top panels, the color scale additionally indicates the mean logarithmic SMR in a bin (set to the minimum of the color scale for particles that were not part of a subhalo). The circles indicate subhalo splashbacks (on the same color scale, point size scaled by SMR). We note streaks in the distribution that are clearly due to disrupting subhalos. In the bottom panels, events with SMR $>0.01$ have been excluded from the distribution. The solid lines show the averaged splashback radius and mass of the halo as given by the mean and various percentiles. The statistical uncertainty around the estimates is shown as a shaded area, but it is too small to be visible at most times (because this particular halo is resolved by a large number of particles). The effects of subhalo infall on the $\rrsp$ distribution are still visible, but significantly reduced compared to the full distribution shown in Figure~\ref{fig:halo_rm}. The insets highlight the turbulent early assembly of the halo.}
\label{fig:halo_rm}
\end{figure*}

After a tracer has fallen into a halo, we begin looking for splashback, that is, the apocenter of the tracer's first orbit in the halo. For simplicity, we ignore all angular information and consider only the tracer's radius and radial velocity. Ideally, one would record the entire history of a tracer's orbit, but a significant fraction of the particles in a simulation are being tracked at any given time, meaning that storing many time slices of their trajectories becomes extremely memory-consuming. We thus restrict ourselves to storing only the past four values of the time $t_{\rm i}$ (in Gyr), the tracer radius $r_{\rm i}$ and halo radius $R_{\rm i}$ (in physical $\kpch$), and the radial velocity $v_{\rm i}$ (in physical $h^{-1} {\rm kpc}/{\rm Gyr}$), where $i$ represents indices running from $0$ to $3$. We begin our analysis once all four time slices have been set, that is, when the tracer has been in existence for four snapshots.

First, we seek the tracer's closest approach to the halo center, or pericenter. We detect such minima by finding times when $\vr$ switches from negative to positive values. Due to the fast velocity of the tracer near its pericenter, this switch happens rapidly and can usually be detected robustly. In particular, we look for an upward zero crossing in velocity where $v_0 < 0$, $v_1 < 0$, $v_2 > 0$, and $v_3 > 0$. Demanding two positive and negative time bins on each side of the crossing makes the algorithm robust to noise. However, some trajectories are unresolved in time, that is, have an orbital timescale smaller than four time bins. To detect such orbits, we also record the number of ``invalid'' upward crossings where only one time bin on each side is positive and negative. If we find more than one such crossing but no valid minimum, we abort the trajectory. 

Otherwise, if either $r_1$ or $r_2$ is a minimum (smaller than the adjacent radii), the velocity and radius information are consistent. We could assume a linearly changing velocity and extract the point where $v = 0$, but this method would take into account only velocity but not radial information. Instead, we extrapolate both forward and backward from the two points, writing
\begin{equation}
\label{eq:rperi}
r_{\rm peri} = r_1 + v_1 (t_{\rm peri} - t_1) = r_2 + v_2 (t_{\rm peri} - t_2)
\end{equation}
and thus
\begin{equation}
t_{\rm peri} = \frac{r_1 - r_2 - v_1 t_1 + v_2 t_2}{v_2 - v_1} \,.
\end{equation}
Because of the the two equalities in Equation~(\ref{eq:rperi}), the result tells us whether the radial and velocity information are consistent: if $t_1 < t_{\rm peri} < t_2$, the interpolation has given a valid result, and we adopt $t_{\rm peri}$ and compute $r_{\rm peri}$ from Equation~(\ref{eq:rperi}). Otherwise, we set $r_{\rm peri}$ to either $r_1$ or $r_2$, whichever has the lower value, and $t_{\rm peri}$ to the corresponding time. We linearly interpolate the halo radius $R(t)$ to $t_{\rm peri}$, and record the ratio $r_{\rm peri} / \rtom(t_{\rm peri})$.

Once the pericenter has been established, we begin to look for an apocenter, a maximum in $r$ where $\vr$ changes from positive to negative. As before, we demand that $v_0 > 0$, $v_1 > 0$, $v_2 < 0$, and $v_3 < 0$. If, for example, $v_2$ is negative but $v_3$ positive, we abort the trajectory because it is likely noisy. We apply a similar algorithm as for the pericenter: if either $r_1$ or $r_2$ is a maximum, we use the equivalent of Equation~(\ref{eq:rperi}) for $r_{\rm sp}$ and find
\begin{equation}
t_{\rm sp} = \frac{r_2 - r_1 + t_1 v_1 - t_2 v_2}{v_1 - v_2}
\end{equation}
and
\begin{equation}
r_{\rm sp} = r_1 + v_1 (t_{\rm sp} - t_1) \,.
\end{equation}
We accept this solution if $t_1 < t_{\rm sp} < t_2$. This is almost always the case if the trajectory has good time resolution, but in reality, the snapshot spacing can be a significant fraction of the halo's dynamical time, the halo centers and velocities (as determined by the halo finder) suffer from noise, and $r$ and $\vr$ often seem inconsistent due to accelerations that occurred on timescales smaller than the snapshot spacing. In such cases, we have to weigh multiple options. First, we check whether $r_0$ or $r_3$ are maxima (which happens relatively frequently). However, we also want to consider the information given by the $v$ trajectory, particularly the point where it crosses zero:
\begin{equation}
t_{\rm cross} = t_1 - \frac{v_1}{v_2 - v_1} (t_2 - t_1) \,.
\end{equation}
Thus, we assign the far-away maximum, say $r_0$, a merit function:
\begin{equation}
f_0 = \frac{ \frac{r_0}{\max{(r_1, r_2)}} - 1 }{\frac{|t_0 - t_{\rm cross}|}{\tdyn}} \,.
\end{equation}
This number weighs how much larger the radius of this new maximum is than $r_1$ and $r_2$ and compares the difference to how far in time the maximum strays away from the crossing point where $\vr = 0$. If $r_0$ or $r_3$ is a maximum, and the corresponding merit factor $f_0$ or $f_3$ exceeds $0.2$, we accept this maximum. If the merit factor is below $0.2$, we set $\tsp = t_{\rm cross}$ and interpolate to find the corresponding radius:
\begin{equation}
r_{\rm sp} = r_1 + (r_2 - r_1) \frac{t_{\rm cross} - t_1}{t_2 - t_1} \,.
\end{equation}
Once $\tsp$ and $r_{\rm sp}$ have been determined, we find the enclosed mass $\mmsp$ by interpolating a two-dimensional grid in $\log({r/\rtom})$--$t$ space which is saved for the last four snapshots in each halo, and for 50 radial bins, spaced logarithmically between $0.01 \rtom$ and the maximum tracking radius $r_{\rm delete}$. We have verified that increasing the number of radial bins does not systematically change the inferred splashback masses.

\begin{figure*}
\centering
\includegraphics[trim = 6mm 2mm 12mm 2mm, clip, scale=0.57]{\figdir/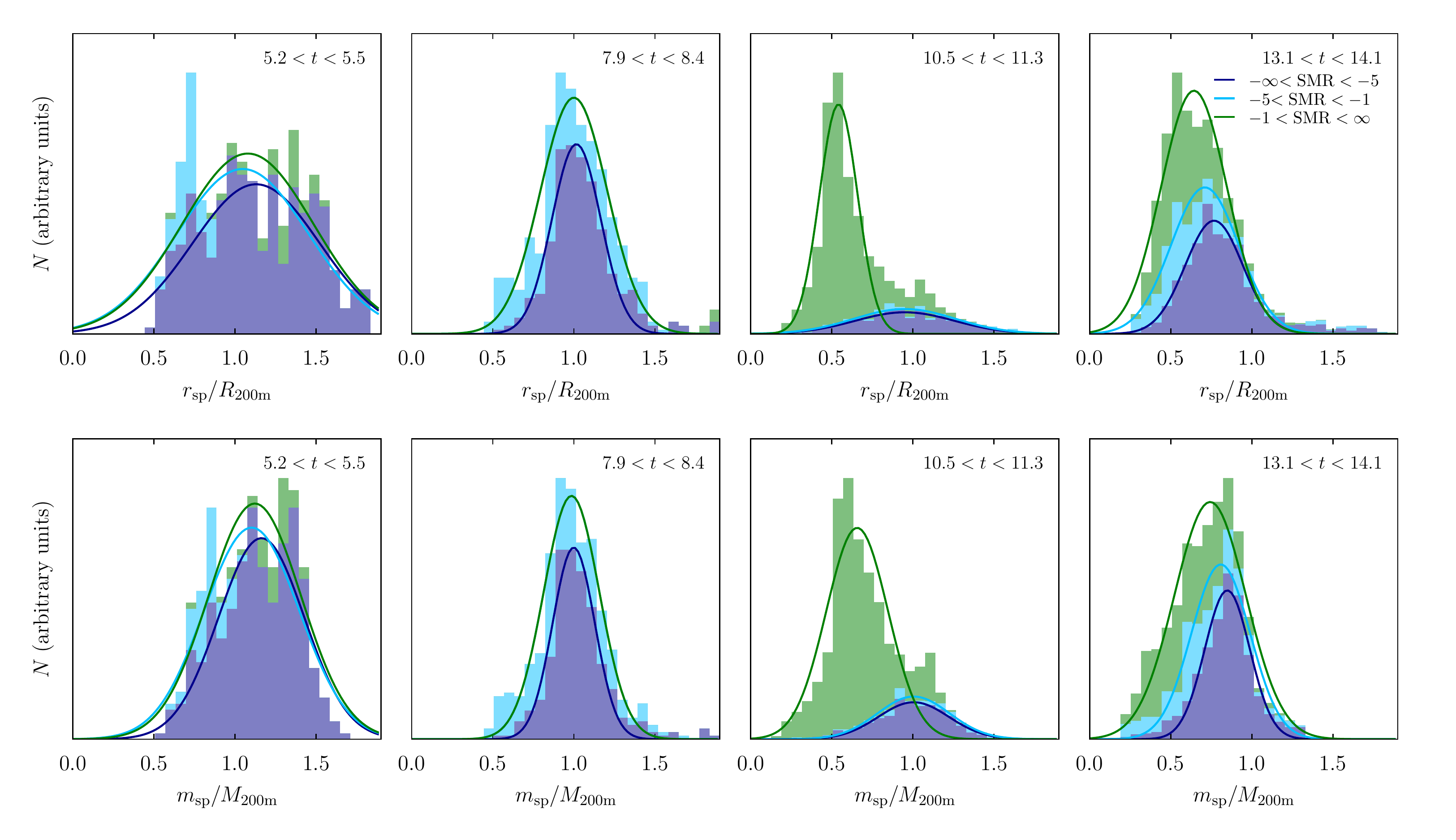}
\caption{Distribution of the splashback radius ($\rrsp/\rtom$, top row) and enclosed mass ($\mmsp/\mtom$, bottom row) of individual particles. The width of the four time bins (columns) was chosen to correspond to $0.2$ dynamical times, the smoothing time scale over which $\rrsp$ and $\mmsp$ are averaged (Section~\ref{sec:sparta:rsp:average}). The distributions were drawn from an arbitrary halo in TestSim100, but are characteristic for the different types of distributions observed. In the second time bin, a few events lie beyond the plotted range in radius. The stacked histograms differentiate the contributions from particles that entered the halo without belonging to a subhalo or as part of a subhalo smaller than $10^{-5}$ times the host mass (dark blue), those from subhalos with mass ratios between $10^{-5}$ and $0.1$, and those from major mergers with mass ratios greater than $0.1$. The solid lines show the best-fit Gaussians to the particle distribution including the contributions up to a given subhalo mass ratio, demonstrating that major mergers can impose a significant bias on the distribution. See Section~\ref{sec:sparta:rsp:selection} for a detailed discussion.}
\label{fig:rsp_dist}
\end{figure*}

A few examples of the results of our algorithm are shown in Figure~\ref{fig:traj_halo}. The particles shown in the left column fall into their halo at different times and with different velocities. In comparison, the subhalo trajectories in the center column exhibit more noise because they are based on two halo positions identified by the halo finder. Furthermore, we observe that the subhalo orbits gradually shrink due to dynamical friction before the subhalo is eventually disrupted. Finally, the right column shows another set of particle orbits from a significantly smaller halo. These orbits suffer from poorer snapshot time resolution and larger halo finder noise. In one of the four cases, the algorithm aborts because it cannot reliably identify a pericenter, due to multiple adjacent zero crossings of the radial velocity. 

The algorithm described in this section identifies a splashback event for about 90\% of the particle trajectories in TestSim100 (and for about 85\% of the subhalo trajectories). We discuss the convergence of our algorithm with snapshot spacing and mass resolution in Section~\ref{sec:res}.

\subsection{Determining the Halo Splashback Radius}
\label{sec:sparta:rsp}

While the infall and splashback events of individual particle and subhalo tracers contain a wealth of information, we are usually interested in a more compact description of the splashback radius, namely the mean, median, or some other percentile of the $r_{\rm sp}$ events as a function of time. Whenever a halo's history ends (because it merges or reaches the final snapshot of the simulation), we analyze all particle infall and splashback events to find such averages. This process consists of two main steps: selecting a sample of representative splashback events, and averaging them as a function of time.

\subsubsection{Selecting Valid Splashback Events}
\label{sec:sparta:rsp:selection}

Figure~\ref{fig:halo_rm} shows the raw data entering this analysis, namely the density of particle splashbacks. Some disturbing structures are apparent: streaks of relatively high event density that migrate outward with time. These features are the signatures of mergers. When a subhalo is tidally disrupted (often near the pericenter of its orbit), the orbits of its particles diverge. Those particles that are most bound to the host halo reach their apocenter soonest and at the smallest $\rrsp$, whereas relatively unbound particles are flung out, resulting in a late apocenter and large $\rrsp$. As expected, many of the streak features in Figure~\ref{fig:halo_rm} can be associated with a particular subhalo (round points). One might imagine that the contributions from relatively bound and unbound particles would cancel out, but subhalos suffer from dynamical friction \citep{chandrasekhar_43, vandenbosch_99, boylankolchin_08, adhikari_16_df}. The more massive the subhalo, the more its contribution is biased toward low $\rrsp$. In other words, subhalos are not faithful dynamical tracers of the host halo potential. Thus, we do not consider subhalo splashback events at all when computing $\rsp$, and furthermore we exclude splashback events from particles that entered as part of a massive subhalo.

In order to visualize the importance of this exclusion, Figure~\ref{fig:rsp_dist} shows the distribution of $\rrsp$ and $\mmsp$ at a fixed time for an example halo, with the different colors indicating splashbacks that originated from subhalos with different SMRs. The distributions vary hugely from time to time and halo to halo, but those shown in Figure~\ref{fig:rsp_dist} demonstrate a few typical cases. At the earliest times, the distribution exhibits significant shot noise due to the small number of particles. In the second time bin, no major mergers occur, and the distributions in both radius and mass are well fit by a Gaussian, though with a tail toward high radii. In the third time bin, a major merger has occurred, and subhalo particles dominate the splashback events. Due to dynamical friction, they splash back at smaller radii than the non-subhalo particles. Even the sample with $\mathrm{SMR} < 0.1$ exhibits a significant dynamical friction bias.

In order to address this issue, we exclude all particle splashbacks that originated from subhalos with $\mathrm{SMR} > 0.01$. Below this value, the resulting averaged $\rsp$ and $\msp$ do not change significantly (see Section~\ref{sec:res:smr} for a more formal discussion). The bottom panels of Figure~\ref{fig:halo_rm} show the splashback distribution after the SMR exclusion has been applied. Some of the streak features are still apparent (due to particles that were erroneously not tagged as belonging to a subhalo), but the density of such particle events is sufficiently reduced.

\subsubsection{Averaging over Splashback Events}
\label{sec:sparta:rsp:average}

\begin{figure}
\centering
\includegraphics[trim = 0mm 4mm 5mm 0mm, clip, scale=0.62]{\figdir/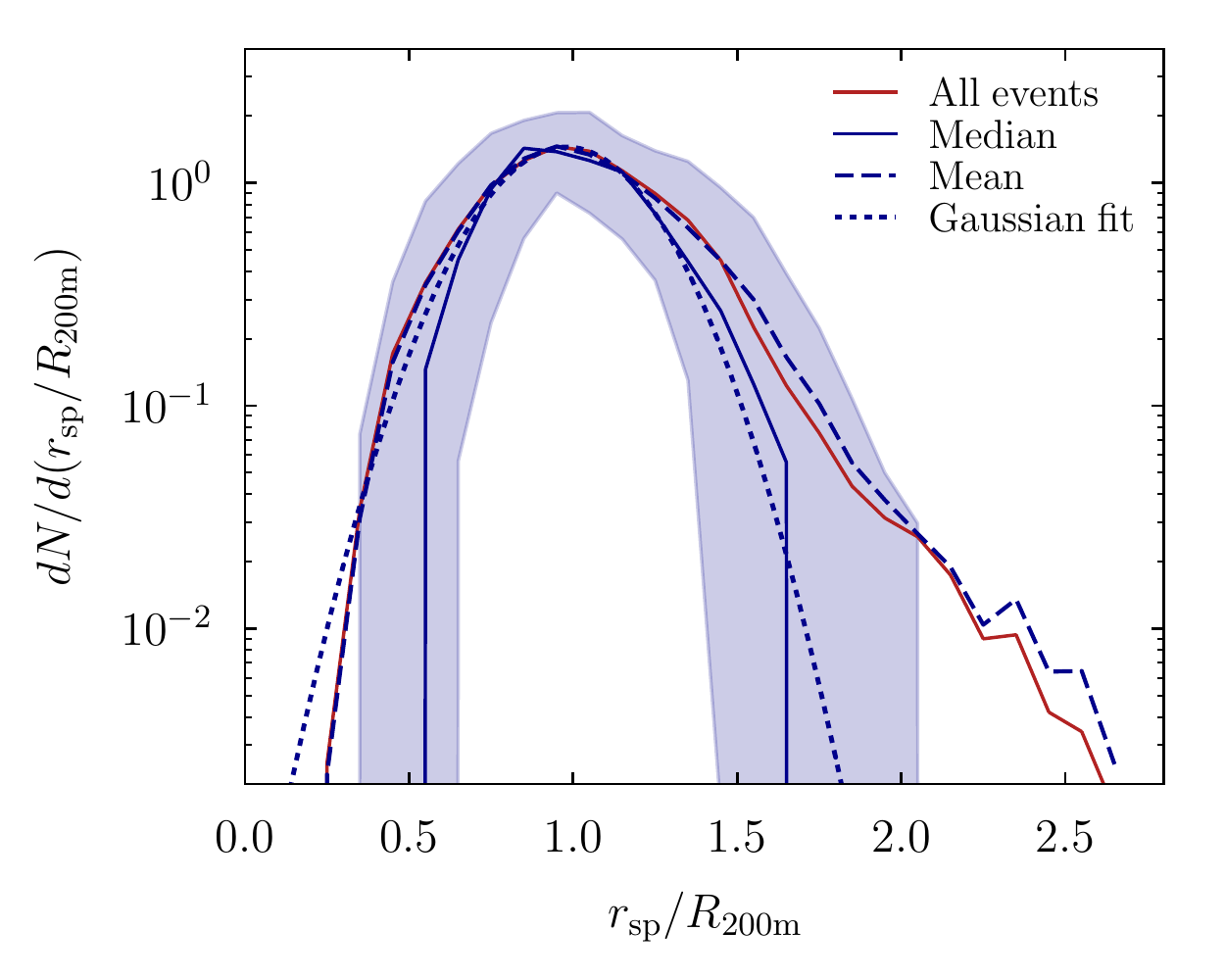}
\includegraphics[trim = 0mm 4mm 5mm 0mm, clip, scale=0.62]{\figdir/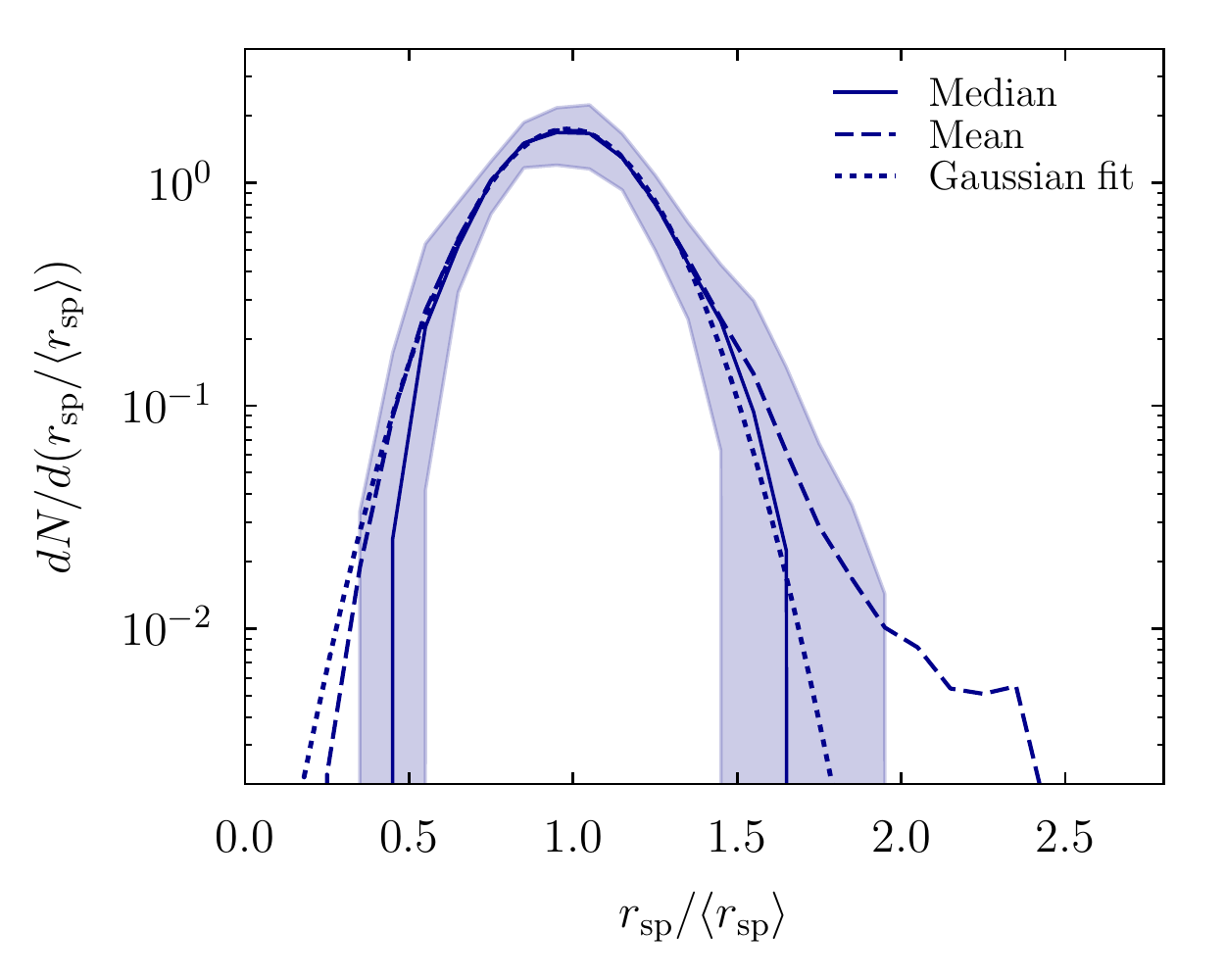}
\caption{Average distribution of particle apocenters. The distribution of $\rrsp$ is computed for each halo in units $\rtom$ (top panel) and the mean $\rrsp$ (bottom panel) of that halo. The solid lines and shaded areas show the median and 68\% scatter of those distributions, and the dashed line shows the mean. While the median plummets to zero around $0.5$ and $1.7$ $\rtom$, the mean shows a small tail toward higher $\rrsp$. In the top panel, we also compare the mean distribution to the total distribution of all $\rrsp$, i.e., the mean weighted by the number of particles in each halo. The two lines are almost identical, indicating that there are no strong trends with mass in this distribution. Finally, the dotted lines in each panel show the best-fit Gaussian to the median distribution that provides a reasonable fit at intermediate radii.}
\label{fig:rsp_dist_av}
\end{figure}

Given a distribution of splashback events such as that in Figure~\ref{fig:halo_rm}, $\rsp(t)$ and $\msp(t)$ are, to some degree, a matter of definition. Figure~\ref{fig:rsp_dist_av} shows the average shape of this distribution as a function of both $\rtom$ and the mean $\rrsp$ of each halo. The figure is based on all halos with \num{10000} or more particles in TestSim100 at $z = 0 \pm 0.25$ Gyr (including lower-mass halos leads to the median dropping to zero closer to the center of the distribution). For comparison, the dotted lines show Gaussian fits to the mean distribution. The mean of $\rrsp / \avg{\rrsp}$ is well described by the Gaussian distribution except for a few-percent tail at $\rrsp \gsim 1.5 \avg{\rrsp}$. As a result, the mean and intermediate percentiles of the distribution are very close to those inferred from a Gaussian fit, whereas the highest percentiles exceed the Gaussian expectation.

Figure~\ref{fig:rsp_dist_av} suggests that the $\rrsp$ distribution is relatively regular, and that taking its mean and percentiles should be well-motivated definitions of a halo's $\rsp$. We wish to avoid binning the $\rrsp$ values in time as that would introduce a number of free parameters such as the bin size. Instead, we smooth each event in time so that the value of $\rsp$ is determined by the $\rrsp$ within some time range around it, and we compute the weighted mean of those events. The smoothing function is arbitrary in principle; we use a Gaussian kernel to give each event a weight 
\begin{equation}
\label{eq:weight}
w_{\rm i} = \exp \left({\frac{-(t - t_{\rm i})^2}{2 \sigma^2}} \right)
\end{equation}
where $t_{\rm i}$ is the time at which the event occurred. This filter is normalized such that a splashback event at $t$ has weight one, rather than such that each event contributes unity total weight. The latter would be correct if we were to integrate the weights over the width of a bin, but we are instead considering the weights at an infinitesimally small time slice. Due to the infinite extent of a Gaussian, splashback events at all times would influence $\rsp$ at all other times. In practice, the contributions become very small at separations of a few snapshots, and we thus truncate the filter at $\Delta t_{\rm max} = 3 \sigma$.

When setting the standard deviation $\sigma$, we recognize that, due to its dynamical nature, the splashback radius can only change significantly over roughly a dynamical time. We thus set $\sigma = \sigmadyn \tdyn$ where $\sigmadyn = 0.2$ is a constant. We compute the weighted mean of the events at time $t$,
\begin{equation}
\label{eq:rspmean}
\rsp^{\rm mean} = \frac{1}{W} \sum_{i=0}^{N-1} w_{\rm i} r_{\rm i}
\end{equation}
where $r_{\rm i} = r_{\rm sp,i}$ is the radius of the $i$th splashback event and $W$ is the sum of all weights $w_{\rm i}$. The statistical uncertainty on the weighted mean can be computed from the weighted variance as
\begin{equation}
\label{eq:rspmeanerr}
\sigma^{\rm mean} = \sqrt{\frac{\sum_{i=0}^{N-1} w_{\rm i} (r_{\rm sp,i} - \rsp^{\rm mean})^2}{W^2-\sum_{i=0}^{N-1} w_{\rm i}^2}} \,.
\end{equation}
We also compute various weighted percentiles of the $\rrsp$ distribution. For this purpose, we order the $\rrsp$ values ascendingly and define the cumulative percentage at value $i$ as
\begin{equation}
\label{eq:cumulative1}
P_{\rm i} = \frac{100}{W_{\rm N}} \left( W_{\rm i} - \frac{w_{\rm i}}{2} \right)
\end{equation}
where $W_{\rm i}$ is the cumulative weight at value $i$. For a given percentile $p$, we find the first value $k$ with $P_{\rm k} > p$ and linearly interpolate to obtain $\rsp^{p}$. If $p < P_{\rm 0}$, we set $\rsp^{p} = r_{0}$, and if $p > P_{\rm N-1}$, we set $\rsp^{p} = r_{\rm N-1}$. The details of this procedure matter only for extreme percentiles and bins with very few contributing splashback events where the statistical uncertainty can be significant. Unfortunately, there is no standard expression for the statistical uncertainty on weighted percentiles of an unknown distribution. Any analytical estimate relies on the derivative of the distribution, which is by definition extremely noisy in the case of small samples. Thus, we perform a bootstrap resampling of the percentile calculation; in other words, we draw $N_{\rm BS} = 200$ random samples from the distribution with replacement and perform the calculation on each sample. For each percentile, the quoted value is the mean of the samples,
\begin{equation}
\rsp^{\rm p\%} = \frac{1}{N_{\rm BS}} \sum_{i=0}^{N_{\rm BS}-1} \rsp^{\rm p\%, i} \,,
\end{equation}
and the uncertainty is estimated from the variance of the samples:
\begin{equation}
\sigma^{\rm p\%} = \sqrt{\frac{1}{N_{\rm BS} - 1} \sum_{i=0}^{N_{\rm BS}-1} (\rsp^{\rm p\%, i} - \rsp^{\rm p\%})^2} \,.
\end{equation}
We restrict the computation to the relatively low number of $200$ samples for performance reasons. We find that the percentile estimates exhibit a variance of less than 1\% between runs, including the extreme tails of the distribution as quantified by the 99th percentile. The mean and a number of percentiles calculated in this manner are shown in the bottom panels of Figure~\ref{fig:halo_rm}.

In principle, the algorithm described above can operate on bins with as few as two particle splashback events, but the results would be extremely noisy. We could introduce a cut on the number of $\rrsp$ events, but those events could have occurred at a time far away from the current time bin (for example, because the halo had recently been a subhalo). Thus, we introduce a minimum weight, $w_{\rm min} = 10$, corresponding to $10$ events exactly at the time in question, or a set of events with equivalent weights. Lowering $w_{\rm min}$ to arbitrarily low values improves the completeness, but at the cost of extremely uncertain $\rsp$ determinations. This minimum does not significantly bias the average $\rsp$ of halos.

\begin{deluxetable*}{lllll}
\tablecaption{Parameters of the \sparta Algorithm
\label{table:params}}
\tablewidth{0pt}
\tablehead{
\colhead{Parameter} &
\colhead{Value} &
\colhead{Introduced} &
\colhead{Convergence Tests} &
\colhead{Explanation (Impact on $\rrsp$ or $\rsp$)}
}
\startdata
$r_{\rm create}$ & $2 \rtom$ & \S\ \ref{sec:sparta:tcrptl} & - & Radius where code starts tracing orbits (must be greater than $\rtom$ to record infall events) \\
$r_{\rm delete}$ & $3 \rtom$ & \S\ \ref{sec:sparta:tcrptl} & \S\ \ref{sec:res:rdelete} & Radius where code stops tracing orbits (an effective maximum on $\rrsp$) \\
$\smrmax$ & $0.01$ & \S\ \ref{sec:sparta:rsp:selection} & \S\ \ref{sec:res:smr}, Figure~\ref{fig:conv} & Maximum sub-to-host mass ratio for $\rrsp$ events (high SMR events are biased low in $\rrsp$) \\
$\sigmadyn$ & $0.2$ & \S\ \ref{sec:sparta:rsp:average} & \S\ \ref{sec:res:sigmadyn}, Figure~\ref{fig:conv} & Smoothing timescale in units of the dynamical time (aggressive smoothing can bias $\rsp$ high) \\
$w_{\rm min}$ & 10 & \S\ \ref{sec:sparta:rsp:average} & - & Minimum weight in a time bin necessary to compute $\rsp$ (noisy $\rsp$ if this weight is too small) \\
$w^*_{\rm max}/w_{\rm tot}$ & $0.5$ & \S\ \ref{sec:sparta:rsp:correction} & \S\ \ref{sec:res:correction}, Figure~\ref{fig:conv} & Maximum weight of the correction at the last snapshots (if too large, can overcorrect $\rsp$)
\enddata
\tablecomments{Parameters of the \sparta algorithm that can, if set to inappropriate values, influence $\rrsp$ or $\rsp$. Each parameter's value was set such that it does not bias $\rsp$ in a significant way, as explained in the listed sections.}
\end{deluxetable*}

Finally, we wish to compute $\msp$ as well as $\rsp$. We follow the same procedure as for $\rsp$, meaning we compute it from the enclosed mass within the splashback radii of the individual particles $\mmsp$. This choice means that we do not exactly preserve the relation $\msp = M(<\rsp)$. On the other hand, inaccuracies in $\rsp$ (e.g. due to smoothing at times of sharp changes) do not necessarily translate into errors in $\msp$. In other words, we treat $\msp$ as an independent aspect of the distribution of splashback events rather than as a secondary consequence of $\rsp$.

\subsubsection{Correction for the Final Snapshots}
\label{sec:sparta:rsp:correction}

Finally, we need to correct for two biases that occur at the end of the simulation (typically at $z = 0$). First, the number of splashback events in the final time bin (between the second-to-last and last snapshots) is drastically lower than in the previous time bins, which is to be expected given the algorithm described in Section~\ref{sec:sparta:ressbk}. Thus, we ignore any splashback events that have occurred after the time of the second-to-last snapshot as they are likely biased in some nontrivial way. Second, the distribution of splashback events considered in the final snapshots becomes asymmetric due to the smoothing discussed above: the Gaussian filter is sensitive to events at earlier times, but there are no events at later times. This asymmetry can lead to a significant and systematic bias because $\rsp$ is, in most cases, increasing with time. Thus, ignoring splashback events that would have occurred in the future leads to an underestimate of $\rsp$ at late times.

We correct for this asymmetry as follows. In the absence of any information about the ``missing'' events, we must extrapolate the past evolution of $\rsp$ into the future. We recognize that the evolution of $\rsp$ (or $\msp$, hereafter summarily called $q(t)$), is given by some true, underlying function $f(t)$, convolved with the Gaussian smoothing kernel such that 
\begin{equation}
q(t) = \int_{t - \Delta t_{\rm max}}^{t + \Delta t_{\rm max}} f(t) w(t) dt \,.
\end{equation}
If the last time at which we have recorded splashback events is $t_{\rm f}$, we need to make a correction for all time bins where $t + \Delta t_{\rm max} > t_{\rm f}$. In particular, we correct the mean of the distribution in those bins such that 
\begin{equation}
\label{eq:corr}
q^{\rm corrected} = \frac{w_{\rm q} q + w^* q^*}{w_{\rm q} + w^*}
\end{equation}
where $q$ is $\rsp$ or $\msp$ determined from the distribution up to time $t_{\rm f}$, $q^*$ is the contribution we would obtain from the distribution of events at times later than $t_{\rm f}$, and $w_{\rm q} \equiv w(<t_{\rm f})$ and $w^* \equiv w(>t_{\rm f})$ are the integrated weights before and after $t_{\rm f}$. We base our extrapolation on the assumption that $f(t)$ evolves linearly, $f(t) \approx f_0 + f_1 t$. We perform a least-squares fit for the free parameters $f_0$ and $f_1$ using the four time bins before $t_{\rm f}$ and compute the correction term $q^*$:
\begin{align}
q^* &= \int_{t_{\rm f}}^{t + \Delta t_{\rm max}} e^{-\frac{(t'-t)^2}{2\sigma^2}} (f_0 + f_1 t') dt' \nonumber \\
&= \sqrt{\frac{\pi}{2}} \sigma (f_0 + f_1 t) ({\rm erf(x_1)}-{\rm erf(x_0)}) + \sigma^2 f_1 (e^{-x_0^2} - e^{-x_1^2})
\end{align}
where
\begin{align}
x_0 & \equiv (t_{\rm f} - t) / (\sqrt{2} \sigma) \nonumber \\
x_1 & \equiv \Delta t_{\rm max} / (\sqrt{2} \sigma)
\end{align}
and erf denotes the error function. We perform this correction iteratively, i.e. starting with the first time bin that is affected by the asymmetry of $\rrsp$ events. The next time bin's correction is then based on the previous corrected time bins, and so on. 

The details of how the best-fit slope is determined are unimportant. For example, varying the number of fitted time bins between three and eight has virtually no effect, nor do limits on the $\chi^2/N_{\rm dof}$ of the fit. However, we find that the algorithm described above can over compensate slightly and performs better if the weight given to the correction term, $w^*$ in Equation~(\ref{eq:corr}), is limited to half of the total weight, that is, $w^* \leq w_{\rm q}$. Furthermore, the algorithm works best for the most stable estimates of $\rsp$ and $\msp$, the mean and median of the distribution of splashback events. The bin-to-bin fluctuations grow for the higher percentiles, leading to additional noise in the extrapolation but no significant improvement in the average bias of the $\rsp$ estimates. Thus, we compute the correction factor for the mean, $q^{\rm corrected}_{\rm mean} / q_{\rm mean}$, and apply it to both the mean and all percentile estimates. We discuss the performance of the correction algorithm in Section~\ref{sec:res:correction}.


\section{Results}
\label{sec:res}

\begin{figure*}
\centering
\includegraphics[trim = 5mm 2mm 2mm 0mm, clip, scale=0.52]{\figdir/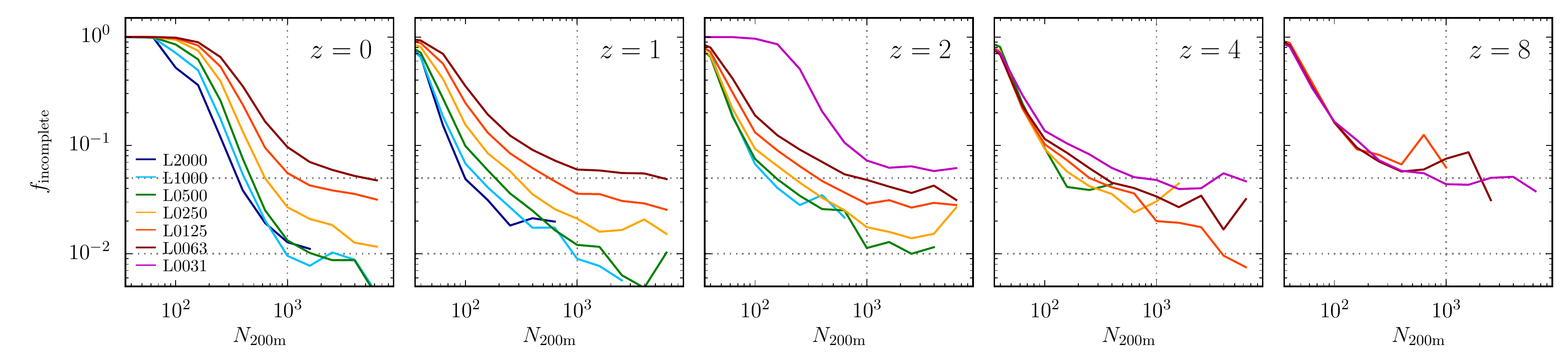}
\caption{Completeness of the $\rsp$ sample at different redshifts. The panels show the fraction of halos above a certain particle number $\ntom$ that do not have a valid $\rsp$ at the indicated redshift (either due to a lack of particle splashback events, or because the halo was recently a subhalo). The gray vertical lines highlight a limit of $1000$ particles, and the horizontal lines highlight $5\%$ and $1\%$ incompleteness. Above $1000$ particles, the incompleteness is almost entirely due to halos that were recently subhalos, an effect that depends on the box size of a simulation. Our algorithm identifies $\rsp$ and $\msp$ in at least 95\% of halos with at least $1000$ particles, with the exception of the smallest boxes at low redshift.}
\label{fig:valid_frac}
\end{figure*}

\begin{figure*}
\centering
\includegraphics[trim = 2mm 0mm 1mm 0mm, clip, scale=0.48]{\figdir/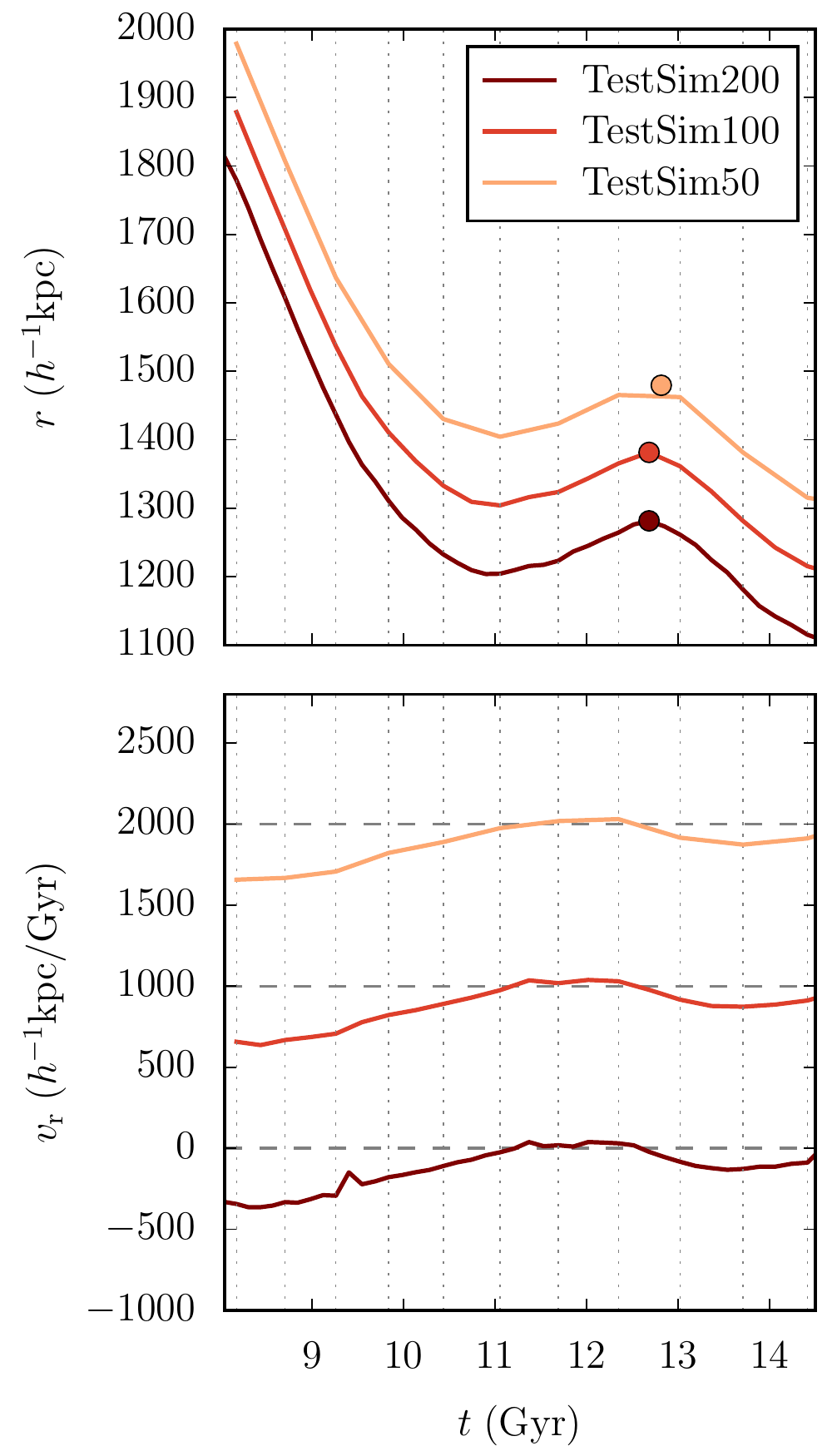}
\includegraphics[trim = 12mm 0mm 1mm 0mm, clip, scale=0.48]{\figdir/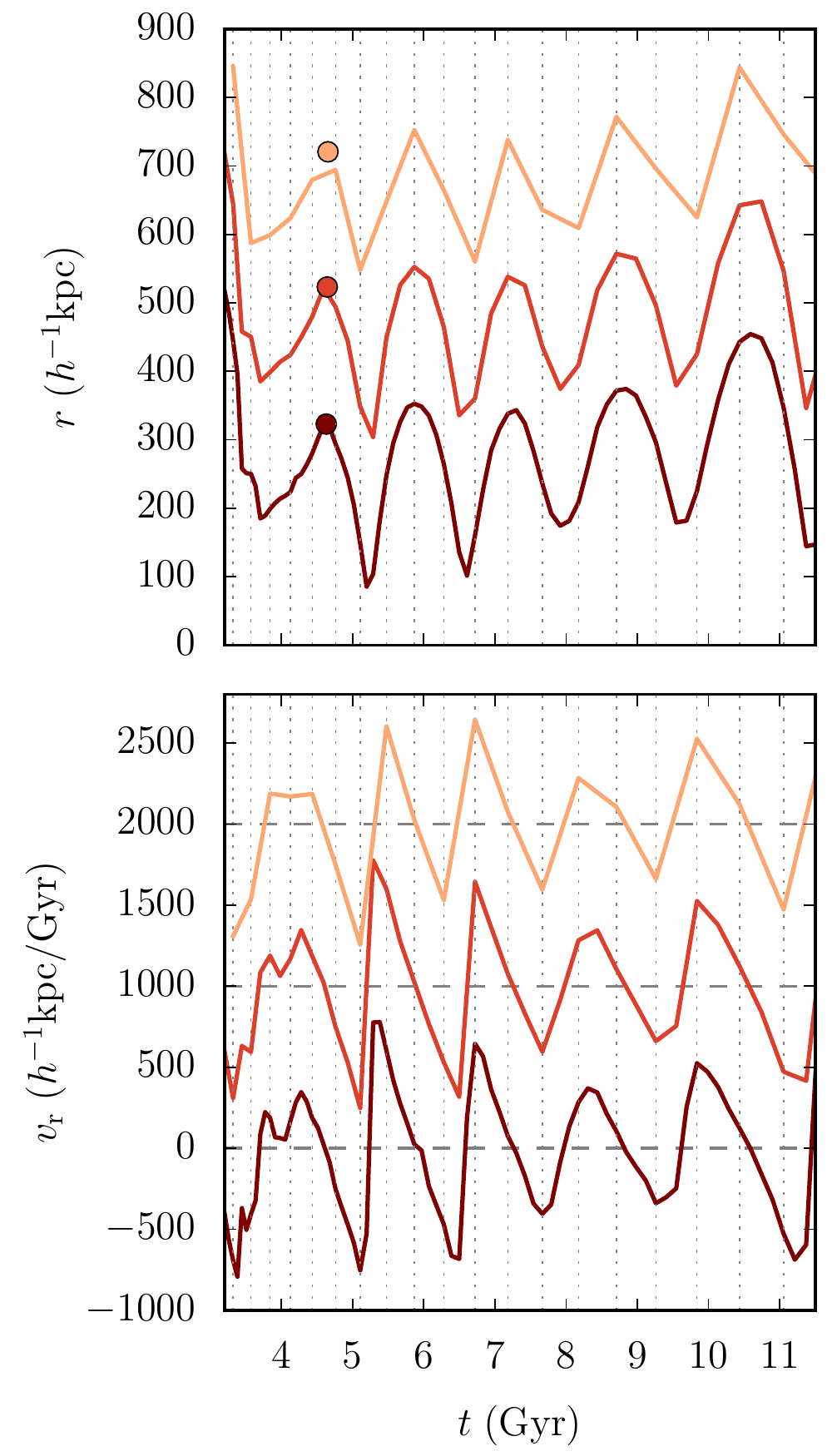}
\includegraphics[trim = 12mm 0mm 1mm 0mm, clip, scale=0.48]{\figdir/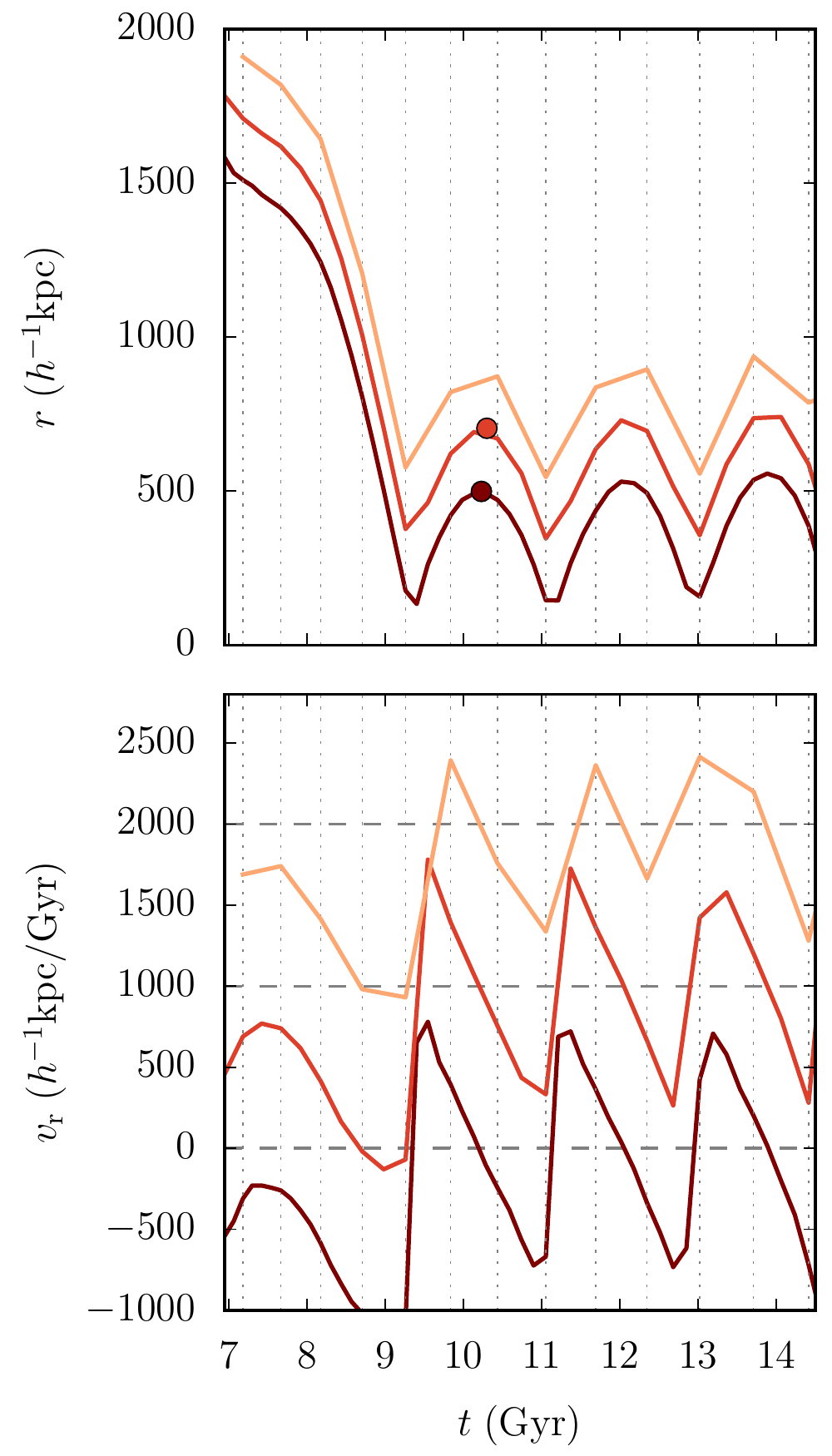}
\includegraphics[trim = 12mm 0mm 1mm 0mm, clip, scale=0.48]{\figdir/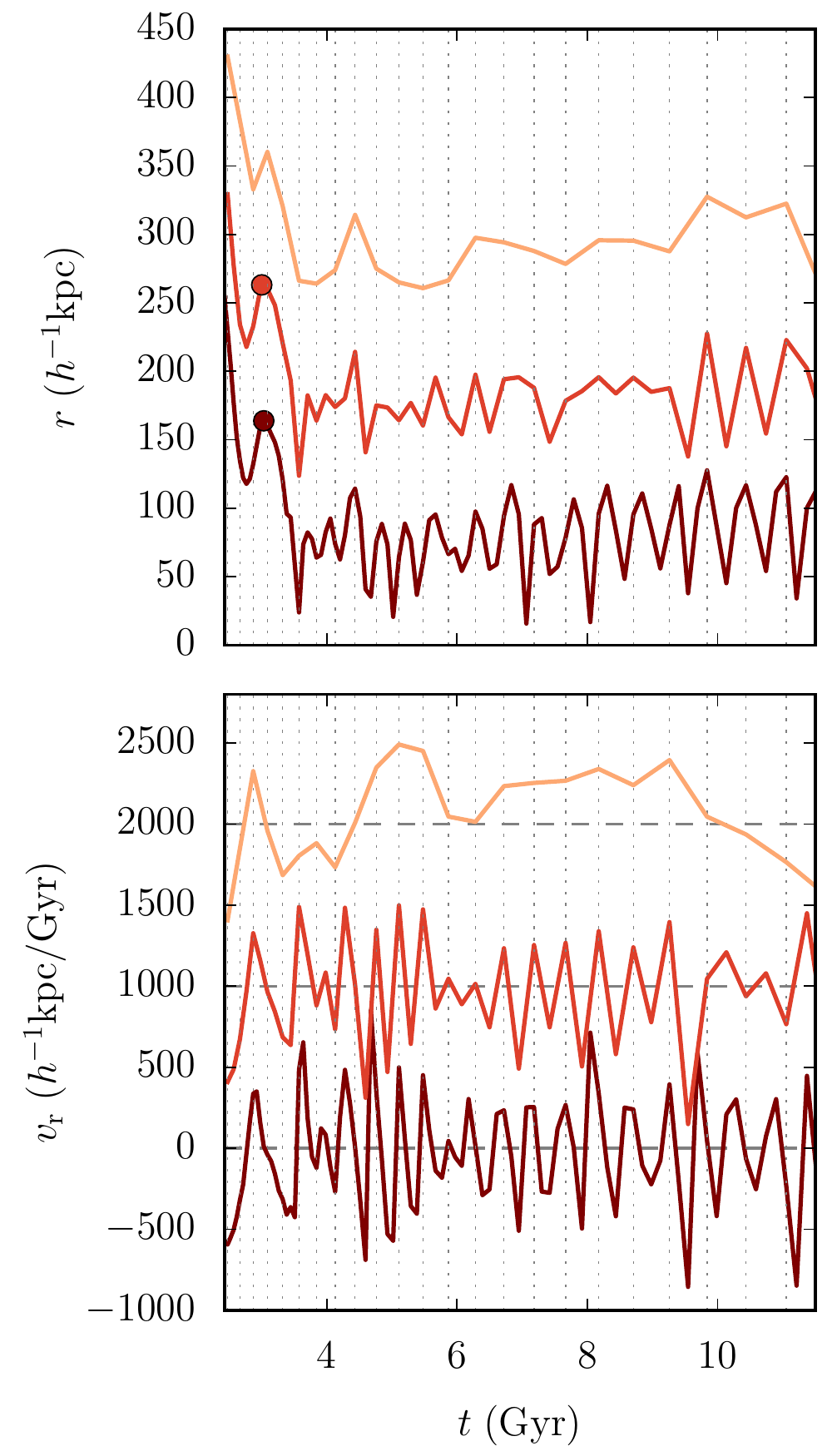}
\caption{Comparison of trajectories of the same particles for different snapshot spacings. The radii (top panels) and radial velocities (bottom panels) are offset from each other for clarity. Splashback events identified by \sparta are shown with solid points. The gray vertical lines show the snapshot times in the lowest-resolution run. The particles were drawn arbitrarily from the same example halo as in Figure~\ref{fig:halo_rm} in order to illustrate characteristic effects of snapshot spacing as described in Section~\ref{sec:res:timeres}.}
\label{fig:conv_timeres_traj}
\end{figure*}

In the previous section, we have described an algorithm to compute a halo's $\rsp$ and $\msp$ throughout its history. In the process, we have introduced a few free parameters that might influence the results (Table~\ref{table:params}). In this section, we discuss the algorithm's performance and convergence with respect to all free parameters as well as mass resolution and snapshot spacing. While we consider the \gammarsp to establish convergence, we leave any analysis of the dependence of $\rsp$ on halo mass, accretion rate, and cosmology for \citetalias{diemer_17_rsp}.

We will establish a number of cuts on our halo sample, namely a limit of $\ntom \geq 1000$ particles and a sub-to-host mass ratio $< 0.01$ for all particle $\rrsp$ events. Wherever the \gammarsp is shown, we additionally require that the halo was a host at the beginning of the interval over which $\Gamma$ is measured, namely a dynamical time ago. We do not, however, exclude halos that were a subhalo at some point during that interval. Halos that experienced such a backsplash event during the last dynamical time make up for about 2\% of the population at $z = 0$, a fraction that drops below a percent at higher redshift where the dynamical time is shorter. Excluding such halos has no measurable impact on the median \gammarsp.

\subsection{Completeness}
\label{sec:res:completeness}

\begin{figure*}
\centering
\includegraphics[trim = 4mm 2mm 61mm 1mm, clip, scale=0.51]{\figdir/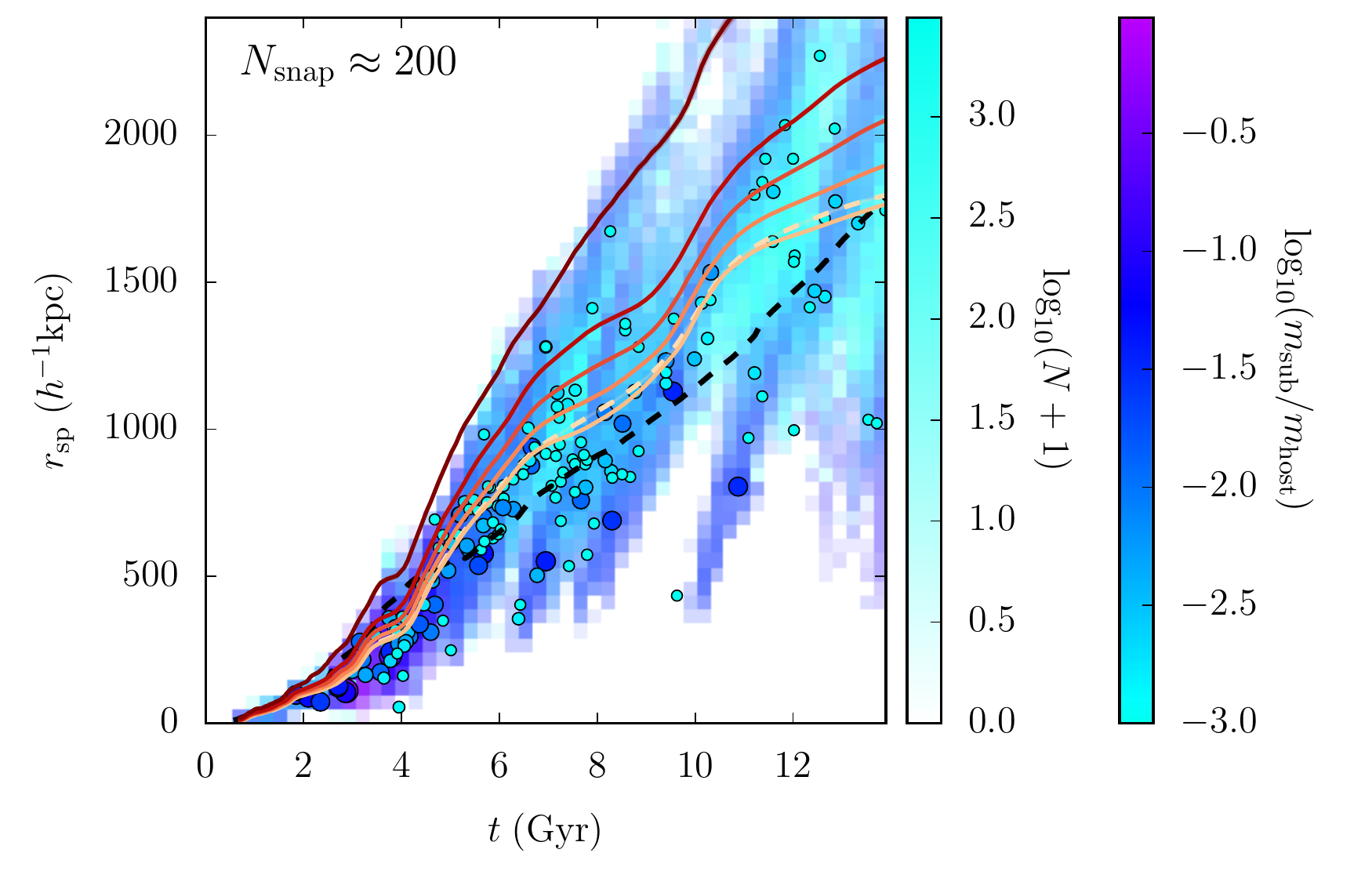}
\includegraphics[trim = 25mm 2mm 61mm 1mm, clip, scale=0.51]{\figdir/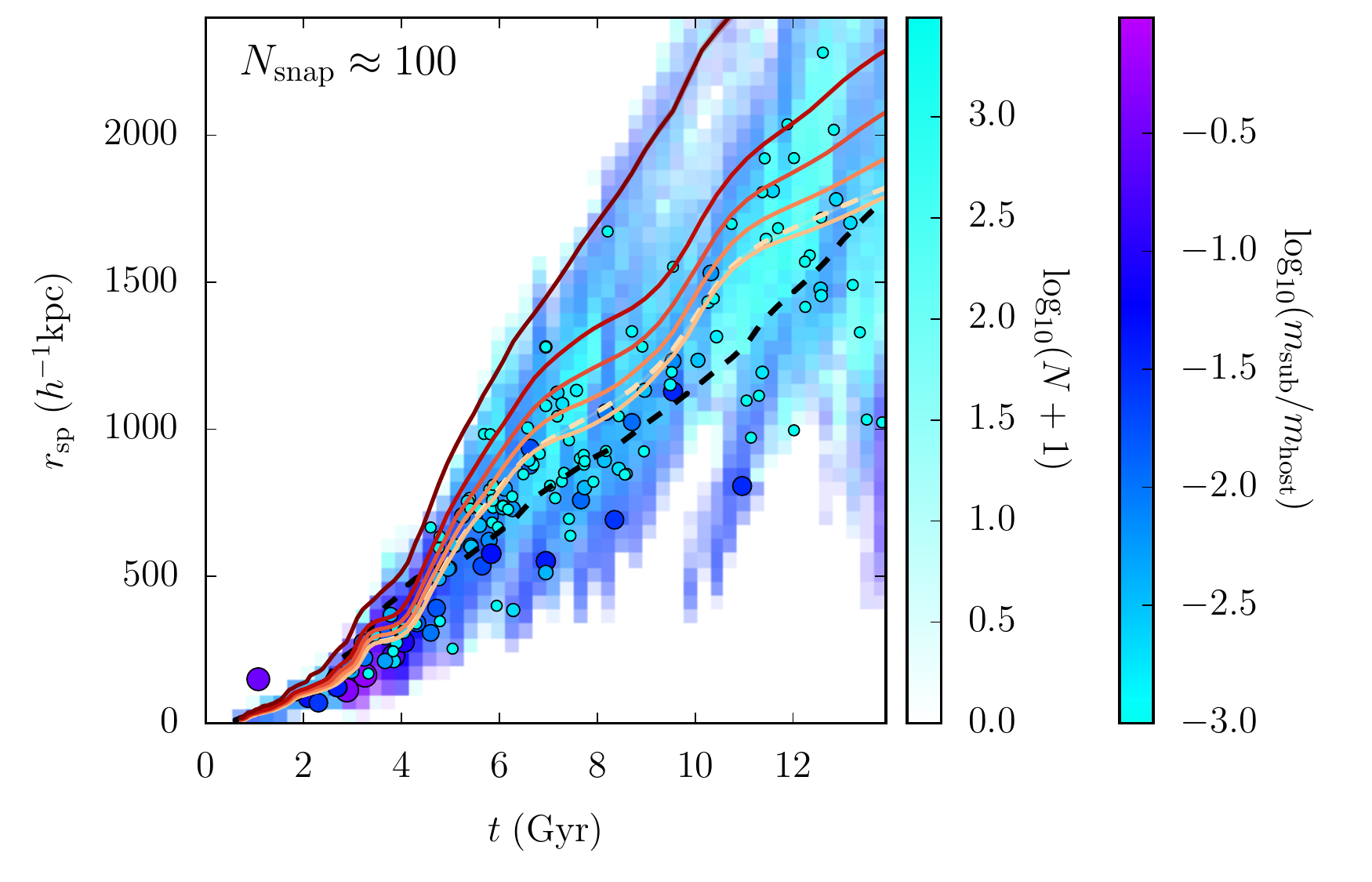}
\includegraphics[trim = 25mm 2mm 8mm 1mm, clip, scale=0.51]{\figdir/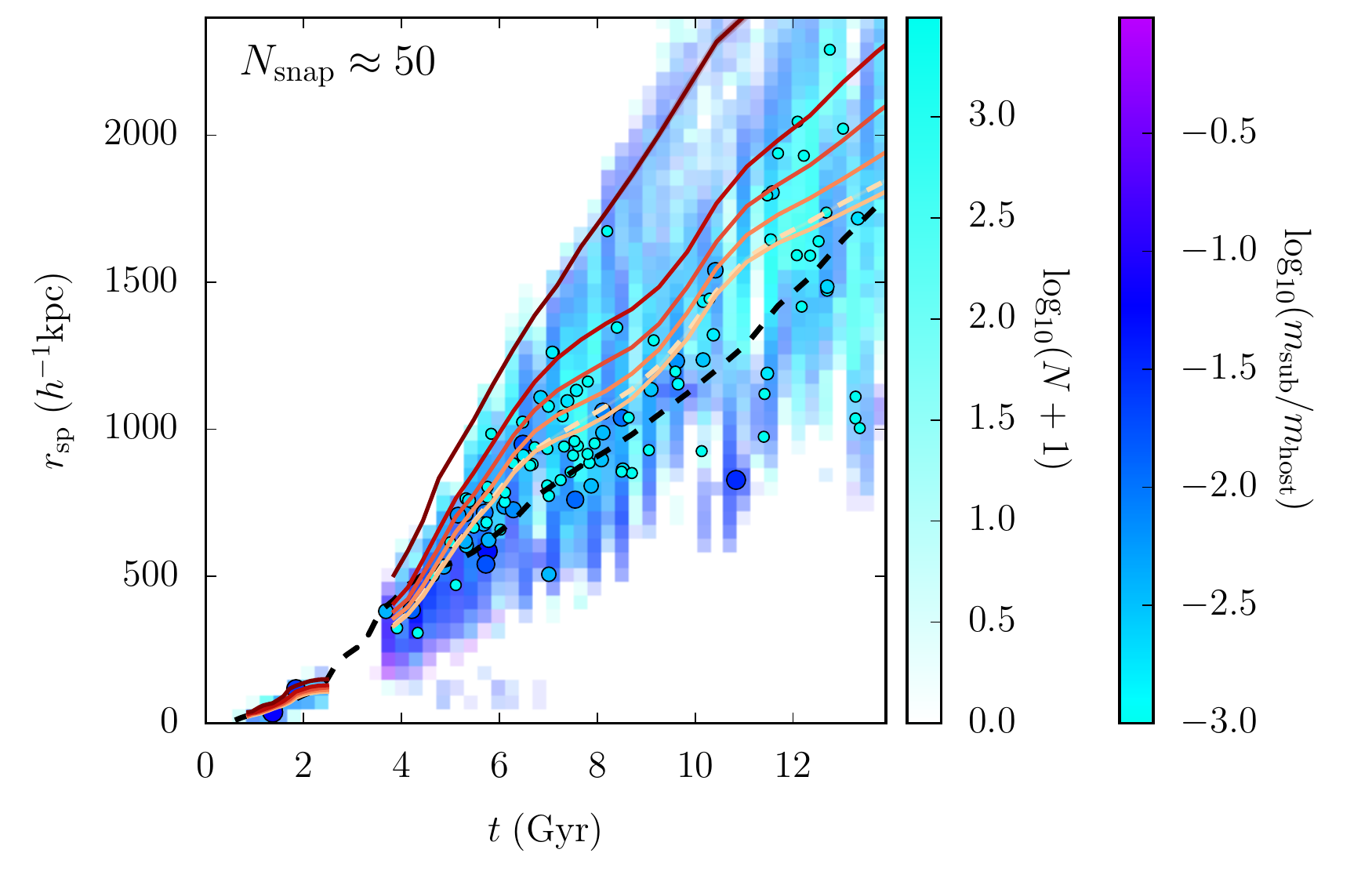}
\caption{Particle and subhalo splashback radii for the same example halo as in Figure~\ref{fig:halo_rm}, but for different snapshot spacings. The left panel shows the same distribution of $\rrsp$ as the left panel of Figure~\ref{fig:halo_rm}, but with the histogram down-sampled to 50 bins. The center and right panels show $\rrsp$ for the same halo, but based on only every other and every fourth snapshot, respectively. It is apparent that the halo finder results vary drastically with time resolution, resulting in different subhalo splashbacks and the halo temporarily becoming a subhalo in the lowest-resolution run (around $3$ Gyr). Despite the different input data from the particle trajectories and halo catalogs, \sparta recovers a very similar $\rsp$.}
\label{fig:conv_timeres_halo}
\end{figure*}

\begin{figure}
\centering
\includegraphics[trim = 1mm 2mm 4mm 1mm, clip, scale=0.50]{\figdir/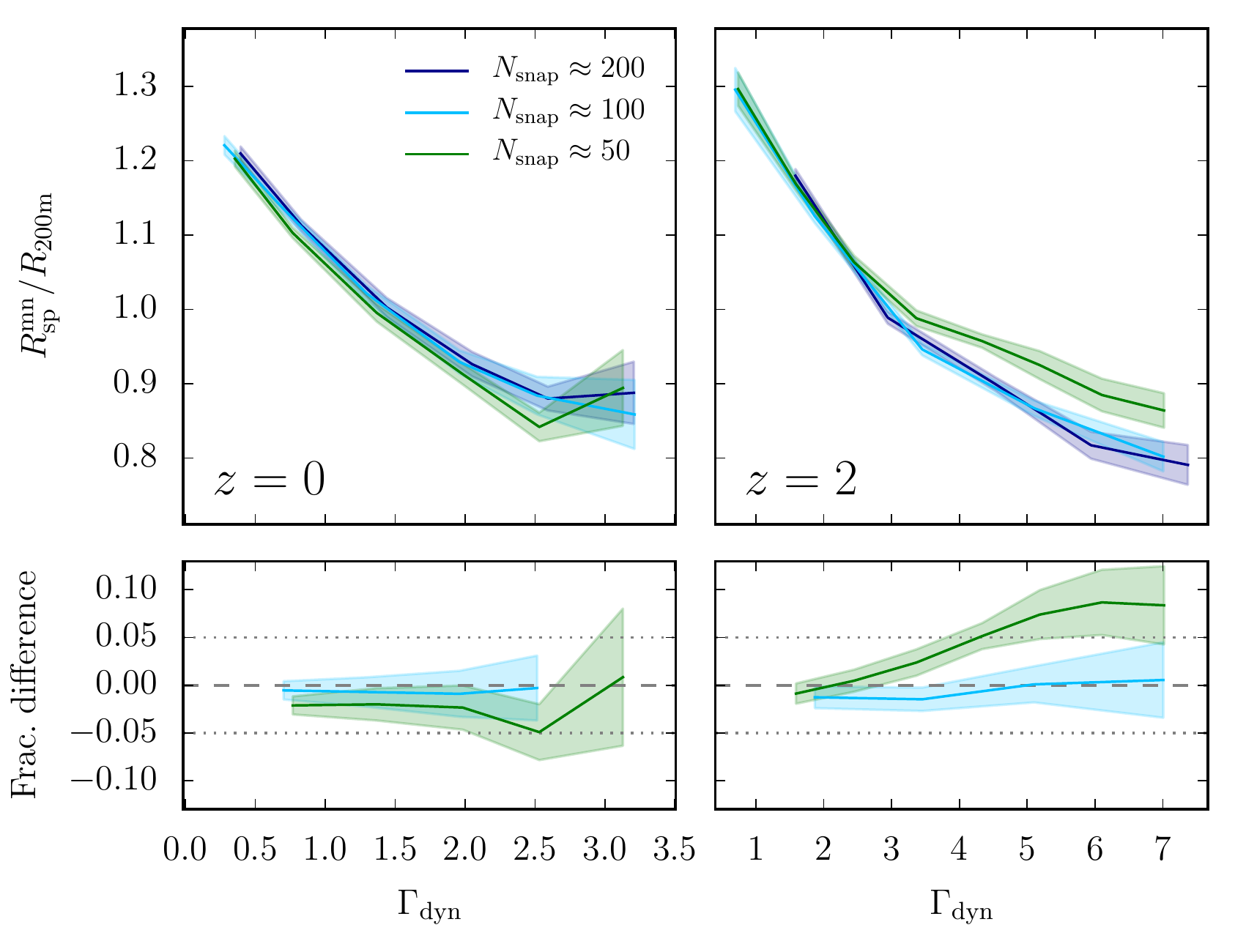}
\caption{Convergence of the mean $\Gamma$-$\rsp$ relation with snapshot spacing. For simplicity, only redshifts $0$ and $2$ are shown, and the mass $\msp$ is omitted as it converges in a fashion very similar to $\rsp$. While the simulation with $50$ snapshots shows differences of up to $30$\% in the mean relation, the simulation with $100$ snapshots has converged to the same relation as that with $200$ snapshots within the statistical uncertainty.}
\label{fig:conv_timeres_gammarsp}
\end{figure}

As a first test, we consider how successful the algorithm is in assigning splashback radii and masses to halos. The computation can fail either because there are not enough splashback events close to a given time (i.e., the weight falls below $w_{\rm min}$; this issue obviously becomes more common for lower-mass halos) or because the halo was recently a subhalo. For such backsplash halos, the computation is resumed, but the algorithm needs to accumulate a few snapshots before the first splashback events are recorded.

Figure~\ref{fig:valid_frac} shows the completeness as a function of the number of particles in a halo for different simulations. We note that only halos that reached $\ntom \geq 200$ at some time were written to disk. The incompleteness in poorly resolved halos ($\ntom \lsim 500$) is predominantly caused by a lack of splashback events. In better resolved halos, virtually all failures are due to the halo recently having been a subhalo. This effect is most important for the smallest box sizes because physically less massive halos are more likely to be subhalos.

Most importantly, however, we find that about 95\% of halos with at least $1000$ particles are assigned an $\rsp$ and $\msp$. The only exception are the smallest boxes at low redshift, for example, L0031 at $z = 2$ and L0063 at $z < 2$. 

\subsection{Convergence with Snapshot Spacing}
\label{sec:res:timeres}

\begin{figure*}
\centering
\includegraphics[trim = 0mm 6mm 5mm 0mm, clip, scale=0.56]{\figdir/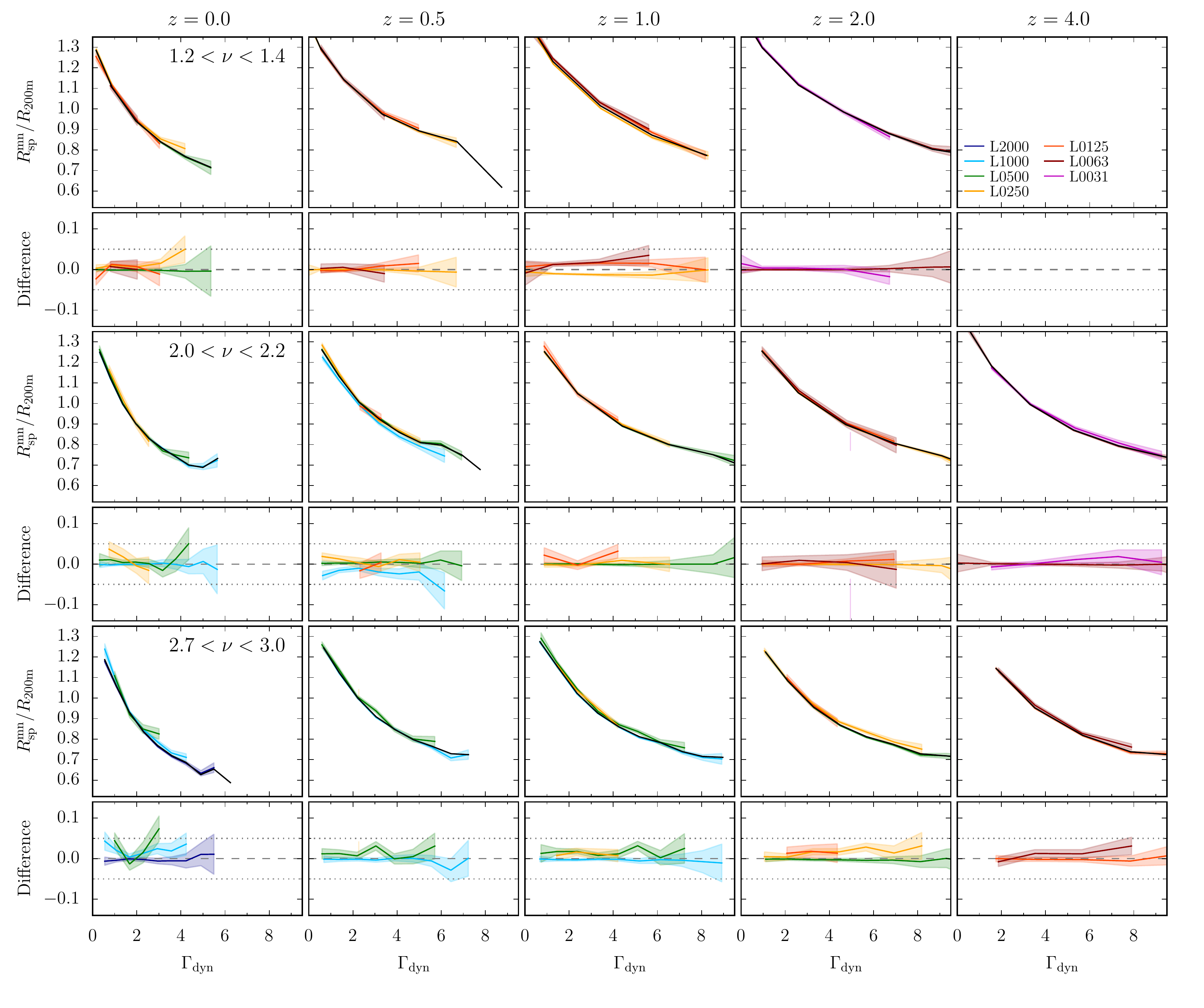}
\caption{Convergence of $\rsp$ with mass resolution in bins of redshift (columns) and peak height (rows). Each set of panels shows the mean \gammarsp for halos with $N_{\rm 200m} = \mtom / m_{\rm p} \geq 1000$. Each colored line represents the results from one of the simulations, which differ from each other by a factor of $8$ in mass resolution at fixed halo mass and redshift. Differences in $\rsp$ would indicate a dependence on mass resolution. The bottom panels in each set show the residual of each simulation's results from the average over halos in all simulations (black line). Given the statistical uncertainty on the mean relations (shaded areas), we do not find any statistically significant mass resolution effects at $N_{\rm 200m} \geq 1000$.}
\label{fig:conv_mass}
\end{figure*}

As described in Section~\ref{sec:sparta:ressbk}, the \sparta algorithm uses four time bins to determine the splashback radius of particles, corresponding to about $0.6\ \tdyn$ at early times (Figure~\ref{fig:conv_timeres}). Given this large fraction of the dynamical time, reducing the time resolution of the trajectories might lead to systematically different estimates of $\tsp$ and $\rrsp$. In this section, however, we demonstrate that this effect is not significant as long as the evolution of a simulation is represented by about 100 or more snapshots.

Figure~\ref{fig:conv_timeres_traj} illustrates a few characteristic ways in which individual particle trajectories are affected by snapshot spacing. The trajectory on the left is well resolved in time even in the low-resolution run, and the algorithm has no problem in reconstructing the correct splashback despite the relatively weak feature in radius and velocity. The second trajectory illustrates the importance of interpolation: even though the snapshot where the radius is largest is missing from the low-resolution version, the splashback event is reconstructed correctly. The third trajectory illustrates a common reason why a splashback event might fail to be found in lower-resolution trajectories: the velocity becomes positive for only one snapshot around splashback, not enough for the algorithm to confidently identify a maximum in radius. Finally, the fourth trajectory illustrates a pathological case: the orbital period of the particle is in resonance with the snapshot output frequency, meaning that the true orbit is entirely misrepresented in the lowest-resolution run.

Having understood the effects of snapshot spacing on individual trajectories, we now consider the $\rrsp$ history of the example halo from Figure~\ref{fig:halo_rm}. The left panel of Figure~\ref{fig:conv_timeres_halo} shows the same data as in the left panel of Figure~\ref{fig:halo_rm}, but using a lower-resolution histogram resolution. The center and right panels show the same plot for the subsampled versions, TestSim100 and TestSim50. The most notable differences between the three panels stem not from the \sparta results, but from the halo finder output: even between the $200$- and $100$-snapshot versions, the subhalo splashbacks are distributed somewhat differently. Moreover, in the $50$-snapshot run the merger tree code identified the example halo as a subhalo for a period around $3$ Gyr, leading to the gap in splashback events. We note that the merger tree code is not designed to operate with such few snapshots in the first place \citep{behroozi_13_trees}. Despite the drastic differences in the merger tree history, the \sparta results are very similar to the runs with more snapshots. 

The right panel of Figure~\ref{fig:conv_timeres_halo} exhibits other visibly different features. For example, some particles splash back at very low radii between $4$ and $8$ Gyr. We have verified that these events stem from pathological trajectories with orbital times that are in resonance with the snapshot frequency, as shown in Figure~\ref{fig:conv_timeres_halo}. While the algorithm will (correctly) fail to identify a splashback in the majority of such trajectories, some randomly mimic a pericenter and apocenter a long time after the particle has entered the halo. Furthermore, the run with the fewest snapshots lacks splashback events at very late times because \sparta can only identify splashback events that happen within a few snapshots of the end of a trajectory. Given that neither of these issues is observed in the run with $100$ snapshots, we need not worry about them.

However, the example halo shown in Figures~\ref{fig:halo_rm} and~\ref{fig:conv_timeres_halo} is resolved by many particles, and the snapshot spacing could have more severe effects on halos with fewer particles. We investigate a statistical halo sample in Figure~\ref{fig:conv_timeres_gammarsp} that shows the \gammarsp for the test simulations with $200$, $100$, and $50$ snapshots. We compare the mean of the halo samples in each simulation because it is more sensitive to outliers than the median. As expected, the run with the lowest number of snapshots shows clear deviations from the other runs, up to $30$\% at high redshift and mass accretion rate. The run with $100$ snapshots, however, has converged to within the statistical uncertainty. We thus conclude that simulations with about $100$ snapshots or more, like those used in this paper, are suitable for the dynamical analysis of trajectories.

\subsection{Convergence with Mass Resolution}
\label{sec:res:massres}

\begin{figure*}
\centering
\includegraphics[trim = 2mm 2mm 0mm 1mm, clip, scale=0.65]{\figdir/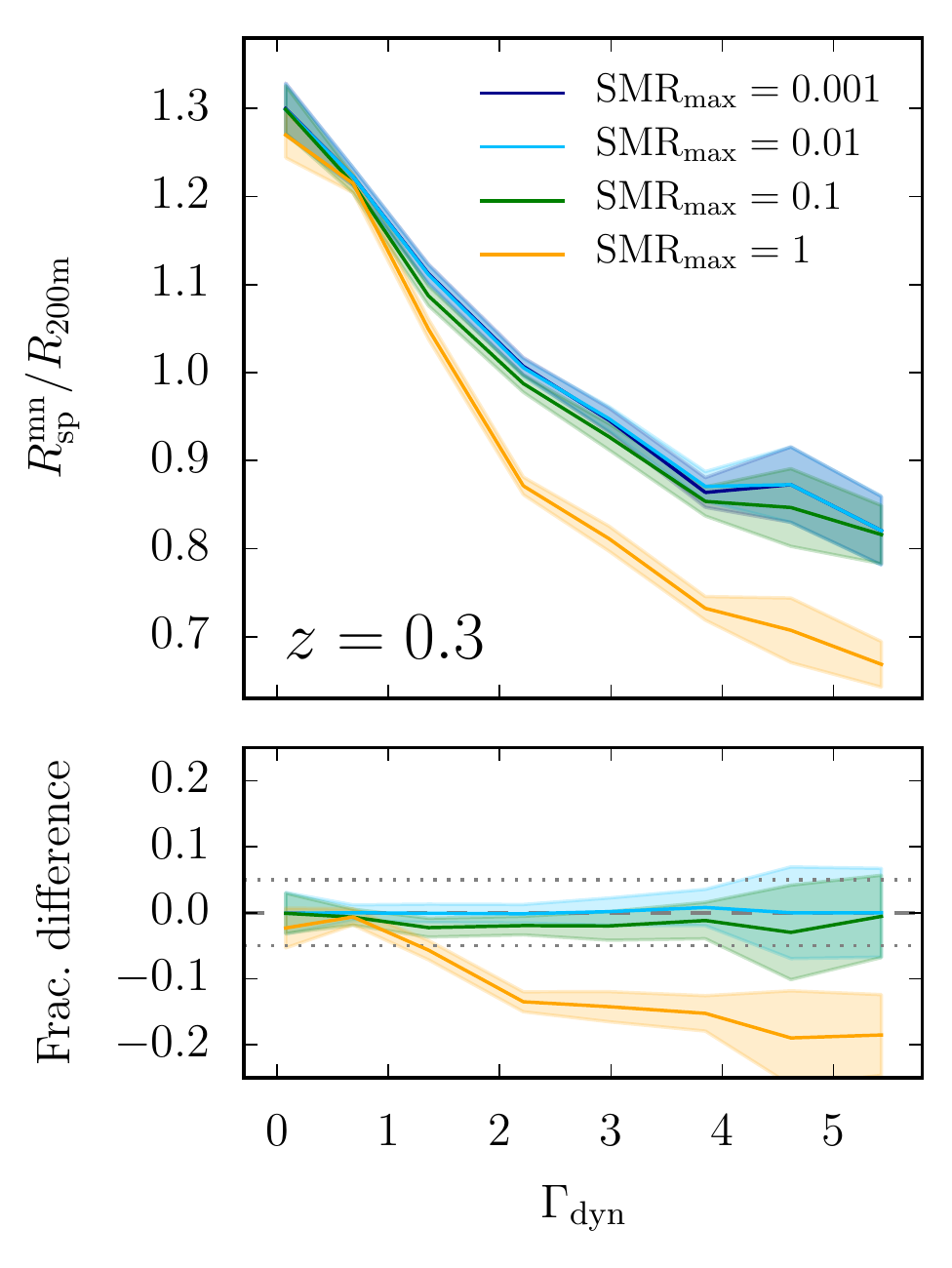}
\includegraphics[trim = 22mm 2mm 0mm 1mm, clip, scale=0.65]{\figdir/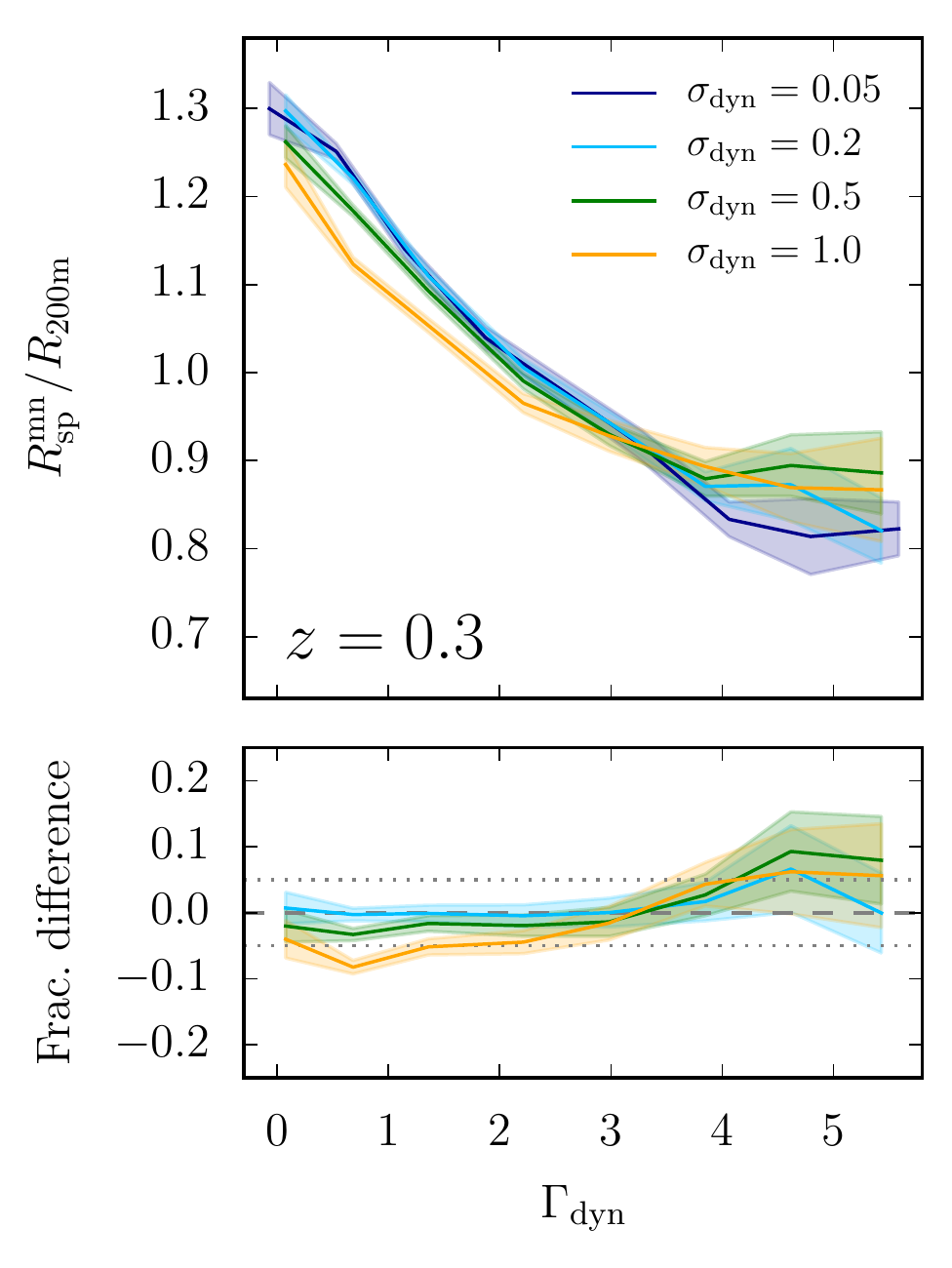}
\includegraphics[trim = 22mm 2mm 0mm 1mm, clip, scale=0.65]{\figdir/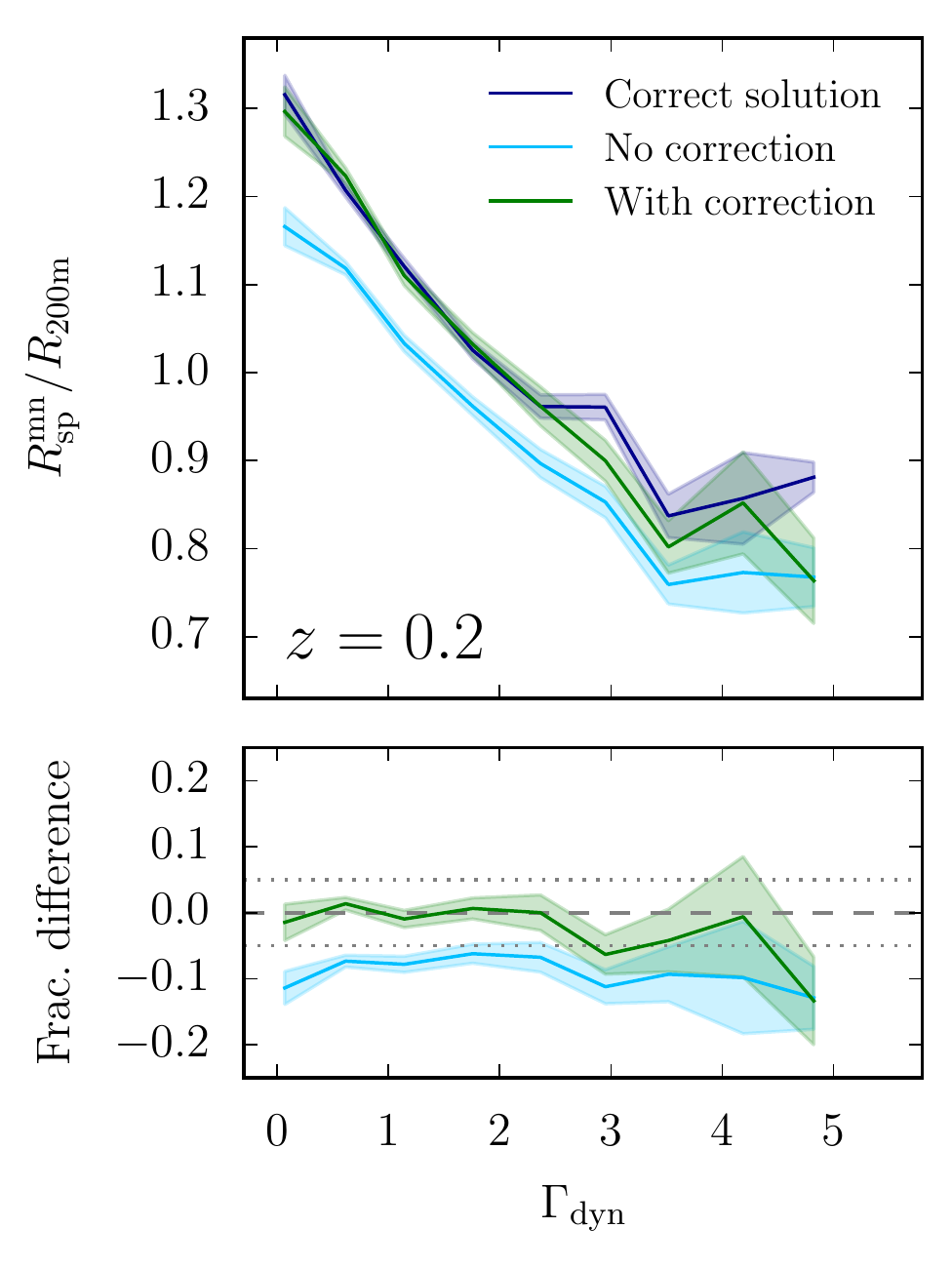}
\caption{Convergence of $\rsp$ with maximum sub-to-host mass ratio (left), smoothing timescale (center), and late-time correction (right). The plots show $\rspmean$, but the convergence properties are similar for the median and higher percentiles. Left column: When all splashback events are included in the calculation (yellow), $\rsp$ is biased low by up to 20\% due to dynamical friction. Excluding particles from subhalos larger than $\smrmax = 0.1$ reduces this bias significantly (green), but a systematic difference of a few percent remains. The light blue line showing the fiducial value of $\smrmax = 0.01$ is indistinguishable from the dark blue reference solution with $\smrmax = 0.001$. Center column: same as the left column, but for different values of the smoothing timescale $\sigmadyn$. With aggressive smoothing such as $\sigmadyn \geq 0.5$, a bias compared to the unsmoothed $\rsp$ is apparent. At our fiducial value of $\sigmadyn = 0.2$, this bias has disappeared. Right column: the effect of the correction applied to the last few snapshots of a simulation where the $\rrsp$ distribution becomes asymmetric in time. The dark blue line shows the true evolution of $\rsp$, taking future splashback events into account. The light blue line shows the same simulation but stopped at the redshift shown, $z = 0.2$, without applying a correction. Due to the asymmetry in the final bins, $\rsp$ is underestimated by about 10\%. The green line shows the results after the correction term has been applied, recovering the correct solution on average (the deviations at high $\Gamma$ are not statistically significant or systematic).}
\label{fig:conv}
\end{figure*}

In principle, there is no reason to assume that the splashback radii of individual particles should depend on the number of particles with which a halo is resolved, but one could imagine that the averaging procedure might introduce a bias when only a few $\rrsp$ events are present. We test the effects of mass resolution statistically by considering the \gammarsp in simulations with different particle masses. At fixed mass, each smaller box size corresponds to an eight times higher number of particles (Table~\ref{table:sims}). However, we need to be careful: the halo mass functions in the different simulation boxes are not the same, which could lead to confusion between resolution effects and a potential mass dependence of $\rsp$. We avoid this issue by considering narrow bins in peak height. 

Figure~\ref{fig:conv_mass} shows a comparison between the \gammarsp in different simulations for halos with $N_{\rm 200m} \geq 1000$. Each row corresponds to a bin in peak height and each column to a redshift. Given the uncertainties on the mean relations, we cannot find any statistically significant mass resolution effects. Moreover, the 68\% scatter in the relations (not shown in Figure~\ref{fig:conv_mass}) does not increase appreciably in the lower-resolution simulations, indicating that the lower number of particles in each halo does not introduce a significant random error into the measurements of $\rsp$. Lowering the minimum particle number to $500$ causes $\approx 10\%$ deviations in the mean relations, which persist to particle numbers as low as $200$. While Figure~\ref{fig:conv_mass} shows the relations for the mean $\rsp$, the convergence is equally good for $\msp$ and for the higher percentiles. Only the 99th percentile $\rsp$ of low-$\nu$ halos shows about 10\% deviations between the simulations.

We conclude that the mass resolution limit should depend on the purposes of a given investigation. If 10\% errors in the averaged $\rsp$ are acceptable, even halos with a few hundred particles can be considered. However, at such low numbers, the completeness decreases significantly (Figure~\ref{fig:valid_frac}). Thus, we stick with a limit of $N_{\rm 200m} \geq 1000$ as it guarantees a completeness of about $95\%$ and a convergence of better than $5\%$ with mass resolution.

\subsection{Convergence with Subhalo Mass Ratio}
\label{sec:res:smr}

In Section~\ref{sec:sparta:rsp:selection}, we discussed the importance of dynamical friction in subhalos, and we excluded all particles that belonged to a subhalo with a sub-to-host mass ratio greater than $\smrmax = 0.01$ at infall. The left column of Figure~\ref{fig:conv} justifies this choice by comparing the mean $\rsp$ for halos with $\ntom \geq 1000$ in the TestSim100 for different values of $\smrmax$. Including splashback events from all massive subhalos leads to a severe bias of 20\% (yellow line), and excluding particles from subhalos $1/10$ the size of the host and greater reduces the bias to a few percent (green line). We conservatively choose $\smrmax = 0.01$ (light blue line), which leads to values of $\rsp$ that are statistically indistinguishable from an even stricter cut of $\smrmax = 0.001$. The convergence is equally good for the median and higher percentiles, as well as at all redshifts. 

\subsection{Convergence with Smoothing Timescale}
\label{sec:res:sigmadyn}

There is one free parameter in the \sparta algorithm for which there is no obviously ``correct'' value: the smoothing timescale over which $\rrsp$ events are averaged to compute $\rsp$, $\sigma = \sigmadyn \tdyn$. Increasing $\sigmadyn$ eliminates unphysical changes in $\rsp$ due to shot noise or times when few particle splashback events were recorded, but it can also smooth out physically meaningful features in a halo's $\rsp$ history. Physically, one would expect that the smoothing timescale should be in the range $0 < \sigmadyn < 0.5$ because $\rsp$ reacts to changes in the halo potential in roughly half a dynamical time (the time for a particle to travel from the center to its apocenter). 

We have visually inspected the $\rsp$ histories of a number of halos and found $\sigmadyn = 0.2$ to be a good compromise, smoothing out noise without altering the overall evolution of $\rsp$ and $\msp$. However, while this value may seem to work well for individual halos, we need to confirm that the smoothing does not introduce a systematic bias in $\rsp$. The center column of Figure~\ref{fig:conv} shows the \gammarsp for several values of $\sigmadyn$. For high values, $\sigmadyn \geq 0.5$, there is a small but systematic bias toward lower values, but for our fiducial value of $\sigmadyn = 0.2$ this bias disappears. 

While the bias is insignificant for the mean and median, there is a positive bias at higher percentiles. In particular, the range up to the 85th percentile experiences a bias of 5\% or less at all redshifts, whereas higher percentiles such as the 99th can be biased up to 10\%. We have checked that this is not a resolution effect: the bias persists even for highly resolved halos with $\ntom > 10000$ and is noticeable when inspecting the $\rsp$ histories of individual halos computed with different values of $\sigmadyn$. 

In summary, the fiducial value of $\sigmadyn = 0.2$ leads to unbiased estimates of the mean and median, biases of less than 5\% up to the 85th percentile, and increasingly biased measurements of the highest percentiles. However, reducing $\sigmadyn$ to much smaller values is not physically sensible and leads to significant fluctuations of $\rsp$ with time.

\subsection{Convergence of the Correction at the Final Snapshots}
\label{sec:res:correction}

In Section~\ref{sec:sparta:rsp:correction}, we applied a multiplicative correction term to the averaged $\rsp$ at the last snapshots of a simulation in order to counteract the effects of missing splashback events that would have occurred in the future. The right column of Figure~\ref{fig:conv} compares the results with and without this correction term. In particular, the dark blue line shows the \gammarsp for the mean as derived from the TestSim100 snapshots, where the simulation was run into the future and thus presents the ``correct'' solution to compare to. The light blue line shows the results for the last snapshot of the same simulation, but only run to $z = 0.2$. As expected, $\rsp$ is underestimated by about 10\%. The green line shows the same snapshot but with the correction term applied. 

Comparisons for mass instead of radius, as well as other statistics such as the median or higher percentiles, look similar, with a slight increase of the bias toward the highest percentiles. On average, the correction term reduces the bias in TestSim100 to less than 1.2\% (in both $\rsp$ and $\msp$, in the mean and up to the 85th percentile). The highest percentiles, such as the 99th, can still suffer biases of up to 8\%. Due to the large bin-to-bin noise in the highest percentiles, it is not easy to remove this bias (see Section~\ref{sec:sparta:rsp:correction}). Finally, we note that while the correction term debiases $\rsp$ and $\msp$ on average, it leads to slightly increased scatter around the mean $\rsp$ and $\msp$. Compared to the uncorrected distribution, the standard deviation increases from between $6\%$ and $9\%$ to between $10\%$ and $12\%$ for the mean and up to the 85th percentile, and from about $11\%$ to about $14\%$ for the highest percentiles. 

In summary, the correction recovers the correct $\rsp$ and $\msp$ values at the last snapshots on average, but at the cost of adding a random scatter to the estimates for individual halos. However, the correction has a strong effect only at the very latest snapshots. For example, if a simulation with our fiducial cosmology is run to $z = 0$, the correction term has an effect after $z = 0.28$, and a strong effect only within one $\sigma$, after $z = 0.08$. Thus, uncertainties due to the correction can be avoided by considering a time slightly before the end of the simulation.

\subsection{Dependence on the Maximum Tracer Radius}
\label{sec:res:rdelete}

As explained in Section~\ref{sec:sparta:tcrptl}, we have set the maximum radius to which \sparta tracks particles to $r_{\rm delete} = 3\rtom$. This radius sets an effective maximum $\rrsp$ because we would miss apocentric passages at larger radii. In a TestSim100 run with an extreme value of $r_{\rm delete} = 6\rtom$, \sparta finds additional splashback events in about $4\%$ of halos (depending on redshift and mass resolution), which would indicate a serious bias. The mean $\rsp$ is lower by about 3\% in the run with our fiducial value of $r_{\rm delete} = 3\rtom$, and the highest percentiles can be lower by up to 5\%. However, the median of the $\rsp$ distribution is virtually unaffected, with a bias of less than $0.1\%$. 

Moreover, a visual inspection of the affected halos shows that their high-$\rrsp$ events are not physically meaningful. Virtually all such events are caused by disruption, that is, particles that have left the halo and caused spurious splashback events far away. In a large fraction of the cases, the disruption is due to the tidal forces from a larger host that a halo is about to merge with. We have experimented with excluding all halos that will become subhalos within one dynamical time, but while such a cut does, indeed, remove a significant fraction of the halos for which $r_{\rm delete}$ matters, it also removes many halos that do not suffer from disruption effects whatsoever. Moreover, the cut merely reduces the bias on the mean to $1.5\%$, meaning that disruption due to host halos is only responsible for part of the effect. The rest is presumably due to fly-by events, close encounters where the center of a halo does not enter $\rtom$ of the larger halo. Such events are common and can easily remove particles from the smaller halo, but are not recorded in the halo catalogs. Another argument against a cut on future subhalos is that it cannot be performed toward the end of a simulation because we do not know which halos would merge in the future.

For these reasons, we refrain from cutting out future subhalos and conclude that a finite value of $r_{\rm delete}$ is physically sensible. We caution, however, that the mean $\rsp$ of a sample of halos can be affected by outlier values due to unphysical $\rrsp$ distributions. This is one of the reasons why we focus on the median rather than mean $\rsp$ in \citetalias{diemer_17_rsp}.

\section{Discussion}
\label{sec:discussion}

\begin{figure*}
\centering
\includegraphics[trim = 6mm 2mm 5mm 0mm, clip, scale=0.61]{\figdir/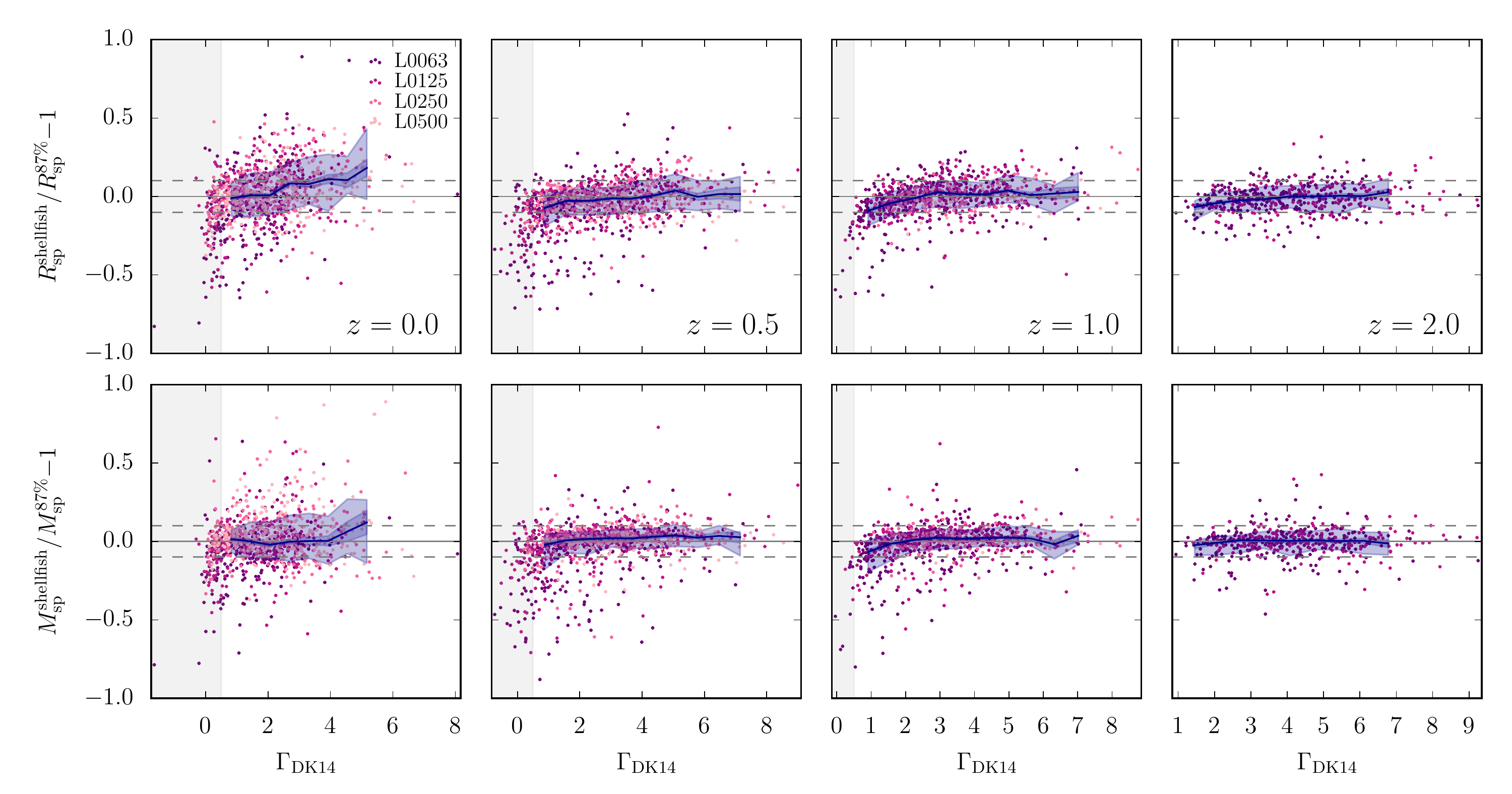}
\caption{Comparison between \shellfish and \sparta estimates of $\rsp$ (top row) and $\msp$ (bottom row), where the \sparta estimate corresponds to the 87th percentile of the $\rrsp$ distribution. The fractional difference between the estimates is shown for individual halos (points colored according to the simulation box) and as a function of $\Gamma$ in order to highlight potential trends with mass accretion rate. The blue line and shaded areas show the median, statistical uncertainty, and 68\% scatter of the distribution. As explained in \citet{mansfield_17}, \shellfish is not expected to give reliable results for slowly accreting halos with $\Gamma < 0.5$ that have been excluded from the median relations (gray shaded areas).}
\label{fig:shellfish}
\end{figure*}

Having demonstrated the numerical convergence of our algorithm, we now discuss topics related to the physical interpretation of our results. We establish a connection with the sharp drop in density associated with $\rsp$ by comparing our results to those from the \shellfish code, and we discuss the impact of the orbital parameters of particles, resolution effects, as well as the significance of extremely low and high accretion rates.

\subsection{Comparison with \shellfish}
\label{sec:discussion:shellfish}

All previous approaches to measuring $\rsp$ have relied not on particle dynamics but on the sharp drop in density associated with the splashback caustic \citep{diemer_14_profiles, adhikari_14, more_15, more_16, mansfield_17}. Moreover, the density drop is observable in the real universe, whereas the splashback radii of DM particles are not. Thus, it is paramount that we establish the connection between our results and those based on the density structure of halos. Here, we compare to the only measurement of $\rsp$ in individual halos undertaken so far, namely the \shellfish algorithm of \citet{mansfield_17}. We restrict ourselves to a halo-by-halo comparison of $\rsp$ and $\msp$, and we leave an analysis of the average \gammarsp for \citetalias{diemer_17_rsp}. The \shellfish algorithm operates on a fundamentally different principle than \sparta: it finds sharp density drops in a large number of random sight lines and derives a (not necessarily spherical) $\rsp$ shell that delineates the drop radii. Unlike \sparta, the algorithm can extract $\rsp$ from a single snapshot alone, but it demands a somewhat higher resolution of \num{50000} particles per halo \citep{mansfield_17}. 

The \shellfish results are most closely approximated by $\rspes$, a relatively high percentile. Figure~\ref{fig:shellfish} shows a comparison of this definition and the \shellfish results for halos that fulfill the \shellfish resolution requirement. At low $\Gamma$, the distribution was subsampled in order to achieve more even coverage of all mass accretion rates. The relative difference in the mean or median is, on average, less than $3.3\%$ at all redshifts, and less than $1\%$ when all redshifts are combined. The 68\% scatter is largest at $z = 0$ (about $18\%$) and decreases to about $10\%$ at higher $z$, with an average of $14\%$ when all redshifts are combined. The increased scatter at $z = 0$ is partially due to \sparta's correction for the final snapshots. 

One important difference between the algorithms is that \sparta (in its current incarnation) works in spherically symmetric coordinates. As expected, \shellfish gives smaller $\rsp$ values than \sparta for the most aspherical halos, which can exhibit major-to-minor axis ratios of up to $2.5$. For such objects, \shellfish infers $\rsp$ values that are up to $\approx 30\%$ lower than those of \sparta. In other words, \shellfish computes a volume-weighted spherical $\rsp$, whereas \sparta's results are mass-weighted. However, the effects of asphericity can only account for a small fraction of the scatter in Figure~\ref{fig:shellfish} because the vast majority of halos are only moderately aspherical, with axis ratios less than $1.5$, where the difference between \shellfish and \sparta becomes negligible on average.

Attempting to establish as close a correlation between the two methods as possible, we have also experimented with definitions that combine two of \sparta's percentile measurements, such as 
\begin{equation}
\rsp^{\rm combined} = \sqrt{\rspmed \times \rspnn} \,.
\end{equation}
This definition was motivated by the finding that $\rspmed$ corresponds to the inner edge of the ``steepening region'' according to \shellfish, that is, the part of the density profile where the slope begins to steepen beyond the values expected from a profile without a splashback feature \citep{mansfield_17}. However, the outer edge of the steepening region is not approximated to sufficient accuracy by any percentile of the splashback distribution, perhaps because of the relatively unreliable determination of the highest percentiles of the apocenter distribution (Section~\ref{sec:res}). Thus, we failed to find a definition that agrees with \shellfish better than $\rspes$ does, and we will use $\rspes$ for further comparisons in \citetalias{diemer_17_rsp}.

\subsection{Dependence on Orbital Parameters}
\label{sec:discussion:orbital}

\begin{figure*}
\centering
\includegraphics[trim = 3mm 5mm 29mm 1mm, clip, scale=0.58]{\figdir/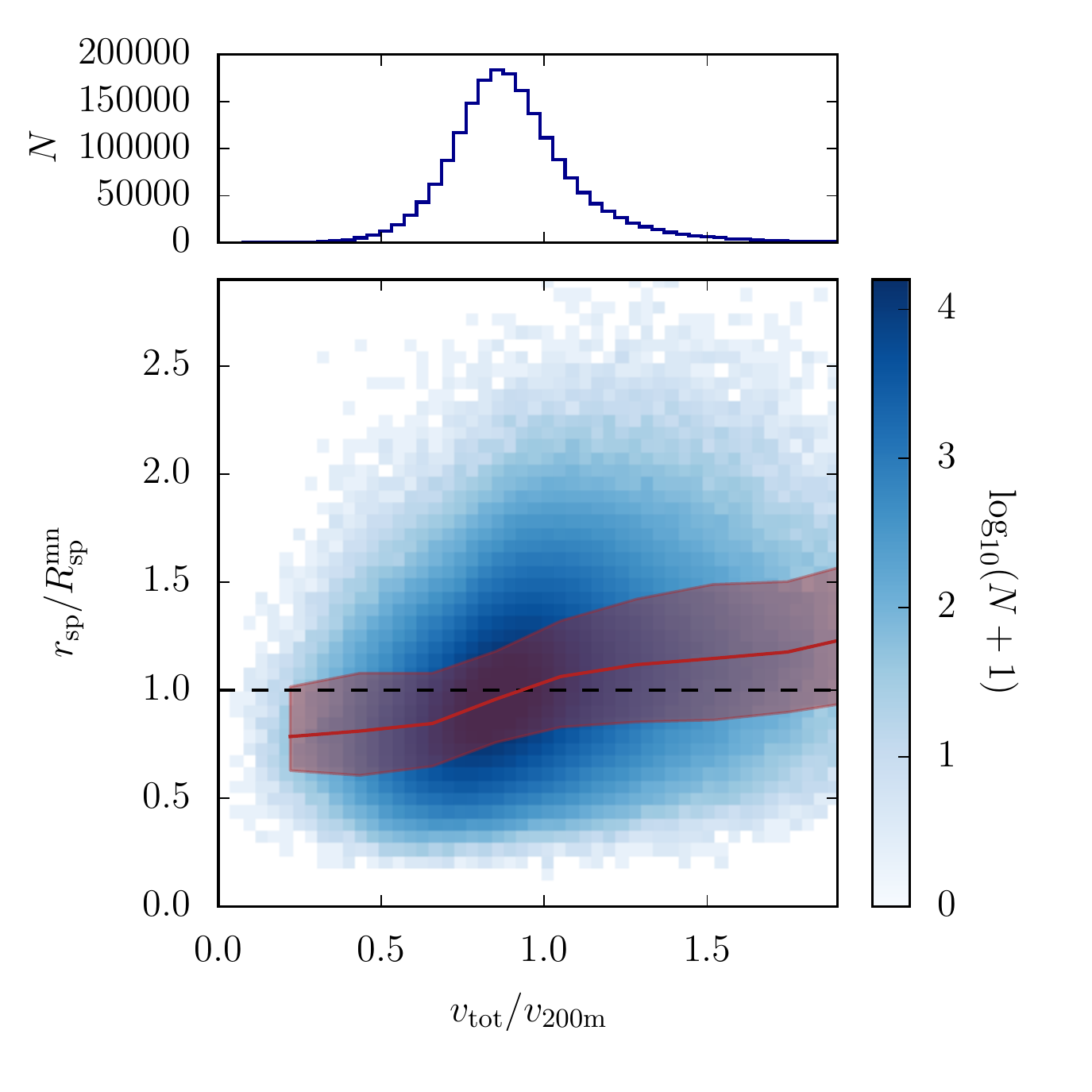}
\includegraphics[trim = 25mm 5mm 29mm 1mm, clip, scale=0.58]{\figdir/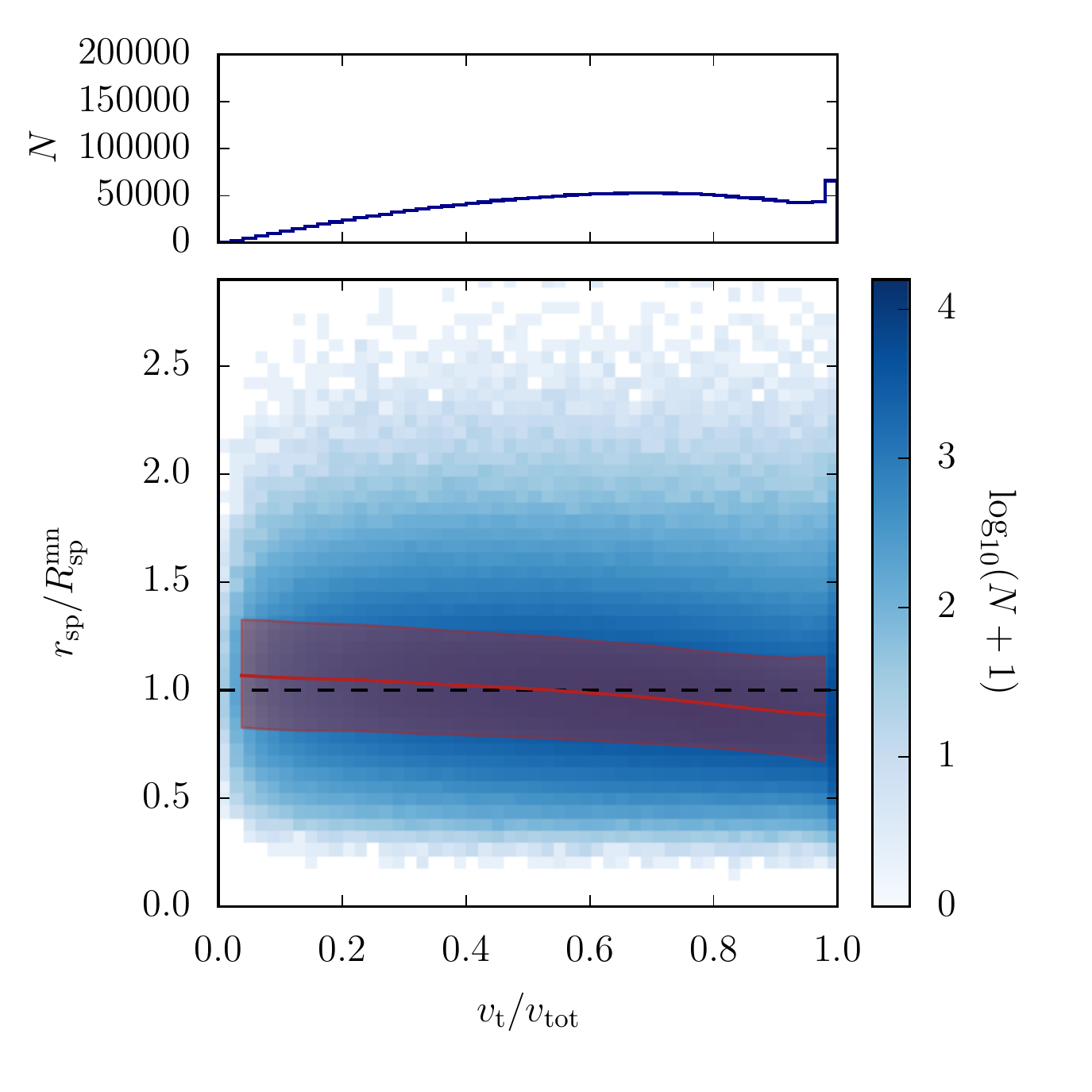}
\includegraphics[trim = 25mm 5mm 8mm 1mm, clip, scale=0.58]{\figdir/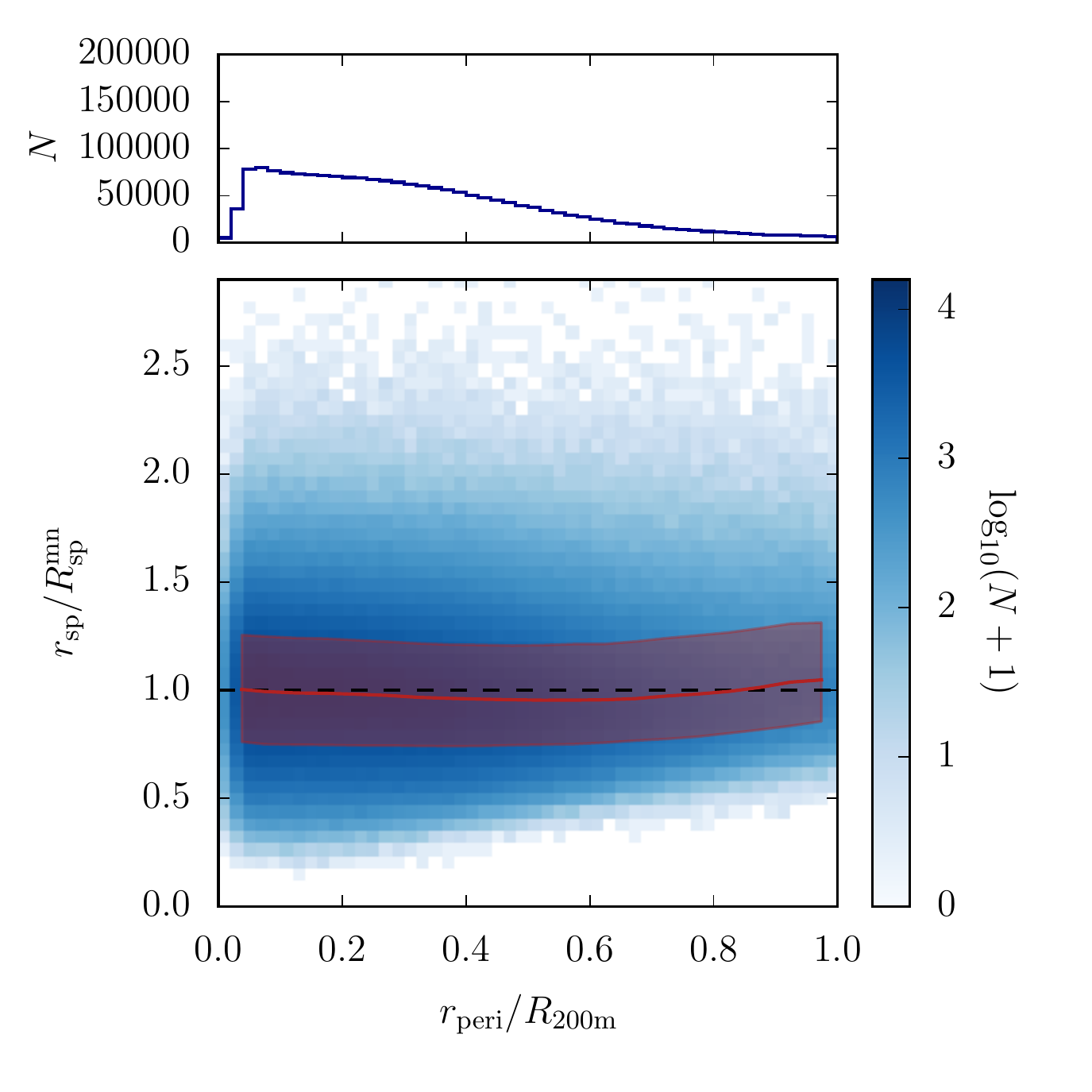}
\caption{Dependence of the splashback radius of individual particles on their orbital parameters. The figure shows $\rrsp$ relative to the halo's $\rspmean$ for all splashbacks in TestSim100 that are used for the $\rsp$ computation (i.e. halos with more than $1000$ particles, SMR $<0.01$). The distribution is shown as a function of the total velocity at infall (left), the circularity of the orbit at infall (center), and the pericentric distance to the halo center during the first orbit (right). The blue histograms show the logarithmic density of particle events in these parameter spaces, while the top panels show the one-dimensional distribution as a function of the orbit parameters. The red line and shaded area show the median and 68\% scatter in the distribution. The most important factor in determining $\rrsp$ is the kinetic energy a particle has at infall (left panels), with slower particles splashing back at smaller radii. The circularity at infall (center panels) is less important in determining $\rrsp$. The pericentric distance (right panels) correlates poorly with the relative $\rrsp$ because it conflates two effects: a very close pericentric approach can be the result of a generally low velocity at infall or of a low tangential component (and vice versa).}
\label{fig:rrm}
\end{figure*}

We have treated particles as a set of unbiased dynamical tracers of the halo potential, but in reality they enter the halo with certain initial conditions, namely a radial and tangential velocity. Figure~\ref{fig:rrm} shows how $\rrsp$ is influenced by the total velocity at infall, the circularity at infall, and the pericentric distance to the halo center. For this figure, we have considered all splashback events in TestSim100 that satisfied the bound on the subhalo mass ratio, that occur in halos with $\ntom \geq 1000$, and for which $\rspmean$ could be computed. We divide $\rrsp$ by the halo's $\rspmean$ at the time of infall in order to scale out the halo's overall radius.

The first impression is that the scatter in the distributions is large. Clearly, the total velocity at infall (or kinetic energy, left panels) has a significant effect on $\rrsp$: as expected, particles with higher energy splash back at larger radii. The circularity (quantified as the fraction of the velocity that is in the tangential direction at infall, center panels) has a smaller effect where more radial orbits lead to slightly higher $\rrsp / \rsp$. This difference may seem surprising at first, since in a spherically symmetric halo we would expect only the kinetic energy to matter. However, at fixed kinetic energy and infall time, particles on more circular orbits will splash back later than their radial counterparts. If $\rsp$ is growing, they will thus be assigned a smaller relative $\rrsp$.

Finally, the right column of Figure~\ref{fig:rrm} shows the impact of the pericentric distance of the first orbit, that is, how close the particle came to the halo center after its first infall. This distribution is cut off at $r_{\rm peri} / \rtom = 1$ as we do not consider trajectories that never entered within $\rtom$. The pericentric radius has a surprisingly small effect on $\rrsp$. One might imagine that $r_{\rm peri}$ would be a good proxy for circularity and thus show a correlation similar to $v_{\rm tot}$. However, $r_{\rm peri}$ is itself positively correlated with both the kinetic energy and circularity at infall. The two effects appear to more or less cancel out so that $\rrsp$ is within 5\% of the mean at all $r_{\rm peri}$.

The cumulative influence of orbital parameters on a halo's $\rsp$ will, of course, be much smaller because the particles in a halo represent a mixture of radial and tangential orbits. We note that the distribution of pericentric radii depends somewhat on time resolution at small radii, which have to be interpolated if the snapshot spacing is coarse (Section~\ref{sec:sparta:ressbk}). Otherwise, the distributions shown in Figure~\ref{fig:rrm} are not significantly different when TestSim200 is used instead of TestSim100.

\subsection{Baryonic and Resolution Effects}
\label{sec:discussion:baryonic}

The results presented in the previous section allow us to assess the potential importance of baryonic effects. After all, the inner density profile and angular momentum distribution can change significantly between DM-only and hydrodynamic simulations \citep[e.g.,][]{velliscig_14, zhu_17_illustrisorbits}. However, the kinetic energy of particles at infall should not be altered much by the presence of baryons, meaning that baryonic effects are unlikely to influence $\rrsp$ significantly (on average; the orbits of individual particles might change dramatically). We plan to test this conjecture directly by running \sparta on cosmological simulations with DM-only and hydrodynamic incarnations, such as Illustris \citep{vogelsberger_14}.

A similar argument applies to some of the well-known resolution effects in $N$-body simulations. The orbits of particles in halos almost invariably suffer from numerical inaccuracies due to the finite mass and force resolutions of simulations, particularly near the halo center, where many particles are clustered within a few force resolution lengths \citep{moore_98, klypin_01, power_03, diemand_04_convergence}. Thus, the exact orbital trajectories of individual particles may not be reliable. As long as energy is conserved, however, their splashback radii should be unaffected by such issues. We note that two-body relaxation should only play a role at early times when halos are resolved by few particles \citep{diemand_04_2body}.

\subsection{Extreme Mass Accretion Rates}
\label{sec:discussion:extremegamma}

We have inspected some halos with extremely low or high mass accretion rates in order to ascertain whether the $\rsp$ measurements in those regimes are trustworthy. We find that negative accretion rates are virtually always caused by a disruption due to mergers, a situation in which $\rtom$ may not be well defined in the first place. While the halo finder may suddenly assign a lower $\rtom$ to a merging pair of halos, the splashback radius does not react in the same way. However, $\rsp$ does not necessarily capture the effects of a merger either, because it would take on the order of a dynamical time for the increased mass to translate into a larger $\rsp$. The splashback mass, however, does ``feel'' a merger instantaneously because it is simply defined as the mass within $\rsp$. Thus, $\msptom$ can reach very large values during mergers, describing a physical reality.

Surprisingly, very high accretion rates also correspond to mergers in most cases, but they indicate a situation where a large subhalo is accreted without disrupting the host sufficiently to lead to a spurious decrease in $\mtom$. In such situations, $\rsptom$ seems to approach a constant, and $\msp$ also tracks $\mtom$ closely. Thus, the flattening of the $\rsptom$ and $\msptom$ relations at high $\Gamma$ appears to be physical.


\section{Summary}
\label{sec:conclusion}

We have described a code framework and specific algorithm to compute the splashback radius of dark matter halos in cosmological $N$-body simulations from particle dynamics. Our main conclusions are as follows:
\begin{enumerate}
\item We have introduced \sparta, a versatile, parallel analysis framework for particle-based simulations that will be described in detail in a future publication.
\item Within this framework, we have presented an algorithm that tracks the orbits of particles and subhalos about their host halo centers. Based on only four stored time bins, \sparta can reliably determine the time and radius of infall, pericenter, and first apocenter (splashback).
\item We find that particles in large subhalos (mass ratio $0.01$ or greater) have lowered splashback radii due to dynamical friction. After excluding such particles, we smooth the splashback distribution in time with a Gaussian filter to obtain the halo's splashback radius, $\rsp$.
\item Our algorithm can determine $\rsp$ for 95\% of all host halos with more than $1000$ particles within $\rtom$, while the completeness drops at lower particle numbers and for the smallest simulation boxes at low redshifts. The remaining halos are backsplash halos that recently became a host halo again.
\item We test the convergence properties of our algorithm with respect to mass resolution, snapshot spacing, and a number of other parameters. We find that the average splashback radii of halos are converged to better than 5\% for halos with at least $1000$ particles if the simulation has about $100$ or more saved snapshots. This convergence extends to roughly the 85th percentile of the particle splashback distribution and degrades somewhat for the highest percentiles.
\item We compare our measurements of $\rsp$ to those from the \shellfish code of \citet{mansfield_17}, and we find the best agreement if the 87th percentile of the apocenter distribution is used to define $\rsp$. The algorithms agree to a few percent on average, with about 15\% halo-by-halo scatter.
\end{enumerate}
We are planning a number of improvements to the \sparta code, as well as further investigations into the physical meaning of the apocenter distribution. For example, it remains to be explored how the \sparta results are related to other methods of measuring caustics in the DM density field, such as the algorithm of \citet{vogelsberger_11_caustics} or ORIGAMI \citep{falck_12, neyrinck_12}. Inspired by the results of \citet{mansfield_17}, we aim to extend our analysis to the distribution of splashback events in angular space to measure three-dimensional splashback surfaces. We intend to apply \sparta to particle-based hydrodynamical simulations such as Illustris \citep{vogelsberger_14} in order to investigate how $\rsp$ correlates with the properties of gas and stars in galaxies and clusters. The \sparta code will eventually become publicly available.


\vspace{0.5cm}

We thank Peter Behroozi for making his halo finder and merger tree code publicly available, for fixing certain issues related to mass definitions for the purpose of this work, and especially for his tree code, which was adapted for \sparta. We are grateful to Andrey Kravtsov, Surhud More, Philip Mansfield, Nicola Amorisco, Andi Burkert, Neal Dalal, Lars Hernquist, and Laura Sales for helpful discussions. We are especially indebted to Surhud More and the Kavli IPMU in Tokyo for their hospitality during a visit when key parts of the \sparta code were designed. B.D. gratefully acknowledges the financial support of an Institute for Theory and Computation Fellowship. This work made extensive use of the \textsc{Midway} computing cluster at the University of Chicago Research Computing Center. 


\bibliographystyle{aasjournal}
\iflocal
\bibliography{../sf.bib}
\else
\bibliography{sf.bib}
\fi

\end{document}